\documentclass[a4paper,11pt]{article}
\usepackage{jheppub} % for details on the use of the package, please see the JINST-author-manual
\usepackage{lineno}
\usepackage{epsfig,amsfonts}
\usepackage{graphicx}

\usepackage{tikz}
\usepackage{mathtools}

\usepackage{physics}
\usepackage{mathrsfs}
\usepackage{amsthm,amssymb}
\usepackage[scr=stixtwoplain]{mathalpha}
\usepackage[most]{tcolorbox}
\usepackage{hhline}
\usepackage{braket}
\usepackage{upgreek}
\usepackage{subcaption}
\usepackage[ruled,vlined]{algorithm2e}
\usepackage[linktocpage=true]{hyperref}
\usepackage[dvipsnames]{xcolor}
\usepackage{musicography}
\usepackage[normalem]{ulem}

\usepackage{tikz}
\usetikzlibrary{positioning,calc,automata}

\usepackage{amssymb}   % or amsfonts
\usepackage{dsfont}    % for \mathds{1}

\tikzset{
  vertex/.style={circle, fill=black, inner sep=1.5pt},
  edge/.style={line width=0.6pt}
}

\definecolor{purple_nice}{rgb}{0.4,0.2,0.7}
\definecolor{fuel_blue}{RGB}{42,162,185}
\definecolor{YInMn_blue}{RGB}{46, 80, 144}
\definecolor{ultramarine}{RGB}{63, 0, 255}
\definecolor{KLEIN_blue}{rgb}{0, 0.18, 0.65}
\hypersetup{
    colorlinks=true,
    linkcolor=ultramarine,
    citecolor=ultramarine,
    filecolor=fuel_blue,
    urlcolor=ultramarine,
}
\usepackage[T1]{fontenc}
\usepackage{amsmath}
\usepackage{mdframed}
\usepackage{graphics}
\usepackage{amsfonts}
\usepackage{amssymb}
\usepackage[multiple]{footmisc}
\usepackage{bm}
\usepackage{amsthm}
\usepackage{slashed}
\usepackage{enumerate,enumitem}
\usepackage{amsbsy}
\usepackage{bbm}
\usepackage{url}
\usepackage{bbm} 
\usepackage{color}
\usepackage{framed}
\usepackage{caption}
\usepackage{float}
\usepackage{cancel}
\usepackage{hyperref}
\usepackage{fnpct}
\usepackage{comment}

\arxivnumber{2511.03791} % if you have one

\title{\boldmath Graded S-Matrices, 
Generalised Gibbs Ensembles and  Fractional-Spin CDD Deformations}

\author[a]{Nicolò Brizio,}

\author[b]{Tommaso Morone,}

\author[b]{Nicolò Primi,}

\author[b]{Roberto Tateo}

\affiliation[a]{Dipartimento di Fisica e Astronomia, Università degli Studi di Padova and \\ INFN, Sezione di Padova, Via Marzolo 8, 35131,
Padova, Italy}

\affiliation[b]{Dipartimento di Fisica, Università di Torino and \\ INFN, Sezione di Torino, Via P. Giuria 1, 10125, Torino, Italy}

\emailAdd{nicolo.brizio@unipd.it}

\emailAdd{tommaso.morone@unito.it}

\emailAdd{nicolo.primi@unito.it}

\emailAdd{roberto.tateo@unito.it}

\abstract{We introduce and study a class of two-dimensional integrable quantum field theories that carry an internal $\mathbb{Z}_n$ structure. These models extend factorised scattering beyond the conventional framework, featuring both the usual hierarchy of integer-spin conserved charges and an additional tower of fractional-spin ones.
Our construction relies on a reparametrisation of rapidity space that lifts standard scattering amplitudes to a multiplet related by an internal cyclic symmetry.  This construction is naturally embedded within a generalised Gibbs ensemble, which provides the natural framework for a consistent graded Thermodynamic Bethe Ansatz. This leads to new Y-systems encoding the graded spectrum. In a special case, these functional relations match those obtained via the ODE/IM correspondence from the monodromy analysis of the quantum cubic oscillator. Even in the simplest models, for one sign of the auxiliary temperature, the finite-volume ground-state energy spectrum undergoes an infinite sequence of level crossings as the coupling strength increases. A preliminary analysis also suggests that these theories exhibit structural connections with cyclic orbifolds. Within this setup, one can consistently include extra CDD factors that realise fractional-spin analogues of the $T \overline{T}$ deformation. In analytically tractable cases, a Hagedorn-like behaviour is observed for a sign of the flow parameter, and the deformed spectrum develops a finite limiting temperature.}

\begin{document}
\maketitle
\flushbottom

\section{Introduction and motivations}
There are two complementary ways to think about \textit{Quantum Field Theory} (QFT).
In one approach, one begins by writing down a Lagrangian -- often guided by symmetry and the hope that nature favours simplicity -- and builds upward using perturbation theory or numerical methods to extract physical predictions. In the other, one starts from a set of minimal assumptions and asks what kinds of theories could exist if all we demand is overall consistency. This second perspective, introduced by W. Heisenberg \cite{Heisenberg1943a, Heisenberg1943b, Heisenberg1944} and later refined by G. Chew \cite{Chew1961Book}, laid the foundations for what became known as the S-matrix \textit{bootstrap program}. At its heart was the idea that a few simple principles -- unitarity, analyticity, and crossing symmetry -- could already fix the structure of all particle interactions. In the 1960s, when new colliders revealed an unexpected outbreak of hadronic states that defied any simple hierarchy, this view evolved into the idea of \textit{nuclear democracy}: if the same consistency conditions apply to all particles, then none should be considered more fundamental than another \cite{ChewFrautschi1961}.

The original S-matrix bootstrap was eventually overshadowed by the rise of \textit{Quantum Chromodynamics} (QCD) \cite{GrossWilczek1973, Politzer1973}, but its overall philosophy -- that consistency and symmetry might replace microscopic dynamics -- never disappeared (see \cite{Kruczenski:2022lot} for a modern perspective). During the same period, the search for a consistent scattering theory led to the discovery of the Veneziano amplitude \cite{Veneziano1968}, whose analytic structure captured many features of hadronic physics and hinted at what would eventually become string theory. In two dimensions, however, the bootstrap program found an exact and lasting form. Certain quantum field theories were discovered to possess infinitely many conserved quantities, resulting in so many constraints that every scattering process would break down into a sequence of two-body collisions \cite{ZZ1979AnnPhys, Iagolnitzer1978}. In these theories, known as \textit{Integrable Quantum Field Theories} (IQFTs), scattering events are purely elastic \cite{Parke1980}, and the whole theory is determined by the two-body S-matrix. Once the latter is known, the finite-size spectrum can be accessed via \textit{Thermodynamic Bethe Ansatz} (TBA) equations \cite{Zamolodchikov1990TBA, Martins:1991hw, Fendley:1991xn, Dorey:1996re, Bazhanov:1996aq, Fioravanti:1996rz}.

\subsection{Why graded QFTs? Hints from ODEs}

Scattering data in \textit{Integrable Models} (IMs) can often be organised into a set of functional relations known as the Y-system, which captures the analytic properties of the finite-volume spectrum \cite{Zamolodchikov1991ADE, Zamolodchikov1991RSOS, Klumper:1992vt} -- see also \cite{Kuniba:2010ir} for a collection of short reviews. A similar kind of structure appears in a remarkably different context: the study of \textit{Ordinary Differential Equations} (ODEs) with polynomial potentials \cite{sibuya1975global}. As one moves around infinity, the asymptotic behaviour of a single solution undergoes a series of transitions between distinct angular sectors. The corresponding connection data, or \textit{monodromy}, provide a compact representation of this global analytic structure. Instead of following each local expansion separately, one may describe the entire solution space using a finite set of Stokes multipliers and connection coefficients, specifying how the local solution bases are glued together across the complex plane. The relations constraining the monodromy data in ODEs mirror those organising the scattering and spectral data in IQFTs \cite{Bazhanov:1994ft, Bazhanov:1996dr, Bazhanov:1998dq}. In both contexts, local analytic information is tied together by global consistency, and the dynamics can be expressed as a set of functional equations. The realisation in the late 1990s that these two frameworks are in fact connected marked the beginning of what is now known as the ODE/IM correspondence, a bridge between integrable models and the analytic theory of differential equations \cite{Dorey:1998pt, Bazhanov:1998wj, Bazhanov:1998dq, Dorey:1999uk} -- see \cite{Dorey:2007zx} for a pedagogical introduction, as well as \cite{Ito:2020htm, Fioravanti:2021bzq, Ito:2025pfo} for recent developments in the field.

A clear example of this correspondence comes from the cubic Schrödinger oscillator. Its integrable counterpart is the scaling Lee–Yang model, the simplest non-trivial minimal model of two-dimensional conformal field theory perturbed by its unique relevant operator. This model is integrable, admits an exact S-matrix \cite{Cardy:1989fw, Smirnov:1989hh}, and the ground-state energy in finite volume is determined by a single non-linear integral equation derived from Zamolodchikov’s TBA \cite{Zamolodchikov1990TBA}. Excited states can be accessed via analytic continuation of the TBA equations~\cite{Bender:1969si, Dorey:1996re}. The TBA equation can be recast into a Y-system, which consists of one functional relation,
\begin{equation}\label{LY_Y}
    Y(\vartheta^+) Y(\vartheta^-) = 1+ Y(\vartheta)\,,
\end{equation}
where $\vartheta$ is a rapidity variable that parametrises particle momenta, and $\vartheta^\pm = \vartheta \pm i\pi/3$. The same functional relation governs the monodromy of the cubic oscillator, with $\vartheta$ now related to the spectral parameter of the ODE \cite{10.1007/BFb0060056}. In this way, the analytic continuation of asymptotic solutions in a one-dimensional quantum-mechanical problem mirrors the consistency conditions on energy levels in a two-dimensional interacting QFT. What initially appeared to be a curious coincidence turned out to be a deeper structural equivalence between two \textit{a priori} independent frameworks, with the cubic oscillator providing the minimal setting for studying this correspondence.

Beyond its physical relevance, equation \eqref{LY_Y} -- as well as its generalisations -- possesses an intrinsic mathematical beauty. Once the functional relation is imposed, the functions $Y(\vartheta)$ rearrange themselves into rigid algebraic patterns, often with hidden periodicities. In the book ``\textit{Mathematicians: An Outer View of the Inner World}'', D. Zagier illustrates this behaviour with a charming example \cite{Cook2009-al}: 

\begin{quote}
    ``Imagine you have a series of numbers such that if you add $1$ to any number, you get the product of its left and right neighbors. Then this series will repeat itself at every fifth step! For instance, if you start with $3$, $4$, then the sequence continues: $3$, $4$, $5/3$, $2/3$, $1$, $3$, $4$, $5/3$, etc. The difference between a mathematician and a non-mathematician is not just being able to discover something like this, but to care about it and to be curious why it’s true, what it means, and what other things in mathematics it might be connected with.''
\end{quote}

\noindent \textbf{Deformed oscillators and cyclic symmetries}. The cubic potential exhibits a high degree of symmetry: its Stokes diagram is composed of five equally spaced sectors with an opening angle of $2\pi/5$, and the associated monodromy data reduce to a minimal set of independent constants. Owing to this regularity, the functional relations collapse to a single non-linear integral equation, which can be identified with the spectral problem of the scaling Lee–Yang model. When the potential is perturbed by a subleading linear term, the analysis becomes significantly more involved. The cyclic symmetry is broken, and the global monodromy must now be described by a network of constraints. The natural question is whether this network has any structure at all or degenerates into a complicated list of coefficients. The structure underlying this more general case was clarified by Masoero \cite{Masoero:2010is}, motivated by its relation to the first Painlevé equation and, in particular, to the pole structure of the tritronquée solution \cite{Masoero_2010, Masoero:2010um}. Appendix \ref{masoero_appendix} contains a brief overview of the results presented in \cite{Masoero:2010is}. The strategy adopted in his work was to describe the monodromy data in intrinsic geometric terms, thereby avoiding the ambiguities associated with a particular choice of basis in the space of solutions to the ODE. The key idea is to consider asymptotic values: by following a normalised ratio of solutions along each Stokes sector, one obtains five limiting values arranged cyclically at infinity. These values, permuted by the residual $\mathbb{Z}_5$ symmetry of the leading cubic term, provide a natural geometric dataset for the problem in terms of points on the Riemann sphere. From these asymptotic values, one can construct projective invariants -- specifically, cross-ratios -- which capture the monodromy information in a basis-independent manner. Expressed in terms of these invariants, the problem displays a hidden order: under deformations of the potential, the invariants evolve through a cyclic set of functional relations, 
\begin{equation}\label{deformed_Y}
    Y_{k+1}(\vartheta^+)Y_{k-1}(\vartheta^-) = 1+ Y_k(\vartheta)\,,
\end{equation}
with $\vartheta^\pm$ defined as in equation \eqref{LY_Y}, and $k$ taking values in $\mathbb{Z}_5$. Once analyticity and asymptotics are imposed, this system is equivalent to a set of five coupled nonlinear integral equations, referred to as the deformed TBA in \cite{Masoero:2010is}, which encodes the global monodromy of the anharmonic cubic oscillator.

The appearance of a deformed Y-system already hints at a scattering-theoretic interpretation. In integrable field theories, relations of this type are not arbitrary: they follow from the analytic properties of the S-matrix and from the constraints imposed by the infinite tower of conserved quantities. It is therefore natural to expect that the geometric structure uncovered in the ODE context can be reproduced directly from the scattering description. In the rest of this paper, we take up this task. One of the central questions guiding the present work was to understand the relationship between the five-component system \eqref{deformed_Y} and the single-component equation \eqref{LY_Y}. The connection turns out to be remarkably simple. Introducing the conformal parametrisation
\begin{equation}\label{conf_1}
     f_k(\vartheta) = -\frac{\vartheta}{5}+ \frac{2 \pi i k}{5}\,,
\end{equation}
which unwraps the underlying $\mathbb{Z}_5$ symmetry into five sheets of the spectral plane, a single analytic function $Y(\vartheta)$ satisfying \eqref{LY_Y} can be uplifted to the deformed system \eqref{deformed_Y} by setting $Y_k(\vartheta) =  Y(f_k(\vartheta))$.
The consistency of this construction follows from the way shifts in the 
$\vartheta$-variables are mapped under the reparameterisation. From this viewpoint, the deformed Y-system admits two complementary interpretations: geometrically, as a system of relations among cross-ratios of asymptotic values in the monodromy problem; and algebraically, as the pullback of each Y-function along the branches of a conformal covering of the rapidity plane. Already in the scaling Lee–Yang model, one can see that the consistency of the Y-system is not restricted to the specific five-fold covering we just described, and the same functional structure is preserved under a wider class of conformal reparametrisations of the form
 $f_k(\vartheta) = \pm (\vartheta/n -2\pi i w k/n)$, where the positive integer $w$ and the overall sign specify the covering map, and the rank $n$ of the system is fixed by the relation $n = 6w\pm 1$.  The robustness of the Y-system under such reparametrisations indicates that the mechanism at play is not an accident of the scaling Lee–Yang model, but rather a reflection of a more general compatibility between conformal coverings of the spectral plane and
the functional relations underlying integrable quantum field theories.

Our guiding principle is therefore that the deformed Y-system and its corresponding TBA do not constitute peculiar features of the cubic oscillator, but rather they encode the thermodynamic properties of a broader class of models -- that we call \textit{Graded Integrable Quantum Field Theories} -- where additional discrete symmetries merge with the standard integrable structure. Two ingredients are essential to make this correspondence precise. The first is the use of conformal maps in complex rapidity space: the leading growth of the potential enforces a residual cyclic order, and Masoero's equations reflect this $\mathbb{Z}_5$ structure. In scattering theory, a similar cyclic organisation arises when one partitions the rapidity plane into angular sectors related by discrete rotations. Via reparametrisation, one can make this cyclic grading explicit: rapidities are mapped into $n$ congruent sectors, and the kernels and pseudoenergies inherit the corresponding $\mathbb{Z}_n$ structure. The second ingredient, essential in the TBA construction, is the framework of \textit{Generalised Gibbs Ensembles} (GGEs) \cite{Rigol_2007, Mossel:2012vp}. In an integrable model, there are infinitely many conserved charges, and thermal ensembles can sometimes be too restrictive. The GGE formalism extends the density matrix by introducing generalised temperatures coupled to each conserved charge. This provides a natural way to encode deformations: what appear in the ODE language as parameters tilting the cubic potential can, from the scattering perspective, be understood as turning on new terms in the GGE ensemble, corresponding to modifying the source terms in the TBA equations. 
\subsection{Summary of the main results}
This work introduces and explores a new class of $\mathbb{Z}_n$-graded integrable quantum field theories, where the usual analytic and algebraic structures of two-dimensional integrable QFTs are enriched by an internal cyclic symmetry. The grading leads to new functional relations, generalised TBA equations, and distinctive thermodynamic behaviours.

In Section \ref{Majorana:SEC}, we introduce graded free theories, which serve as exactly solvable benchmarks. Starting from the Ising field theory, we construct GGEs that incorporate higher-spin conserved charges. Within this framework, the $\mathbb{Z}_n$ structure emerges naturally through a simple reparametrisation of rapidity space. These models provide a controlled setting to test analytic continuations, excited-state quantisation conditions, and level-crossing phenomena. We also discuss how this framework automatically accommodates generalised $T\overline{T}$-type flows generated by charges carrying fractional Lorentz spin, and study how these deformations affect the energy spectrum.

In Section \ref{SEC:ADE}, we review integrable S-matrices in $1+1$ dimensions, and recall how a large family of reflectionless scattering theories can be classified in terms of Lie-algebraic data, leading to the so-called ADET classification. We conclude this section by presenting a simple but quite striking self-factorisation property: each S-matrix can be written as a product of shifted and rescaled replicas of itself.

In Section \ref{SEC:SMAT}, we extend the graded construction to interacting theories from the S-matrix perspective. We analyse internal consistency conditions, obtain graded bootstrap equations, and show that -- beyond the usual tower of infinite tower of local, conserved charges -- graded integrable QFTs support an additional, infinite set of integrals of motion with fractional Lorentz spin.

In Section \ref{SEC:TBA}, we introduce a graded Thermodynamic Bethe Ansatz, which combines the geometric reparametrization of rapidity space with the GGE formalism. This approach yields a family of coupled integral equations that encapsulate the thermodynamics of graded models. The resulting structure generalises the usual Y-systems, yielding new functional identities that capture the interplay between grading and integrability. We then apply this formalism to the graded Lee–Yang model, deriving explicit graded TBA equations and analysing their ground-state scaling functions.

In Section \ref{SEC:TWIST}, we explore the connection between graded theories, chemical potentials, and twisted sectors. We propose that graded models can be interpreted as cyclic orbifolds of the parent theory. This correspondence naturally extends to interacting cases, suggesting a unifying picture that links graded integrable field theories, orbifold CFTs, and deformations of the ODE/IM correspondence.
\section{An exactly solvable benchmark: graded free theories}\label{Majorana:SEC}
We consider an ensemble of finitely many identical particles on a torus of periods $(L,\beta)$. Taking the side of length $L$ as the space direction, each particle can be parametrised by a rapidity variable $\vartheta$, and the corresponding energies and momenta take the standard relativistic form:
\begin{equation}
    (E,p) = m (\cosh\vartheta, \sinh \vartheta)\,.
\end{equation}
Being space periodic, physical momenta are quantised in units of $2\pi/L$. In the thermodynamic limit, where both the particle number and the system size $L$ are taken large at fixed density, the equilibrium properties of the system are described by the Gibbs ensemble. The corresponding thermal state is encoded in the density matrix
\begin{equation}\label{gibbs}
    \varrho(\beta,L) \propto e^{-\beta H(L)}\,,
\end{equation}
where $H(L)$ is the Hamiltonian of the theory, quantised along a periodic spatial slice of size $L$, and $\beta$ is the inverse temperature. The correct normalisation for the density matrix, $\operatorname{Tr} \varrho(\beta, L)=1$, is fixed dividing the right-hand side of \eqref{gibbs} by the partition function:
\begin{equation}
    \mathscr{Z}(\beta,L)= \Tr e^{-\beta H(L)}\,.
\end{equation}
It is a standard result that the thermal expectation values of a quantum system at inverse temperature $\beta$ can be reinterpreted as correlation functions of a Euclidean field theory compactified on a spatial circle of circumference $\beta$, evolving under the Hamiltonian $H(\beta)$ along a periodic time direction of length $L$. In the large $L$-limit, this expression is dominated by the ground state of $H(\beta)$, with Casimir energy $E(\beta)$. Close to criticality, where the particle mass $m$ vanishes, the system flows to a conformal fixed point, and the exact Casimir energy is obtained via \textit{Conformal Field Theory} (CFT) arguments. In particular, the ground-state energy on a circle of length $\beta$ behaves as \cite{DiFrancesco:1997nk}
\begin{equation}\label{ceff}
    E(\beta) = -\frac{\pi c^{\text{UV}}_{\text{eff}}}{6\beta}\,,
\end{equation}
where $c^{\text{UV}}_{\text{eff}}$ denotes the effective central charge of the theory. For unitary models with periodic boundary conditions (and anti-periodic ones for fermions), $c^{\text{UV}}_{\text{eff}} = c$  coincides with the Virasoro central charge of the underlying conformal field theory. More generally, in non-unitary or twisted sectors, the effective central charge is shifted according to $c^{\text{UV}}_{\text{eff}} = c-24 \Delta$, where $\Delta$ is the (total) scaling dimension of the operator that creates the lowest-energy state above the vacuum. This dual interpretation of the torus partition function lies at the heart of the \textit{Thermodynamic Bethe Ansatz} (TBA), which provides a powerful non-perturbative framework for computing exact quantities in integrable quantum field theories, including the ground-state energy and, via analytic continuation, the excited-state spectra. This framework will be introduced in Section \ref{SEC:TBA}. While the TBA formalism is completely general, its technical machinery can sometimes obscure simple, universal structures that appear more transparently in free theories. For this reason, we will begin our analysis from the Ising model: a free, exactly solvable theory that also easily accommodates an internal grading.
\subsection{Generalised temperatures in the Ising model}
The Ising model, in its various formulations, has long been a cornerstone of the study of critical phenomena and exactly solvable systems. On a lattice, it describes spins arranged on a two-dimensional grid, each taking values $\pm 1$, with nearest-neighbour interactions that favour alignment. At its critical point, the model exhibits a second-order phase transition: long-range order vanishes, and the large-scale behaviour is captured by a unitary conformal field theory with central charge $c=1/2$. Moving away from criticality introduces a finite correlation length. In the continuum description, this corresponds to perturbing the critical Ising CFT by its most relevant operator -- the energy density. The resulting theory is a massive, relativistic quantum field theory, equivalent to that of a single free Majorana fermion with mass $m$ \cite{Kaufman:1949ks, Schultz:1964fv, Pfeuty:1970qrn}. For a free fermionic theory compactified on a spatial circle of circumference $L$, the Hilbert space naturally factorises into independent Fock spaces, one for each momentum mode. Periodic boundary conditions enforce momentum quantisation, so that each mode is labelled by an integer $j \in \mathbb{Z}$:
\begin{equation}
    p_j = m \sinh \vartheta_j = {2\pi j}/{L}\,,\quad j \in \mathbb{Z}\,.
\end{equation}
Each quantised mode behaves as a fermionic oscillator, subject to the Pauli exclusion principle: it can either be unoccupied, contributing zero energy, or occupied exactly once, contributing energy $E_j=m\cosh\vartheta_j$. Because different modes are independent, the full partition function factorises into a product over all momentum levels:
\begin{equation}
    \mathscr{Z}(\beta,L,m) =\prod_{j}\big(1+e^{-\varepsilon(\vartheta_j)}\big)\,,\quad \varepsilon(\vartheta_j) = m\beta \cosh\vartheta_j\,.
\end{equation}
In large volume, the allowed momenta become densely spaced, and correspondingly, the rapidities form a continuum. The density of states in rapidity space follows from $\dd p = m \cosh\vartheta \,\dd \vartheta$, and the discrete product over rapidities can be replaced by an exponential of an integral. The free energy density of the system follows as:
\begin{equation}\label{free_en}
    f(\beta,m) = -\frac{1}{\beta L}\log \mathscr{Z}(\beta,L,m) = -\frac{m}{2\pi\beta}\int_\mathbb{R}\dd\vartheta \, \cosh\vartheta \,\log \big(1+e^{-\varepsilon(\vartheta)}\big)\,.
\end{equation}
While $f(\beta,m)$ is the natural thermodynamic observable in standard statistical mechanics, in the TBA framework it is customary to work instead with a dimensionless quantity, the \emph{ground-state scaling function} (cf. \cite{Zamolodchikov:1989cf}) (or, equivalently, with the so-called \emph{effective central charge}), which makes the dependence on physical scales more transparent. Since the theory possesses a single mass scale $m$, all finite-size and finite-temperature effects in the scaling functions must enter through the combination $r = m \beta$; the ground-state physics therefore depends only on this scaling variable, rather than separately on $m$ and $\beta$. After exchanging the two cycles of the torus, so that $\beta$ is interpreted as the spatial circumference, the Casimir energy of the Hamiltonian $H(\beta)$ is related to the free energy of the theory by
\begin{equation}
E(\beta,m) = \beta\, f(\beta,m)\,.
\end{equation}
In full analogy with \eqref{ceff}, we thus define the ground-state effective central charge as
\begin{equation}\label{en_gr}
    c_\text{eff}(r) =-\frac{6 \beta}{\pi} E(\beta, m)\,.
\end{equation}
This function interpolates between two universal regimes: in the \textit{ultraviolet} (UV) limit, when $r \ll 1$, it reproduces the effective central charge of the underlying conformal field theory; in the opposite \textit{infrared} (IR) regime, with $r \gg 1$, the scaling function vanishes, reflecting the trivial gapped spectrum. 
In other words, $c_{\text {eff }}(r)$ plays the role of a non-perturbative "renormalisation group" flow function, tracking how degrees of freedom decouple across different scales \cite{Zamolodchikov:1986gt}. Combining equations \eqref{free_en} and \eqref{en_gr}, we obtain the expression
\begin{equation}\label{c_e_is}
     c_\text{eff}(r) = \frac{3r}{\pi^2} \int_\mathbb{R}\dd\vartheta  \cosh\vartheta \log \big(1+e^{-\varepsilon(\vartheta)}\big)\,.
\end{equation}
The functions $\varepsilon(\vartheta)$, commonly referred to as \textit{pseudoenergies} in the TBA literature, provide a convenient parametrisation of the statistical weights in the ensemble. It is also customary to introduce the auxiliary Y-function $Y(\vartheta) = e^{\varepsilon(\vartheta)}$. In the present free-fermion setting, it satisfies a simple functional relation: using $\cosh (\vartheta \pm i \pi / 2)= \pm i \sinh \vartheta$, one verifies that
\begin{equation}\label{Ysys_ISING}
Y(\vartheta+i \pi / 2) \; Y(\vartheta-i \pi / 2)=1\,,
\end{equation}
which represents the simplest instance of a Y-system \cite{Zamolodchikov:1991et}. Moreover, one finds that the following periodicity condition is satisfied: $Y(\vartheta + i \pi P) = Y(\vartheta)$ with $ P =2$. Despite its apparent simplicity, this property encodes meaningful physical information, and it reflects that the corresponding perturbation of the Ising conformal field theory is driven by an operator of conformal dimension $\Delta = 1-1/P = 1/2$. \\

\noindent\textbf{Higher-spin charges and GGEs}. So far, equation \eqref{free_en} yielded a complete characterisation of the thermal Gibbs ensemble, where the statistical weight of each state is determined solely by its energy through the Hamiltonian. The special role of the Hamiltonian in ordinary statistical mechanics is a direct consequence of the \textit{principle of maximum entropy}: the correct equilibrium ensemble must maximise entropy while respecting all exact conservation laws of the system. In generic QFTs, the only extensive conserved quantity of genuine relevance is the Hamiltonian itself. In such a situation, the above principle singles out the Gibbs ensemble uniquely: there is no other consistent statistical distribution that is compatible with the known conservation laws.  The Ising model, however, provides a fundamentally different scenario. Being integrable, it possesses an infinite tower of conserved charges  $Q_{s}^\pm(L)$, which can be classified by their Lorentz spin $s \in \mathcal{S}$. In a general integrable QFT, the set of conserved spins $\mathcal{S}$ depends on the model under consideration. For free Majorana fermions, one finds $\mathcal{S} = 2\mathbb{N}+1$. We define the auxiliary charges
\begin{equation}\label{H_s(l)}
    H_s(L)=\frac{1}{2}\left(Q_s^+(L) + Q_{s}^-(L)\right)\,,\quad P_s (L)=\frac{1}{2}\left(Q_s^+(L) - Q_{s}^-(L)\right)\,.
\end{equation}
The lowest ones (for $s=1$) reproduce the familiar energy and momentum operators, whereas their higher-spin analogues are constructed as spatial integrals of local fermionic operators. Importantly, they are all mutually commuting, and thus every many-body state in the theory can be simultaneously labelled by their eigenvalues. In particular, one defines the charges $Q_s^\pm(L)$ so that their action is diagonal: a single fermion of rapidity $\vartheta$ carries an eigenvalue $q^\pm_s(\vartheta)\propto m^s e^{\pm s\vartheta}$, and multi-particle eigenvalues follow additively. The presence of these higher-spin charges in the theory implies that one can consistently generalise the statistical ensemble beyond the standard case. Just as the Gibbs ensemble maximises entropy subject to energy conservation, one can introduce new ensembles that also fix the values of these additional charges. This leads naturally to the \textit{Generalised Gibbs Ensemble} (GGE) \cite{Rigol_2007, Mossel:2012vp}, defined by the density matrix:
\begin{equation}\label{gge}
    \varrho(\{\beta_s\},L) \propto \exp \Big(\!-\sum_{s \in \mathcal{S}} \beta_s H_s(L)\Big)\,. 
\end{equation}
Here, the Lagrange multipliers $\beta_s$ play the role of generalised temperatures. In particular, $\beta_1 = \beta$ corresponds to the usual thermodynamic inverse temperature, conjugate to the Hamiltonian, while the set $\{\beta_s\}_{s\geq 1}$ governs the statistical weight of the higher conserved charges. The definition \eqref{gge} is conventionally restricted to symmetric combinations of $Q_s^\pm(L)$. This is directly analogous to the standard Gibbs case: even though translation invariance ensures that total momentum is conserved, one does not usually introduce such a term in the density matrix, as it would describe a boosted thermal state. The same logic applies to the higher-spin charges. Under parity, the charges $Q^\pm_s$ are exchanged, so that any antisymmetric combination of the two must result in directed currents. Turning on such couplings produces current-carrying stationary states, rather than parity-invariant states at equilibrium. Finally, normalisation of the density matrix is ensured by the generalised partition function $\mathscr{Z}(\{\beta_s\},L)$, from which one defines the free energy density
\begin{equation}
    f(\{\beta_s\}) = -\frac{1}{\beta L}\log \mathscr{Z}(\{\beta_s\},L) = -\frac{m}{2\pi\beta}\int_\mathbb{R}\dd\vartheta  \cosh\vartheta \log \big(1+e^{-\varepsilon(\vartheta)}\big)\,,
\end{equation}
where we introduced the pseudoenergies $\varepsilon(\vartheta)$ such that, acting on a single-particle state $\ket{\vartheta}$, one has
\begin{equation}\label{gge_pseudo}
    \sum_{s \in \mathcal{S}}  \beta_s H_s(L)\ket{\vartheta} \propto  \sum_{s \in \mathcal{S}}  \beta_s (q_s^+(\vartheta)+q_s^-(\vartheta))\ket{\vartheta}=\varepsilon(\vartheta)\ket{\vartheta} \,.
\end{equation}
We observe that the one-particle eigenvalues of the conserved charges scale as $q^\pm_s(\vartheta) = \mathcal{A}_{s} m^s e^{\pm s\vartheta}$, and that the proportionality constants $\mathcal{A}_{s}$ can be absorbed into a redefinition of the generalised inverse temperatures. To easily probe the theory at different regimes, we introduce the parameter $r=m \beta$, together with the dimensionless ratios $\gamma_s = \beta_s/\beta^s$, so that the pseudoenergies \eqref{gge_pseudo} can be written as:
\begin{equation}
      \varepsilon(\vartheta) = \sum_{s \in \mathcal{S}}\gamma_s r^s \cosh(s\vartheta)\,.
\end{equation}
In this form, $r$ controls the overall scale, while the ratios $\gamma_s$ specify the relative strength of higher-spin contributions in the ensemble.
As in the Gibbs ensemble, the torus partition function admits an alternative interpretation upon exchanging the roles of $L$ and $\beta$. In the large-volume limit, the free energy density encodes the Casimir energy of the Hamiltonian $H(\beta)$, leading to the ground-state energy $E(\{\beta_s\}) = \beta f(\{\beta_s\})$. Likewise, the ground-state scaling function of the system can be expressed in terms of the dimensionless parameters $(r, \{\gamma_s\})$ as:
\begin{equation}\label{cc_gge}
     c(r, \{\gamma_s\}) = \frac{3r}{\pi^2} \int _\mathbb{R}\dd\vartheta  \cosh{\vartheta} \log \big(1+e^{-\varepsilon(\vartheta)}\big)\,.
\end{equation}
Finally, we observe that the functional relation \eqref{Ysys_ISING} characteristic of free fermions persists when generalised temperatures are introduced. Indeed, for each odd spin $s$, one has the identity:
\begin{equation}
    \cosh (s(\vartheta+i \pi / 2))+\cosh (s(\vartheta-i \pi / 2))=2 \cosh (s \vartheta) \cos \left({s \pi}/{2}\right)=0\,,
\end{equation}
which ensures $\varepsilon(\vartheta+i \pi / 2)+\varepsilon(\vartheta-i \pi / 2)=0$ for any pseudoenergy of the form \eqref{gge_pseudo}. 

\subsection{Introducing graded QFTs}
The Ising model is also the simplest setting that allows for an internal $\mathbb{Z}_n$ grading. Since the theory is free, the associated Y-system collapses to the minimal relation \eqref{Ysys_ISING}. Any extension of the free fermion theory with additional structure should therefore reduce to this equation in a suitable limit. The key idea of this section is that the functional relation \eqref{Ysys_ISING} remains stable under certain non-trivial coverings of the rapidity plane. In particular, we consider the family of maps
\begin{equation}\label{ising_map}
    f_k(\vartheta)=\frac{\xi}{n}(\vartheta-2 \pi i w k)\,, \quad k \in \mathbb{Z}_n\,,
\end{equation}
parametrised by an integer $w \in \mathbb{N}$ and an overall orientation $\xi= \pm 1$. In general, one might expect that pulling back the Ising Y-system through these maps should break the closure of the functional relations. Surprisingly,  we instead find that overall consistency survives as long as the following simple arithmetic condition holds:
\begin{equation}\label{condition}
    n = 4w+\xi\,.
\end{equation}
When \eqref{condition} is satisfied, the single function $Y(\vartheta)$ can be unfolded into a $\mathbb{Z}_n$-multiplet $Y_k(\vartheta)=Y(f_k(\vartheta))$, whose components are related by a system of $\mathbb{Z}_n$-graded functional equations:
\begin{equation}\label{ysys_graded_ising}
   Y_{k-\xi}(\vartheta+ i\pi /2) \, Y_{k+\xi}(\vartheta- i\pi /2) = 1\,. 
\end{equation}
The origin of the condition \eqref{condition} can be traced back to the interplay between the half-period shifts of the functional relations \eqref{Ysys_ISING} and the branched structure of the covering map. Indeed, applying the map \eqref{ising_map} to a shifted argument produces
\begin{equation}
    f_k(\vartheta \pm i \pi / 2)=\frac{\xi}{n} \vartheta \mp \frac{i \pi \xi}{2 n}-\frac{2 \pi i w k}{n}\,.
\end{equation}
For the functional relation to close in terms of the graded variables $Y_k(\vartheta)$, the imaginary shift $\mp i\pi \xi/(2n)$ in the argument must be compensated by a shift in the index $k$. This works only if the closure condition \eqref{condition} is satisfied. This observation will serve as our template for constructing more elaborate, graded, integrable models later. The periodic structure of the original Y-system \eqref{Ysys_ISING} is also preserved under the reparametrisations \eqref{ising_map}. In particular, one finds
\begin{equation}\label{ising_periodicity}
    Y_k(\vartheta + i \pi P) = Y_{k+2\xi } (\vartheta)\,,\quad P = 2\,.
\end{equation}
Because $n$ is odd, repeated application of this shift cycles through all components $Y_k(\vartheta)$, returning to the starting point only after $n$ steps. The imaginary shift $i \pi P$ therefore acts as a cyclic permutation of the graded Y-functions, while keeping the periodic structure of the system intact. The effect of the covering map is most transparent at the level of the TBA pseudoenergies. In the standard Ising model, the pseudoenergy is simply $\varepsilon(\vartheta) = r \cosh\vartheta$, with $r = m\beta$ the scaling variable. Pulling back this expression along the maps $f_k(\vartheta)$ introduces fractional rapidity shifts, leading to the expression
\begin{equation}\label{disp_1}
    \varepsilon_k(\vartheta) = \varepsilon(f_k(\vartheta)) = m\beta \cosh\left(\frac{\vartheta}{n}-\frac{2\pi i wk}{n}\right)\,.
\end{equation}
These pseudoenergies still satisfy the graded Y-system derived above, but their physical interpretation requires some care: for example, note that the dispersion relation in \eqref{disp_1} no longer looks relativistic, as the argument of the source term has been rescaled by a factor $1 / n$. The missing ingredient comes from the generalised Gibbs ensemble \eqref{gge_pseudo}. Unlike the standard Gibbs ensemble (which couples only to the Hamiltonian), the GGE introduces independent generalised temperatures, one for each local conserved charge. By turning on a source for the spin-$n$ charge, we generate exactly the additional driving term needed to restore relativistic scaling. The pseudoenergies become
\begin{equation}\label{simpler}
    \varepsilon_k(\vartheta) = m^n\beta_n\cosh\vartheta + m\beta \cosh\left(\frac{\vartheta}{n}-\frac{2\pi i wk}{n}\right)\,.
\end{equation}
Because the closure condition $n=4 w+\xi$ forces $n$ to be odd, a spin-$n$ charge always exists in the tower. Thus, the graded construction is automatically well-defined for every admissible $n$. Finally, note that the example above uses only a single additional generalised source term. In general, one may include any number of GGE contributions. The covering map then reorganises all driving terms into a graded multiplet, producing a much richer pseudoenergy structure. In this framework, it is convenient to introduce the scaling variable $r = \beta_n m^n$, and define the ratios $\alpha_s = \beta_s/\beta_n^{s/n}$. In this way, the general pseudoenergies take the form
\begin{equation}\label{complete}
    \varepsilon_k(\vartheta) = \sum_{s\in\mathcal{S}} m^s\beta_s \cosh\left(\frac{s\vartheta}{n}-\frac{2\pi is wk}{n}\right)=\sum_{s\in\mathcal{S}} \alpha_s r^{s/n} \cosh\left(\frac{s\vartheta}{n}-\frac{2\pi is wk}{n}\right) \,,  
\end{equation}
with $\alpha_n = 1$. This choice ensures that the energy-like term scales linearly with $r$. Interpreting each $\varepsilon_k(\vartheta)$ as the pseudoenergy of a particle of mass $m$ and species $k$, the ground-state scaling function is obtained by summing over $k \in \mathbb{Z}_n$. One finds:
\begin{equation}\label{cc_grad}
     c(r, \{\alpha_s\}) = \frac{3 r}{\pi^2}\sum_{k\in\mathbb{Z}_n} \int _\mathbb{R}\dd\vartheta  \cosh\vartheta \log \big(1+e^{-\varepsilon_k(\vartheta)}\big)\,.
\end{equation}
We emphasise that the scaling prescription \eqref{complete} – and the resulting induced scaling for the function \eqref{cc_grad} – is not unique. In the present context, it follows primarily from dimensional considerations rather than from a fundamental physical principle. Nonetheless, this choice is natural within our framework and, as we will show in Section~\ref{scaling_ISING}, it leads to a well-defined ultraviolet regime for the theory. In particular, when $r \to 0$, the above scaling reduces precisely to the setup analysed in the deformed TBA of \cite{Masoero:2010is} (see also Appendix \ref{masoero_appendix}). A conceptually similar scaling prescription has also been adopted in \cite{Downing:2023lop, Downing:2023lnp, Downing:2024nfb}, although in a slightly different physical context -- see also \cite{Zagier2022}.

\subsection{Graded scaling functions and their asymptotics}\label{scaling_ISING}
\begin{figure}
    \centering
    \begin{subfigure}[b]{0.495\textwidth}
        \centering
        \includegraphics[width=\textwidth]{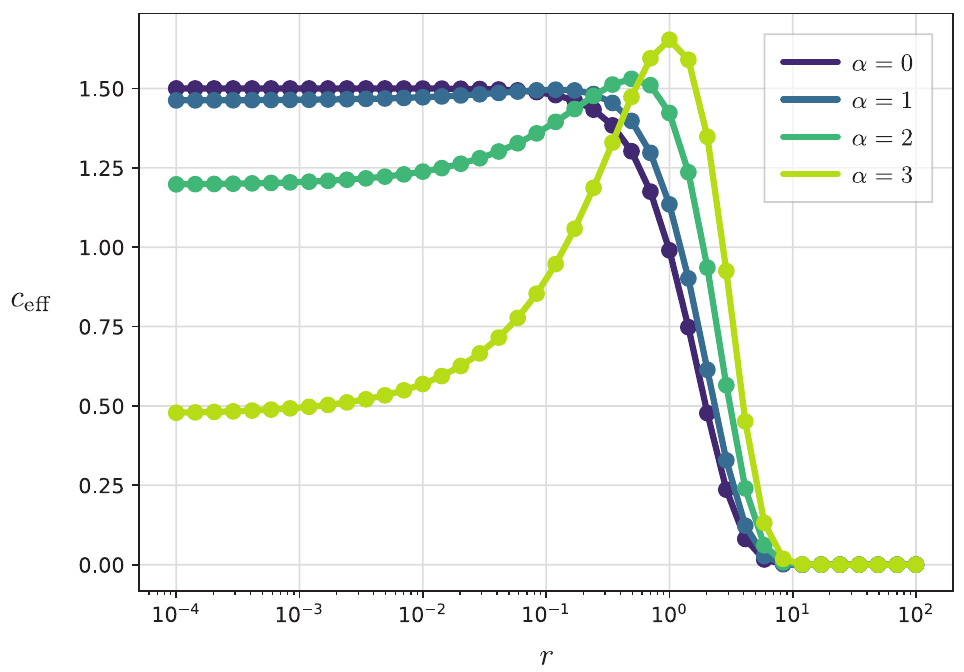}
        \caption{$n=3$.}
        \label{fig:1a2}
    \end{subfigure}
    \hfill
    \begin{subfigure}[b]{0.481\textwidth}
        \centering
        \includegraphics[width=\textwidth]{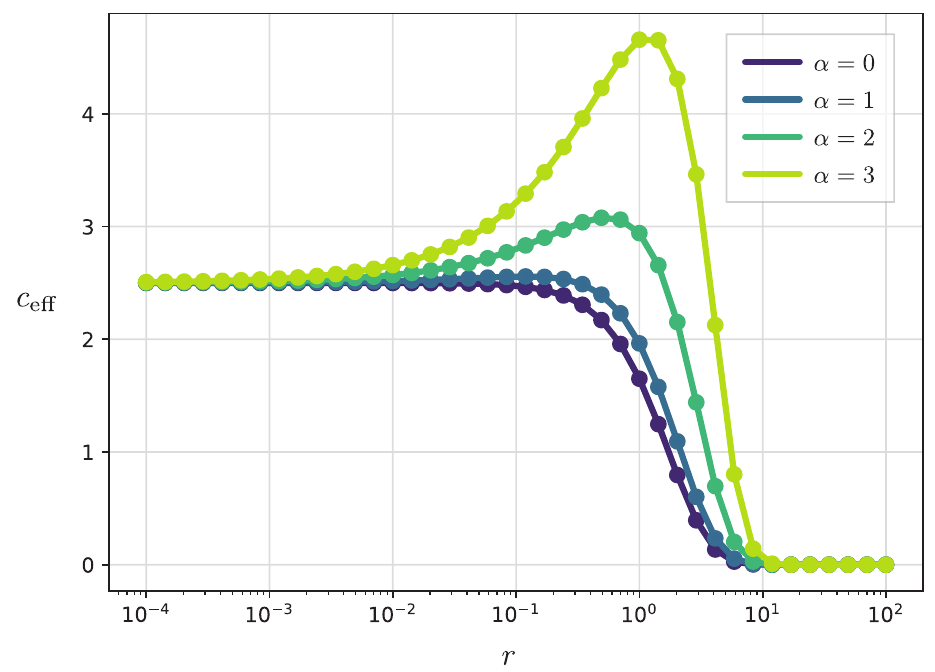}
        \caption{$n=5$.}
        \label{figs:1b}
    \end{subfigure}
    \caption{Ground state scaling function of the graded Ising field theory for sufficiently small values of $\alpha$. When $n=3$, only the trivial case $\alpha = 0$ flows to three independent copies of the Ising CFT. When $n=5$, no ultraviolet corrections are observed, yet a non-trivial structure persists at finite radius. Analogous behaviours are observed for all odd $n>5$.}
    \label{fisg:both}
\end{figure}

Despite its simplicity, the graded Ising model already contains most of the essential features of the general situation, packaged in a few tractable, exact expressions. In this section, we analyse the effect of the reparametrisation maps \eqref{ising_map} on the ground-state scaling function of the theory, restricting, for clarity, to the simpler case \eqref{simpler} in which only the spin-$n$ charge contributes to the generalised Gibbs ensemble beyond the energy term. Using the rescaled parameters introduced in \eqref{complete}, the pseudoenergies take the form
\begin{equation}\label{resc.}
    \varepsilon_k(\vartheta) =  r \cosh\vartheta + \alpha r^{1/n} \cosh\left(\frac{\vartheta}{n}-\frac{2\pi i wk}{n}\right)\,,
\end{equation}
where $k$ takes values in $\mathbb{Z}_n$, $n=4w+\xi$, and we set $\alpha \equiv \alpha_1$. The central charge $c_\text{eff}(r,\alpha)$ is then obtained from \eqref{cc_grad} as a function of the scaling parameter $r$. Since different values of $w$ in \eqref{resc.} simply select different odd integers $n$, and since the analysis only requires $n$ being odd, we can, without loss of generality, fix $w=-\text{sign}\vartheta$, which allows us to express \eqref{cc_grad} as an integral over the half-line:
\begin{equation}\label{ccising_integral}
    c_\text{eff}(r,\alpha) =\frac{6r}{\pi^2} \int_0^\infty \dd\vartheta \cosh\vartheta \log\big(1+e^{-r\cosh\vartheta - \alpha r^{1/n} \cosh(\vartheta/n + 2 \pi i k/n)}\big)\,.
\end{equation}
Expanding the logarithm in the Fermi–Dirac series,
\begin{equation}\label{fermi_dirac}
    \log \left(1+e^{-x}\right)=\sum_{y=1}^\infty \frac{(-1)^{y+1}}{y} e^{-y x}\,,
\end{equation}
and summing over the $n$ graded sectors using the root-of-unity identity \eqref{t=0} derived in Appendix \ref{app:root}, the rapidity integral can be evaluated term by term. Here, we also use the standard identity
\begin{equation}\label{bess_rep}
    \int_0^{\infty} \dd \vartheta z\cosh \vartheta \cosh (j \vartheta) e^{-z \cosh \vartheta}= j\,K_j(z)\,, 
\end{equation}
valid for all $j\in \mathbb{N}$ and $z \in \mathbb{R}$. In equation \eqref{bess_rep}, $K_j(z)$ are the modified Bessel functions of the second kind. Interchanging sums and the $\vartheta$-integral, justified for any positive $r$ by dominated convergence, we obtain a fully explicit (and uniformly convergent) series:
\begin{equation}\label{curves:plot}
    c_{\text{eff}}(r,\alpha) =\frac{6 nr}{\pi^2} \sum_{y= 1}^\infty \frac{(-1)^{y+1}}{y}\Big(I_0(y \alpha r^{1/n}) K_1(yr)+2\sum_{j=1}^\infty\frac{j(-1)^{j n}}{yr} I_{j n}(y \alpha r^{1/n})K_{j}(y r)\Big) \,,
\end{equation}
where $I_j(z)$ are modified Bessel functions of the first kind. This expression naturally separates into two pieces: a \textit{neutral sector}, coming from the $j=0$ term, which generates only even powers of $\alpha$, and a \textit{harmonic sector}, coming from $j\geq 1$, which produces a power series in $\alpha^{n}$. The curves \eqref{curves:plot} can also be obtained via direct numerical integration. The resulting ground-state scaling functions for $n=3,5$ are plotted in Figure \ref{fisg:both} as a function of the scaling parameter $r$ for various values of $\alpha$.\\

\noindent \textbf{Infrared regime}. In the infrared limit, $r \gg 1$, the modified Bessel functions $K_j(z)$ decay exponentially as $e^{-z}$ for every $j\geq 2$. Only the lowest mode $j=1$ survives, so the harmonic sector is exponentially suppressed, and the neutral sector dominates: 
\begin{equation}
c_{\text{eff}}^\text{IR}(\alpha) = \frac{6nr}{\pi^2}I_0(\alpha r^{1/n}) K_1(r)\,.
\end{equation}
Since $I_0(z)$ is even in $z$, the infrared expansion contains only even powers of $\alpha$, independently of $n$. In the ultraviolet limit, the situation changes completely.\\

\noindent \textbf{Ultraviolet regime}. In this regime, which corresponds to $r\rightarrow 0$, the neutral contribution is controlled by
\begin{equation}
    \frac{6 nr}{\pi^2} \sum_{y=1}^\infty \frac{(-1)^{y+1}}{y}I_0(y \alpha r^{1/n}) K_1(y r) \simeq \frac{n}{2} + \mathcal{O}(r^{1/n})\,.
\end{equation}
This yields a constant contribution equal to $n/2$, which is precisely the Virasoro central charge of $n$ decoupled copies of the Ising CFT, each with $c=1/2$. In the UV, this constant gives the leading behaviour of the scaling function, and it validates the idea that, when $\alpha =0$, the graded construction reorganises the free fermion theory into an $n$-fold direct product of the original model. Beyond this universal contribution, one must also consider the effect of the harmonic sector, which encodes the interactions between the copies. The harmonic part is sensitive to the parameter $\alpha$, unlike the neutral contribution. Using integral representations for the Bessel functions, one finds that at any fixed $j$ the leading powers of $r$ cancel, leaving a finite contribution proportional to $\alpha^{j n}$. This yields a systematic expansion for the effective central charge in the ultraviolet,
\begin{equation}\label{cds_ising}
        c_\text{eff}^\text{UV}(\alpha) = \frac{n}{2}+\sum_{j=1}^\infty T_j(n)\alpha^{jn}\,.
\end{equation}
The coefficients $T_j(n)$ are obtained in closed form as
\begin{equation}
    T_j(n)=\frac{12 n}{\pi^2} \frac{j(j-1)!}{(j n)!} 2^{\sigma -3 }(-1)^{j n}(1-2^{1-\sigma}) \zeta(\sigma)\,,
\end{equation}
where we introduced the parameter $\sigma=2-(n-1) j$. Because $n-1$ is even, the parameter $\sigma$ that controls the coefficients $T_j(n)$ is always an even integer whenever $j \geq 1$. For $n>3$, $\sigma$ becomes negative and coincides with a trivial zero of the Riemann zeta function (i.e., the negative even integers), and so $\zeta(\sigma)=0$. As a result, every potential contribution vanishes identically. The single exception is the case $n=3$ with $j=1$, for which $\sigma = 0$ and $\zeta(0)=-1 / 2$. This leads to the final conclusion:
\begin{equation}\label{cc_ds}
      c_\text{eff}^\text{UV}(\alpha) = \frac{n}{2}+\sum_{j=1}^\infty T_j(n) \alpha^{jn} = \begin{cases}
        \dfrac{3}{2} - \dfrac{3}{8 \pi^2} \alpha^3 & \text{if } n=3\,, \\[8pt]
        \dfrac{n}{2} & \text{if } n>3\,.
    \end{cases}
\end{equation}
This exhausts all perturbative UV contributions: apart from the universal constant $n/2$, only in the special case $n=3$ does the harmonic sector generate a finite deformation proportional to $\alpha^3$.  Note that equation \eqref{cc_ds} is in agreement with the numerical results presented in Figure \ref{fisg:both}.

However, this is not the full story. As the parameter $\alpha$ is varied, non-perturbative phenomena can arise, driven by level-crossings in the thermodynamic spectrum. In particular, excited states may overcome the ground state and become energetically favoured in the large-$L$ limit, thereby modifying the effective central charge in a way that cannot be captured by the perturbative expansion. The next section analyses these non-perturbative transitions and shows how the graded structure reshapes the spectrum once they occur.
\subsection{Excited states and level-crossing}
In the standard formulation of the Ising model, the expression for the ground–state scaling function involves an integral of the form
\begin{equation}
    \int_\mathbb{R} \dd\vartheta \cosh\vartheta \log\big(1+e^{-\varepsilon(\vartheta)}\big)\,,
\end{equation}
with $\varepsilon(\vartheta) = r \cosh\vartheta$.  This expression is perfectly well-defined on the real rapidity axis. The only possible non-analyticities come from the logarithm, whose branch points occur when its argument vanishes -- that is, when the quantisation condition $\varepsilon(\vartheta_j) = i\pi(2j+1)$ is satisfied for some $j \in \mathbb{N}$. For real values of the scaling parameter, these branch points lie away from the contour of integration, so the ground state is obtained by simply evaluating the integral as it stands. Excited states appear when we analytically continue $r$ in the complex plane. As $r$ varies, the branch points of the logarithm move through rapidity space. When one of them crosses the real axis, the contour of integration must be deformed. This deformation introduces an additional contribution resulting from the discontinuity of the logarithm across the branch cut.  Physically, these extra terms are interpreted as the energies of one-particle excitations, and the discontinuity precisely reproduces the relativistic dispersion relation $E_j=m \cosh \vartheta_j$, so that each branch point crossing is equivalent to occupying a fermionic mode at rapidity $\vartheta_j$. In this way, the excited spectrum emerges directly from the analytic structure of the ground-state integral: the vacuum corresponds to the original contour, while excited states arise whenever singularities of the logarithm are forced through the integration path \cite{Bender:1969si, Dorey:1996re}. \\

\begin{figure}[h]
    \centering
    \includegraphics[width=1\textwidth]{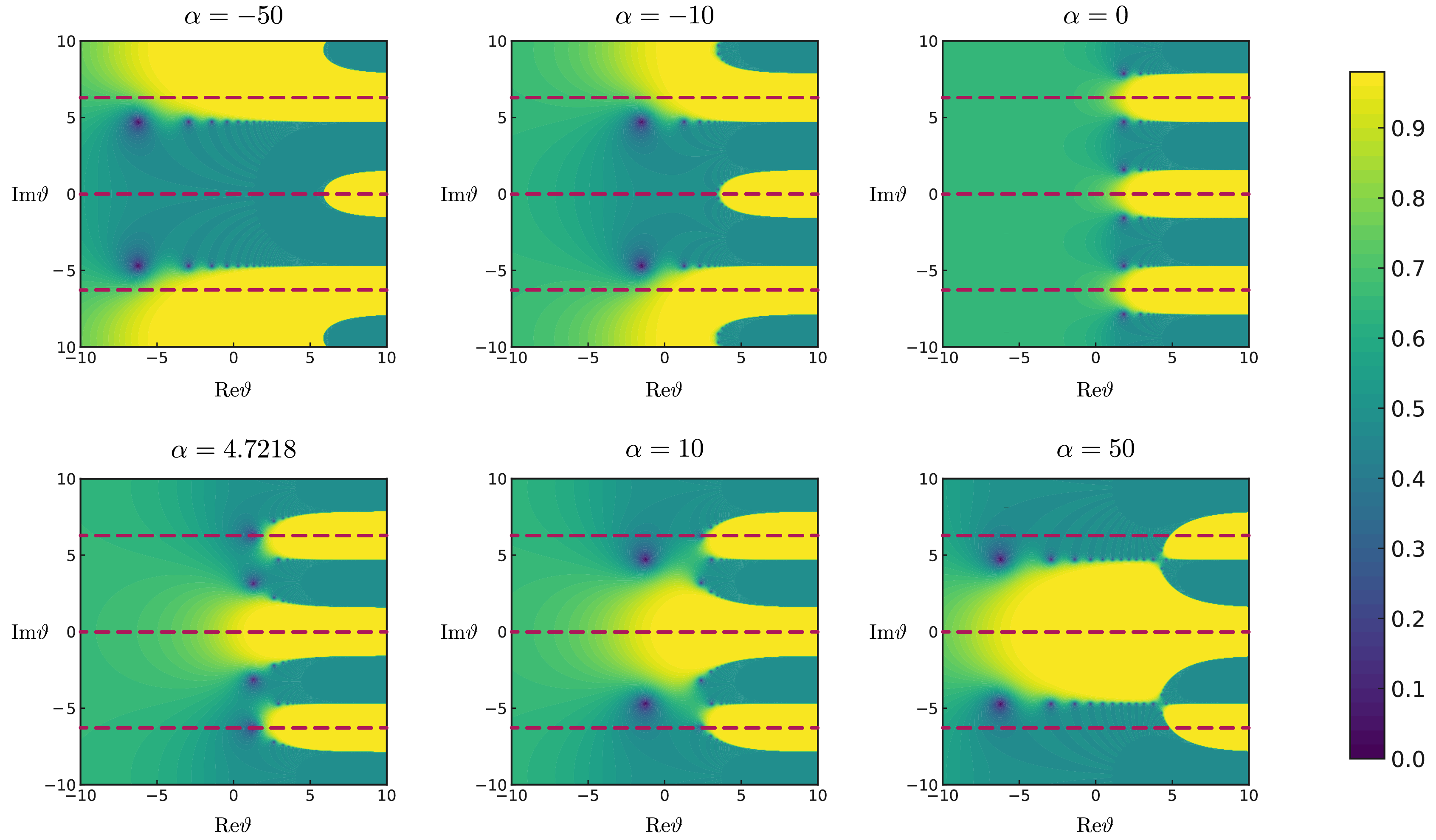}
    \caption{Normalised level sets of the function 
  $1 + \exp(-e^{\vartheta} - \alpha e^{\vartheta/3})$ 
  displayed over a $20 \times 20$ complex strip centred at the origin in the $\vartheta$-plane for various $\alpha$ values. 
  The dashed red horizontal lines indicate $\Im \vartheta\in \{0,\pm 2 \pi\}$. The plots for $k=1,2$ are formally obtained from this picture by translating the imaginary direction, $\Im\vartheta \mapsto \Im \vartheta \pm 2 \pi  $. As $\alpha$ varies, the zeroes of the displayed function move in the complex plane; at the first critical value $\alpha \simeq 4.7218$, a pair of zeroes collides with the $\pm 2 \pi i$ lines, signalling a change in the analytic structure of the ground-state scaling function.}
  \label{fig:levelsets}
\end{figure}

\paragraph{Excited states in graded QFTs.} In the graded theory, the analytic-continuation picture of the Ising model still applies, but with an important refinement. The parameter $\alpha$ modifies the quantisation condition that determines the locations of the branch points of the logarithm. In the ungraded case, these solutions are generically complex and only affect the integral when analytic continuation forces them across the real axis. In contrast, in the graded theory, certain values of $\alpha$ already produce \textit{real} solutions of the quantisation condition. Thermodynamically, this signals a \textit{phase transition} -- an excited configuration that initially lies above the vacuum can become the new ground state as $\alpha$ varies. The generalised quantisation condition reads
\begin{equation}\label{QC}
    \varepsilon_k(\vartheta_j) =  r \cosh\vartheta_j + \alpha r^{1/n} \cosh\left(\frac{\vartheta_j}{n}-\frac{2\pi i wk}{n}\right) = i\pi(2j+1)\,,\quad j\in\mathbb{N}\,.
\end{equation}
Introducing $x = e^{\vartheta/n}$, this becomes a polynomial equation of degree $2n$ in $x$. For generic odd values of $n$, this equation cannot be solved in closed form, and the problem must be handled numerically, reflecting the algebraic complexity of the graded spectrum. In the ultraviolet regime, however, the structure simplifies. Near the edges of rapidity space, one isolates the dominant exponential contributions: the integral is controlled by neighbourhoods of  $\vartheta = \pm \log r$, where the argument of the logarithm is $\mathcal{O}(1)$. Rescaling one edge as $\vartheta\mapsto \vartheta-\log r$ isolates the exponentially dominant term $e^\vartheta$. The opposite edge, governed by the $e^{-\vartheta}$ contributions, decouples. Since the two edges are symmetric, the final ultraviolet scaling function simply acquires a factor of two relative to the single-edge computation. In this limit, scale invariance further allows us to set $r=1$. When $n=3$, the conditions \eqref{QC} reduce to a family of {depressed cubic} equations in terms of the auxiliary variable $x=e^{\vartheta/3}$:
\begin{equation}\label{polynomial_eq}
    x^3 + \alpha e^{2\pi i k/3} x = 2 \pi i(2j+1)\,,\quad j\in\mathbb{N}\,.
\end{equation}
This reduction allows for an explicit study of the full spectrum as $\alpha$ is varied. In particular, solutions to equation \eqref{polynomial_eq} are strongly constrained by the underlying $\mathbb{Z}_3$ symmetry. As $\alpha$ varies, the zeros of the logarithm’s argument trace out the level sets shown in Figure~\ref{fig:levelsets}. At fixed $j$, the roots in the sector $k=1$ are mapped into those in the $k=2$ sector by reflection across the imaginary axis in the complex plane. In contrast, the sector $k=0$ behaves differently: one solution is always purely imaginary, while the remaining two are related by a reflection of their real components. Although this description is accurate, it hides the fact that the three families of solutions are not independent; rather, they are different manifestations of the same algebraic structure. A more symmetric presentation is obtained by factoring out the explicit $\mathbb{Z}_3$ dependence through a rotation of variables. Introducing $y = xe^{2\pi i k/3}$, equation \eqref{polynomial_eq} is mapped into:
\begin{equation}\label{cardano}
    y^3 + \alpha y = 2\pi i(2j+1)\,,\quad j\in\mathbb{N}\,.
\end{equation}
In the $y$-plane, the structure of the roots becomes much easier to visualise. The two complex roots that previously appeared as mirror images in the $k=1$ and $k=2$ sectors now emerge as two distinct, complex solutions of the same cubic. The third solution, on the other hand, remains purely imaginary for all values of $\alpha$. In terms of the auxiliary variable $y$, the real axis $\vartheta \in \mathbb{R}$ is the straight half-line in the $y$-plane that passes through the $k$-th cube root of unity and extends radially to infinity. More generally, an arbitrary contour $\Gamma$ maps as follows: the substitution $x=e^{\vartheta / 3}$ sends imaginary shifts of $\vartheta$ to rotations in the complex $x$-plane, and the multiplication by $e^{2 \pi i k / 3}$ in the definition of $y$ applies an additional rigid rotation of angle $2 \pi k / 3$. We denote the resulting contour by $\Gamma_k$, so that the relevant integral takes the form:
\begin{equation}\label{integral}
    \int_\Gamma \dd\vartheta\, e^\vartheta \log\big(1+e^{-\varepsilon_k(\vartheta)}\big) = 3\int_{\Gamma_k} \dd y \,y^2 \log\big(1+e^{-\varepsilon_k(y)}\big)\,,
\end{equation}
where we denoted $\varepsilon_k(y) \equiv \varepsilon_k(\vartheta(y))$. Integrating by parts splits this expression into two contributions:
\begin{equation}
    3\int_{\Gamma_k} \dd y \,y^2 \log\big(1+e^{-\varepsilon_k(y)}\big)= \left.y^3 \log\big(1+e^{-\varepsilon_k(y)}\big)\right|_{\partial\Gamma_k}- \int_{\Gamma_k}\dd y \,y^3  \partial_y \log \big(1+e^{-\varepsilon_k(y)}\big)\,.
\end{equation}
The first is a boundary term: it is smooth in $\alpha$, does not depend on the location of branch points, and therefore plays no role in the discontinuity. The second term is genuinely meromorphic, and this is where all non-analytic behaviour originates. Its poles sit exactly at the solutions $y_*$ of the cubic \eqref{cardano}, and the corresponding residues are
\begin{equation}
    \operatorname{Res}\left\{-y^3  \partial_y \log \big(1+e^{-\varepsilon_k(y)}\big),y_*\right\}  =-y_*^3\,.
\end{equation}
To extract the non-perturbative contribution, the contour is deformed so that it encloses the poles. As usual, shifting the contour off the real axis picks up the residues, each weighted by $2 \pi i$. One must include contributions from both the left and right sectors, which come in symmetric pairs, as well as an overall factor of $1/2$ that arises from the exponential energy term. Collecting these ingredients, the net effect of each root crossing the integration contour can be summarised schematically as:
\begin{equation}\label{delta_c}
    \Delta c^\text{UV}_\text{eff}(\alpha) = \frac{6 i}{\pi} y_*^3(\alpha, j)\,.
\end{equation}
At $\alpha=0$, the choice of contour $C$ fixes which excited state is being probed. As $\alpha$ is varied away from zero, branch points drift in the complex plane: in this process, a singularity may or may not be encountered, depending on the value of $\alpha$.

\begin{figure}
    \centering
    \begin{subfigure}[b]{0.469\textwidth}
        \centering
        \includegraphics[width=\textwidth]{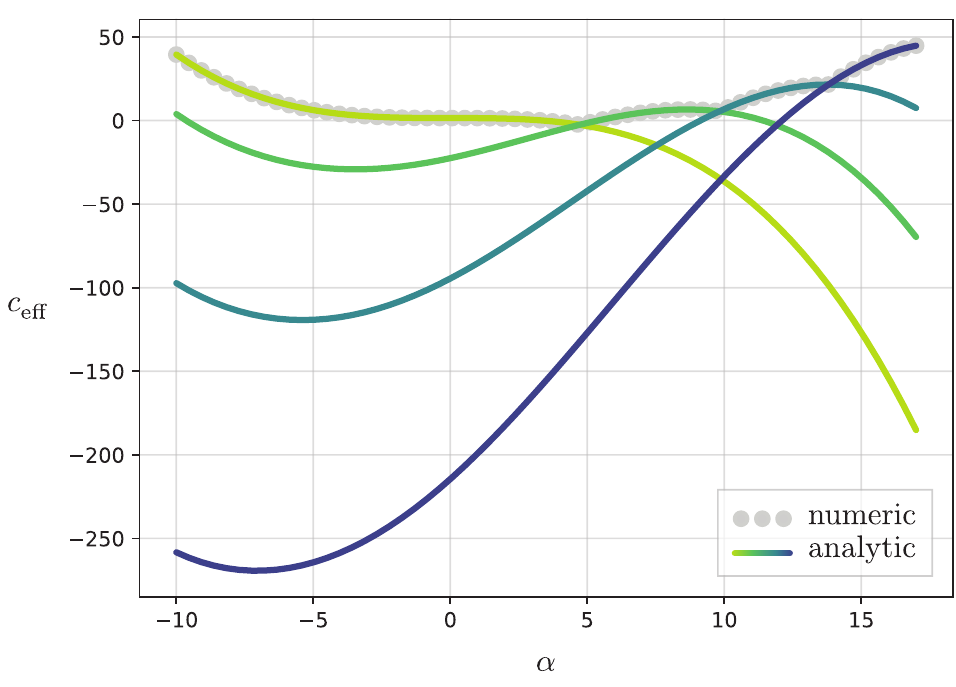}
        \caption{Ground state.}
        \label{fig:13a}
    \end{subfigure}
    \hfill
    \begin{subfigure}[b]{0.495\textwidth}
        \centering
        \includegraphics[width=\textwidth]{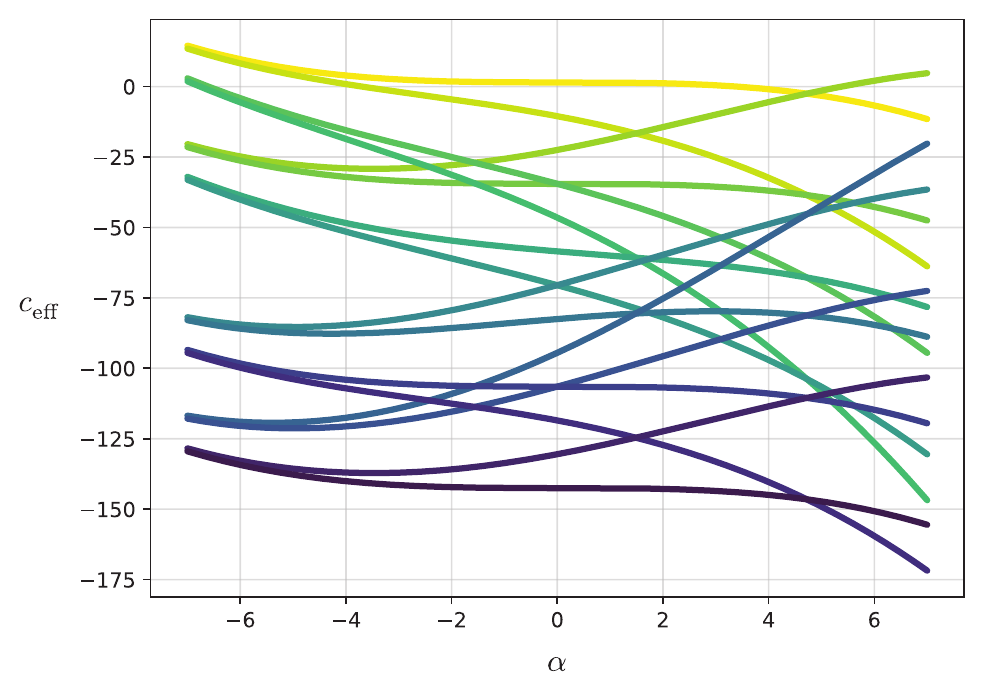}
        \caption{First excited states.}
        \label{figddd:1b}
    \end{subfigure}
    \caption{Plots summarising the behaviour of the first few excited-state scaling functions that populate the spectrum as a function of $\alpha$. In Figure \ref{fig:13a}, the excited states that successively overtake the perturbative ground state as $\alpha$ is increased towards positive values are highlighted. The dotted grey line corresponds to the numerical evaluation of the ground-state scaling function.  Figure \ref{figddd:1b} shows instead all excited states satisfying \eqref{QC} for $j=0,1$.}
    \label{fig:TBA-diagram}
\end{figure}

When the ground state is probed, choosing $\Gamma=\mathbb{R}$, the $k=1$ and $k=2$ sectors produce genuine singularities that eventually cross the integration contour. By contrast, in the $k=0$ sector, the pseudoenergies remain strictly real and positive, so the corresponding quantisation condition never admits a solution for real rapidity $\vartheta$. The crossing points are defined as the special values $\alpha=\alpha_j$ at which equation \eqref{polynomial_eq} develops real roots at fixed $j$ -- or, equivalently, the values at which the roots of \eqref{cardano} lie at an angle of $\pm 2 \pi /3$ radians in the complex $x$-plane. Such values can be determined explicitly, and correspond to: 
\begin{equation}\label{aux_alpha_c}
    \alpha_j = \frac{2}{\sqrt[3]{3}}(2\pi (2j+1))^{2/3}\,.
\end{equation}
In the $y$-plane, the relevant crossings are associated with the two complex solutions of the cubic -- namely, those with a non-vanishing real part. Since these two roots are complex conjugates of one another, they reach the integration contour simultaneously, and their contributions must therefore be summed together.  For $\alpha$ lying between two critical values, $\alpha \in (\alpha_j,\alpha_{j+1})$, the effective central charge receives contributions from the first $j$ pairs of crossing roots. Equivalently, each time a threshold $\alpha_j$ is crossed, the contour encloses an additional pair of poles, producing a discrete jump in the value of $c_{\text {eff}}$. The non-perturbative expression of the ultraviolet ground-state scaling function is therefore:
\begin{equation}
    c^\text{UV}_\text{eff}(\alpha)  = \frac{3}{2} -\frac{3\alpha^3}{8\pi^3} - \frac{6}{\pi} \sum_{\ell = 0}^{j} (y_{1*}^3(\alpha,\ell)+y_{2*}^3(\alpha,\ell)) \in \mathbb{R}\,.
\end{equation}
 Each cubed root depends on $\alpha$, and satisfies the relation \eqref{cardano}, which makes it possible to re-express their residues in terms of the third solution of the cubic using Vieta’s theorem. Thus, once the pair of conjugate solutions $y_1$, $y_2$ is accounted for, the remaining root $y_3$ -- which is purely imaginary, and can be written as $y_3=i v$ with $v \in \mathbb{R}$ -- captures all the necessary information. Combining these considerations, one obtains a compact formula for the ground-state scaling function:
\begin{equation}
    c^\text{UV}_\text{eff}(\alpha)  = \frac{3}{2} -\frac{3\alpha^3}{8\pi^3} - 24(j+1)^2 -\frac{6\alpha}{\pi} \sum_{\ell = 0}^{j} v_*(\alpha,\ell)\,,
\end{equation}
where each $v_*(\alpha,j)$ is determined by the unique real solution to $v^3-\alpha v + 2\pi(2j+1)=0$. The first few excited states that cross the ground state as $\alpha$ is increased are plotted in Figure \ref{fig:13a}. The remaining excited states, which populate the spectrum but do not necessarily overtake the ground state, can be constructed by identifying those solutions $y_*$ to the cubic equation \eqref{cardano} that contribute real terms to \eqref{delta_c}. At fixed $j \in \mathbb{N}$, two possibilities emerge. One can either sum the cubes of the pair of complex-conjugate solutions $y_{1, *}(\alpha, j)$ and $y_{2, *}(\alpha, j)$, related by reflection across the imaginary axis, or instead select the purely imaginary root $y_{3, *}(\alpha, j)$. Each excited configuration is then specified by two binary sequences, $\{\mathsf{E}_\ell\}_{\ell=0,\dots, j}$ and $\{\mathsf{F}_\ell\}_{\ell=0,\dots, j}$, with $\mathsf{E}_\ell,\mathsf{F}_\ell \in \{0,1\}$, which determine which poles contribute to the total discontinuity. The corresponding shift in the ultraviolet effective central charge takes the form
\begin{equation}
    \Delta c_\text{eff}^\text{UV}(\alpha) = \frac{6i}{\pi}\sum_{\ell=1}^j [\mathsf{E}_\ell(y^3_{1,*}(\alpha,\ell)+y^3_{2,*}(\alpha,\ell))+\mathsf{F}_\ell y^3_{3,*}(\alpha,\ell)]\,.
\end{equation}
We plot some of the resulting curves in Figure \ref{figddd:1b}. Note that we have decided to ignore possible excited states with complex energies here, as their physical interpretation is not yet fully understood. However, we do not exclude the possibility that these additional contributions could nonetheless play a role within the full spectrum of the theory. In particular, the analysis of \cite{Maloney:2018hdg, Maloney:2018yrz, Downing:2021mfw, Downing:2023lnp, Downing:2023lop, Downing:2024nfb, Downing:2025huv} shows that imposing modular consistency on GGEs can generate extra contributions in the modular-transformed channel, originating from non-perturbative solutions of the associated TBA-like equations. These additional excitations may appear with complex energies individually, yet they are required to reconstruct the exact modular-transformed GGE. 
\subsection{Generalising $T\overline{T}$ beyond integer spin} 
The $T\overline{T}$ deformation \cite{Cavaglia:2016oda, Smirnov:2016lqw} is a paradigmatic example of an irrelevant, yet exactly solvable, flow in two-dimensional QFTs. See \cite{Jiang:2019epa, He:2025ppz} for a pedagogical introduction to the subject. The deforming operator factorises into a product of conserved currents, which ensures that the finite-volume spectrum and the two-body scattering amplitudes remain under quantitative control throughout the deformation. As noted in \cite{Smirnov:2016lqw},  solvability of the $T\overline {T}$ deformation is not unique, and the same mechanism applies to any bilinear constructed from commuting conserved charges. In principle, charges of arbitrary Lorentz spin can be used to generate analogous deformations, potentially even beyond the integer-spin case. No explicit realisation of such flows was developed there, but the observation makes clear that the stress tensor plays no privileged role -- the key ingredient is current factorisation. A concrete framework for studying these generalised deformations was proposed in \cite{Hernandez-Chifflet:2019sua}. When embedding them into a generalised Gibbs ensemble, sourcing a conserved charge corresponds to an exactly solvable deformation of the finite-volume spectrum. From a thermodynamic point of view, this corresponds to a precise shift in the pseudoenergies. In particular, for a given pair of spin-$s$ conserved charges $Q_{s}^\pm (L)$,  define $H_s(L)$ as in equation \eqref{H_s(l)}. At equilibrium, its corresponding eigenvalue can be computed as:
\begin{equation}\label{prop_const}
    E_s (\beta_s)= -\frac{m^{s}}{2\pi}\int_\mathbb{R} \dd \vartheta \cosh(s\vartheta) \log \big(1+e^{-\varepsilon(\vartheta)}\big)\,.
\end{equation}
For simplicity, here we regard all other temperatures $\{\beta_{s'}\}_{{s'} \neq s}$ as fixed. Then, a generalised $T\overline{T}$ deformation with coupling $\mu \in \mathbb{R}$ corresponds to a shift in the generalised temperatures associated with the spin-$s$ conserved charges: 
\begin{equation}
    E_s(\beta_s;\mu) = E_s( \beta_s + \mu E_s;0)\,.
\end{equation}
Note that the overall proportionality constant in \eqref{prop_const} can always be reabsorbed into a redefinition of the flow parameter $\mu$. Since our primary goal is mainly qualitative, we will not track such numerical factors too closely. In the graded theory, a similar construction holds when pulling back the pseudoenergy $\varepsilon(\vartheta)$ along the maps $f_k(\vartheta)$. Within the setup introduced in \eqref{resc.}, where the generalised Gibbs ensemble involves only two non-zero temperatures, the theory admits two distinct integrable deformations. The first is the familiar $T\overline{T}$ deformation, governed by the energy operator and sourced by the $\cosh\vartheta$ term in the pseudoenergy. The second deformation is genuinely new: it is generated by the fractional-spin charge, whose eigenvalue takes the form
\begin{equation}\label{cons_frac}
     E_{1/n}(\alpha)=-\frac{  m^{1/n}}{2\pi}\sum_{k\in\mathbb{Z}_n}\int_\mathbb{R} \dd \vartheta \cosh\left(\frac{\vartheta}{n}-\frac{2 \pi i wk}{n}\right) \log \big(1+e^{-\varepsilon_k(\vartheta)}\big)\,.
\end{equation}

\begin{figure}
    \centering
    \begin{subfigure}[b]{0.495\textwidth}
        \centering
        \includegraphics[width=\textwidth]{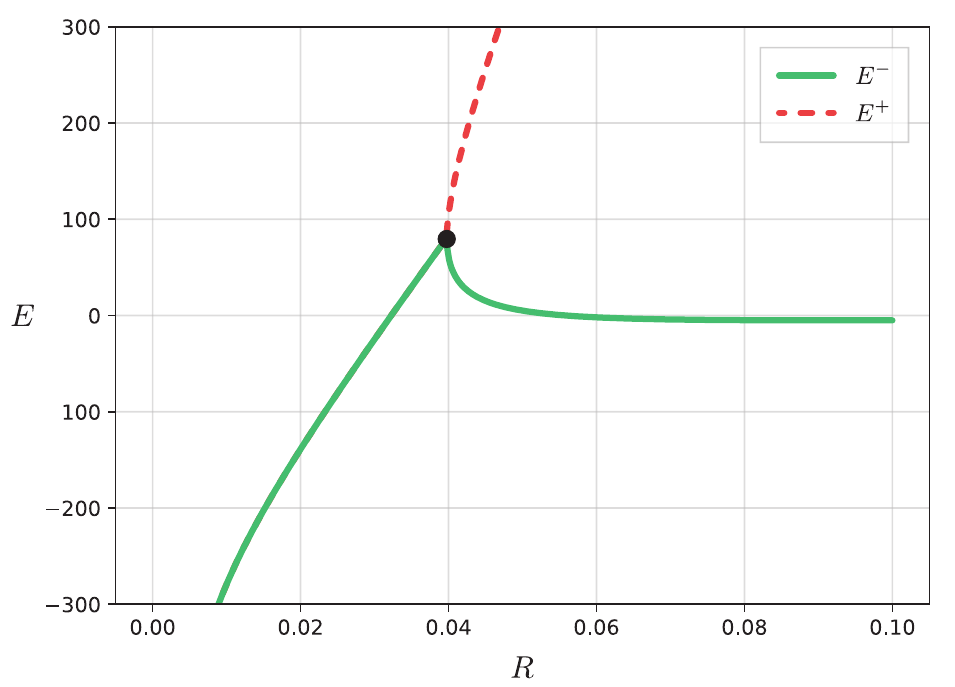}
        \caption{Real part.}
        \label{fig:12ags}
    \end{subfigure}
    \hfill
    \begin{subfigure}[b]{0.495\textwidth}
        \centering
        \includegraphics[width=\textwidth]{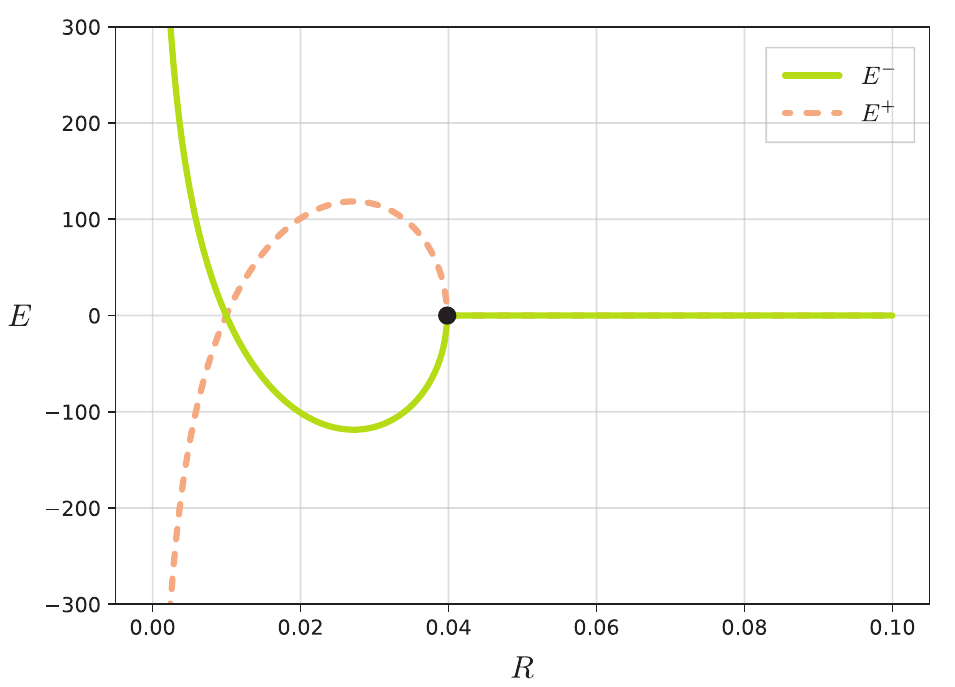}
        \caption{Imaginary part.}
        \label{figdddddd:1bgs}
    \end{subfigure}
    \caption{Ground-state energy of the deformed $\mathbb{Z}_3$-graded Ising model. Both branches $E^{\pm}$ are plotted as a function of $R$ for fixed $\beta = 1$ and $\mu = -1$. At the critical value $R_* = 1/8\pi$, the two branches merge, and the Casimir energy becomes complex. We also observe that, at the isolated point $R \simeq 0.05587$, reality of the ground state is restored. Qualitatively similar behaviours are observed in non-Hermitian, $\mathcal{PT}$-symmetric models of quantum mechanics -- see, for example, Figures 14 and 20 of \cite{Bender:1998gh}.}
\label{figure:ttbar}
\end{figure}

\noindent After performing the rescaling introduced above, we identify $\beta_n \equiv R$, which plays the role of the effective system size in the finite-volume channel, whereas $\beta_1 \equiv \beta$ is now the generalised inverse temperature conjugate to the GGE-type fractional-spin charge \eqref{cons_frac}. In what follows, we focus on the deformation generated by \eqref{cons_frac}, as it produces genuinely new effects associated with the grading. For clarity, we restrict to the ultraviolet regime $r \ll 1$, which we identify with the massless limit $m \rightarrow 0$. In this limit, the fractional-spin term dominates the scaling behaviour, and the integral can be further simplified by shifting the rapidity variable as $\vartheta \mapsto \vartheta-\log r$, as discussed below \eqref{QC}. Then, one verifies that the following identity holds for any value of $\alpha$:
\begin{equation}\label{integral_identity}
 \frac{6}{\pi^2} \sum_{k\in \mathbb{Z}_n} \int_\mathbb{R} \left(e^\vartheta + \frac{\alpha}{n} e^{s\vartheta/n - 2 \pi i swk/n}\right) \log(1+\exp\left(-e^\vartheta - \alpha e^{s\vartheta/n - 2 \pi i swk/n}\right)\!) = \frac{n}{2}\,.
\end{equation}
 Setting $s=1$, the first term in \eqref{integral_identity} corresponds to the integral \eqref{ccising_integral}, and computes the UV central charge of the theory. Ignoring non-perturbative contributions and level-crossing phenomena, it evaluates to \eqref{cc_ds}. The remaining piece instead contributes to the conserved charge \eqref{cons_frac}. Combining these results, and  using that $\alpha = \beta/R^{1/n}$ is proportional to the temperature associated to the spin-$1/n$ charge, we find that
\begin{equation}\label{energy_deformed}
      E_{1/n}(\beta;\mu) =  E_{1/n}(\beta+\mu E_{1/n}(\beta;\mu) ;0) =-\frac{\left(\beta + \mu E_{1/n}(\beta;\mu)\right)^2}{32\pi R} \delta_{n,3}\,,
\end{equation}
with $E_{1/n}(\beta;0) = - \beta^2/32\pi R$. Equation \eqref{energy_deformed} can be solved for $ E_{1/n}(\beta;\mu)$, yielding an exact expression for the deformed charge in the ultraviolet regime. One finds
\begin{equation}\label{sign_1}
    E^\pm_{1/n}(\beta;\mu) = -\frac{1}{\mu^2}\left(\!\sqrt{\beta \mu+8 \pi R}\pm2 \sqrt{2 \pi R} \,\right)^2\delta_{n,3}\,.
\end{equation}
In the graded Ising model, these deformations become trivial in the UV limit for all $n > 3$. Only for $n = 3$ do they generate a non-trivial flow at the CFT point, with ground-state energy
\begin{equation}\label{sign_2}
\begin{split}
  E^\pm(\beta;\mu) &= E(\beta+\mu E^\pm_{1/3}(\beta;\mu);0)  \\ &= -\frac{\pi}{4 R} \left[1+ \frac{16}{\mu^3 \pi^2 R}\left(4 \pi R \pm \sqrt{2\pi R(\beta \mu + 8 \pi R)}\right)^3\right] \,,   
\end{split}
\end{equation}
with 
\begin{equation}
 E(\beta;0)  =  -\frac{\pi }{4 R}+ \frac{\beta ^3}{16 \pi  R^2}\,.
\end{equation}
The $E^+$ branch does not approach the undeformed spectrum when $\mu \to 0$, and instead it diverges as $E^+(\beta ; \mu) \sim \mu^{-3}$. For this reason, we regard $E^-$, which smoothly recovers the original theory, as the ``physical branch''. When $\beta \mu <0$, the square root in the expression above eventually becomes imaginary as $R$ decreases. The vanishing of the square root defines a critical radius $R_* = -\beta\mu/8\pi$. Close to $R_*$, the square root behaves as $\sqrt{R-R_*}$, and the ground-state energy becomes non-analytic:
\begin{equation}
    E^\pm(\beta;\mu) = -\frac{\pi}{4 R_*}\pm\frac{256 \pi^2}{\mu^3}\left(R_*+3 \sqrt{R_*(R-R_*)}\right) + \mathcal{O}(R-R_*)\,.
\end{equation}
Here, $R_*$ acts as a minimal radius: below this value, the ground-state energy becomes complex. Figure \ref{figure:ttbar} shows the two branches of the solution as functions of $R$. This behaviour closely mirrors what happens in $T\overline {T}$-deformed QFTs: the deformation introduces a square-root branch point in the finite-volume energy, beyond which the solution becomes complex. In the thermodynamic picture, this corresponds to a maximal temperature, signalling a Hagedorn-type transition where the density of states grows exponentially and the standard continuation of the spectrum breaks down.

Note that the derivation assumes we remain in the regime where no singularities cross the integration contour. This corresponds to setting $\beta \leq \alpha_0  R^{1/3}\simeq 4.7218 R^{1/3}$, where $\alpha_0$ is the first critical value extracted from the analysis of the auxiliary equation for $\alpha$ (see equation \eqref{aux_alpha_c}). Below this threshold, the UV expression \eqref{sign_2} correctly captures the ground-state branch and its square-root singularity; beyond it, due to level-crossing, additional terms must be included. Moreover, we stress that the results of this section (and, in particular, expression \eqref{sign_2}) apply only to the ground state of the UV theory. For generic excited states, or when the theory is deformed away from the $m \to 0$ fixed point, the resulting deformations lead to significant modifications of the finite-volume spectrum and the associated TBA data. Similar deformations can be introduced in a much more general setting, including fully interacting theories, as we will discuss in Section \ref{SEC:SMAT}. However, the corresponding analysis is typically more involved, as the deformation affects both the scattering data and the finite-volume spectrum in a non-trivial way.
\section{Towards interacting theories}\label{SEC:ADE}
The Ising model served as a free and exactly solvable playground for introducing graded field theories. Its simplicity allowed us to make the key ideas explicit: the effect of the reparametrisation maps, the structure of graded pseudoenergies, and the appearance of non-perturbative corrections. We now move on to interacting theories, where the physics becomes substantially richer. In most relativistic quantum field theories, since interactions among particles are generally assumed to take place in a restricted region of spacetime, scattering processes can be formally described in terms of incoming (respectively, outgoing) asymptotic states, which define quantum states of free excitations long before (respectively, after) the scattering has happened, essentially describing wave packets with approximate positions at given times  \cite{Peskin:1995ev,Weinberg:1995mt}. In two spacetime dimensions, each asymptotic one-particle state can be once again labelled by its rapidity $\vartheta$. Accounting for an additional internal label $a$ which distinguishes among different particle species, energy and momentum are parametrised as:
\begin{equation}
(E_a,p_a) = m_a( \cosh\vartheta,\sinh \vartheta)\,.
\end{equation}
Multi-particle asymptotic states are constructed as ordered tensor products of one-particle states,
\begin{equation}\label{basis}
   \ket{\vartheta_1, a_1; \ldots;\vartheta_j, a_j} = \ket{\vartheta_1, a_1} \otimes \cdots \otimes \ket{ \vartheta_j, a_j}\,,
\end{equation}
with rapidities conventionally ordered as $\vartheta_i>\vartheta_{i+1}$ for in-states, and $\vartheta_i<\vartheta_{i+1}$ for out-states. Note that, because particles in two dimensions cannot bypass one another without interacting, the ordering of rapidities uniquely specifies the spatial sequence of particles in an asymptotic state. The scattering process is fully captured by the \textit{scattering matrix} (or {S-matrix}) $\mathbf{S}$, which maps incoming to outgoing states:
\begin{equation}
\ket{\text{out}} =\mathbf{S}\ket{\text{in}}\,,
\end{equation}
or vice versa, depending on conventions. On the basis of ordered many-particle states \eqref{basis}, S-matrix elements are defined by
\begin{equation}
   S_{a_1\dots a_j}^{b_1\dots b_{k}}(\vartheta_1,\dots, \vartheta_j,\zeta_1,\dots, \zeta_{k}) =  \bra{\zeta_1,b_1;\dots;\zeta_{k}, b_{k}}  \mathbf{S}\ket{\vartheta_1, a_1;\dots;\vartheta_{j}, a_{j}}\,. 
\end{equation}
Note that, in a generic QFT, many-body processes can produce or annihilate particles, redistribute momentum across non-trivial channels, and give rise to scattering amplitudes with highly complicated analytic structures \cite{Eden:1966dnq}. In two dimensions, \textit{Integrable Quantum Field Theories} (IQFTs) lie at the opposite end of this spectrum.
\subsection{Integrable S-matrices in a nutshell}
\begin{figure}
    \centering
    \begin{subfigure}[b]{0.375\textwidth}
        \centering
        \includegraphics[width=\textwidth]{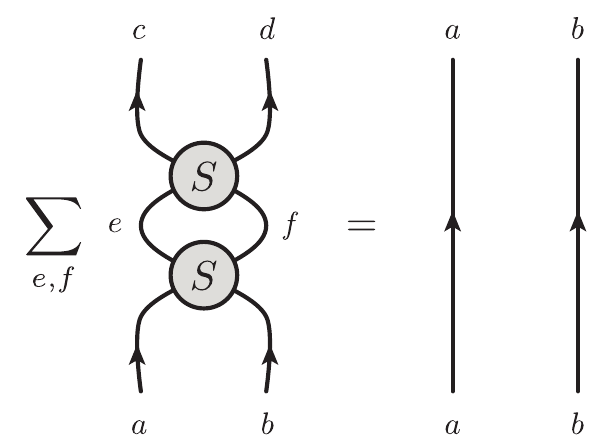}
        \caption{Braiding unitarity.}
        \label{fig:braid}
    \end{subfigure}
    \hspace{1cm} % optional: adjust spacing between the two
    \begin{subfigure}[b]{0.4\textwidth}
        \centering
        \includegraphics[width=\textwidth]{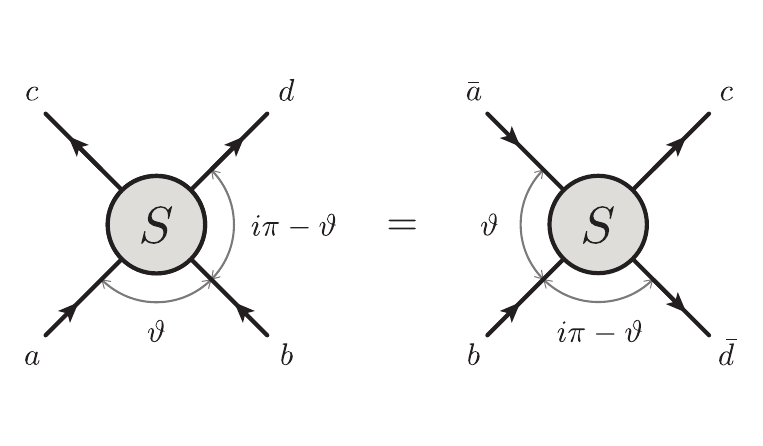}
        \caption{Crossing symmetry.}
        \label{fig:cross}
    \end{subfigure}
    \caption{A diagrammatic representation of the two-particle S-matrix consistency conditions. Here, the vertical direction represents time: incoming particles flow upward into the scattering region, and outgoing particles emerge above it. In Figure \ref{fig:braid}, two particles $a$ and $b$ scatter into intermediate channels $e, f$, and then scatter again into $c, d$. The sum over all allowed intermediate species reproduces two straight worldlines. In other words, performing the exchange twice is equivalent to doing nothing, yielding \eqref{braid}. In Figure \ref{fig:cross}, a two-particle scattering amplitude in the direct channel (or ``$s$-channel'') can be analytically continued into a process in the crossed one (``$t$-channel''). Moving an external leg from the incoming to the outgoing side corresponds to replacing the particle with its antiparticle and shifting the rapidity by $i\pi$, yielding \eqref{cs}.}
\end{figure}

The notion of quantum integrability is related to the existence of an infinite number of \textit{local}, \textit{independent}, \textit{conserved}, and \textit{mutually commuting} spin-$s$ charges $Q_s^\pm$, which act diagonally on the basis of one-particle states,
\begin{equation}
   Q_s^\pm \ket{\vartheta,a} = q^\pm_{s,a}(\vartheta)\ket{\vartheta,a}\,,\quad q^\pm _{s,a}(\vartheta) \propto m_a^s e^{\pm s\vartheta}\,, 
\end{equation}
and, in an additive way, on multi-particle states. Integrability imposes strong constraints on the structure of the S-matrix; for a pedagogical introduction to the subject, we refer the reader to references \cite{Dorey:1996gd, Mussardo:2010mgq, Bombardelli:2016scq}. In spacetime dimensions greater than two, the Coleman–Mandula theorem implies that the presence of even a single, higher-spin global conserved current forces the S-matrix to be trivial, and no scattering can occur \cite{Coleman:1967ad}. In contrast, $2d$ integrable field theories can exhibit non-trivial scattering, and the S-matrix retains a rich structure despite the infinite number of conserved quantities. Still, integrability turns out to be rather constraining, imposing the following properties:
\begin{itemize}

\item \textit{Elasticity}. The number of particles is the same before and after scattering, and the initial and final sets of momenta are equal up to permutations \cite{Zamolodchikov:1978xm}. 
\item \textit{Factorisation}. All multi‑particle amplitudes factorise into a sequence of two‑body scattering processes \cite{Karowski:1978eg}. 
\end{itemize}
While elasticity and factorised scattering ensure that the full S-matrix is completely determined by its two-particle building blocks $S_{ab}^{cd}(\vartheta_1,\vartheta_2)$, relativistic invariance further constrains these amplitudes so that they depend only on the rapidity difference $\vartheta_1-\vartheta_2$. Moreover, the requirement that scattering amplitudes must be independent of the ordering of pairwise collisions leads directly to the \emph{Yang–Baxter equation}, \cite{Yang:1967bm, Baxter:1972hz}
\begin{equation}
    \mathbf{S}_{12}(\vartheta_{12}) \mathbf{S}_{13}(\vartheta_{13})
    \mathbf{S}_{23}(\vartheta_{23}) =\mathbf{S}_{23}(\vartheta_{23})\mathbf{S}_{13}(\vartheta_{13})\mathbf{S}_{12}(\vartheta_{12})\,,
\end{equation}
where $\vartheta_{ij} = \vartheta_i - \vartheta_j$, and the operator $\mathbf{S}_{ij}$ acts non-trivially on the
$i$-th and $j$-th components of the three-particle tensor product space only. While this condition ensures consistency and associativity of scattering, quantum mechanics imposes two additional requirements:
\begin{itemize} 
\item\textit{Unitarity}. Probability must be conserved in every scattering process. At the level of the full S-matrix, this means $\mathbf{S}\mathbf{S}^\dagger = \mathbf{1}$. For $2d$ integrable quantum field theories, where all scattering processes factorise into two-body interactions, this condition translates into a constraint on the two-particle amplitudes,
\begin{equation}\label{phys_un}
S_{a b}^{ef}(\vartheta)(S_{ef}^{ cd}(\vartheta))^*=\delta_a^d \delta_b^ c\,,
\end{equation}
where summation over the indices $e$ and $f$ is implicit (see Figure \ref{fig:braid}). If, in addition, the Hilbert space has a positive-definite inner product, the theory is unitary in the standard quantum mechanical sense. Most $2 d$ integrable models also satisfy a second form of unitarity, called \textit{braiding} unitarity, which reflects the algebraic braid-group structure underlying integrable QFTs:
    \begin{equation}\label{braid}
       S_{ab}^{ef}(\vartheta) S_{ef}^{ cd}(-\vartheta) = \delta_a^d\delta_b^ c\,.
    \end{equation}
    In general, equation \eqref{braid} implies \eqref{phys_un} whenever \textit{Hermitian analyticity} also holds, meaning $(S_{ab}^{ cd}(\vartheta))^* =S_{ba}^{d c}(-\vartheta^*)$. See \cite{Miramontes:1999gd, Takacs:2002fg} for a detailed discussion.
\item \textit{Crossing symmetry}. Crossing expresses the equivalence between particle-antiparticle scattering and ordinary two-particle scattering, related by analytic continuation in rapidity. In terms of two-particle scattering amplitudes, 
\begin{equation}\label{cs}
    S_{ab}^{ cd}(\vartheta) = C_{ae}S_{bf}^{e c}(i\pi-\vartheta)C^{df} \,,
\end{equation}
   where $C_{ab}$ is the  charge-conjugation operator. Concretely, one may write $C_{ab} = \delta_{b\bar{a}}$, with $\bar{a}$ the antiparticle corresponding to species $a$ (see Figure \ref{fig:cross}).
\end{itemize}

\noindent\textbf{The S-matrix bootstrap program}. Simple poles of the two-particle S-matrix within the physical strip $\Im{\vartheta} \in(0, \pi)$ carry a direct physical interpretation, signalling the presence of bound states. The location of each pole determines the binding energy of the corresponding state, whereas its residue fixes the three-point coupling constant for the process. These data are not arbitrary but are constrained by a network of consistency requirements, among which the \textit{bootstrap equations} \cite{Karowski:1978ps, Zamolodchikov:1978xm} play a central role, ensuring the consistency of the spectrum. The classification of integrable scattering theories reduces to the task of characterising those two-particle S-matrices that are meromorphic in the rapidity plane and satisfy the core requirements of the bootstrap program: unitarity, crossing symmetry, and a pole structure consistent with the known particle spectrum and its fusion rules. These conditions severely constrain the analytic form of the scattering amplitudes, but do not determine them uniquely. A large residual ambiguity remains, parametrised by multiplicative meromorphic functions known as Castillejo–Dalitz–Dyson (CDD) factors \cite{Castillejo:1955ed, Smirnov:1992vz}. By construction, CDD factors preserve both unitarity and crossing, while potentially introducing new poles or zeros not enforced by the minimal particle content. To isolate universal information, one defines a minimal S-matrix as the canonical solution of the bootstrap equations with the smallest possible analytic structure. It contains exactly the poles required by the physical bound states and fusion processes, and no further zeros or singularities. CDD factors then generate all other allowed solutions by dressing the minimal one.
\begin{figure}[t]
\centering
\begin{tikzpicture}[every node/.style={font=\small}]
  % Style for nodes
  \tikzset{dyn/.style={circle,fill=black,inner sep=2pt}, lab/.style={font=\scriptsize}}

  % Vertical spacing
  \def\dy{2.0}

  %==== A_n ====================================================
  \node at (-1.0,0) {$A_n :$};
  \node[dyn] (A1) at (0.4,0) {};
  \node[dyn] (A2) at (1.6,0) {};
  \node[dyn] (A3) at (2.8,0) {};
  \node       at (3.9,0) {$\cdots$};
  \node[dyn] (Anm1) at (5.2,0) {};
  \node[dyn] (An)   at (6.4,0) {};
  \draw (A1)--(A2)--(A3)--(3.6,0);
  \draw (4.2,0)--(Anm1)--(An);
  \node[lab,below=3pt] at (A1) {1};
  \node[lab,below=3pt] at (A2) {2};
  \node[lab,below=3pt] at (A3) {3};
  \node[lab,below=3pt] at (Anm1) {$n\!-\!1$};
  \node[lab,below=3pt] at (An) {$n$};
  \node[anchor=west] at (7.4,0)
    {$\;\;\;\bar a = n+1-a,\quad a=1,\dots,n$.};

  %==== D_n ====================================================
  \node at (-1.0,-\dy) {$D_n :$};
  \node[dyn] (D1) at (0.4,-\dy) {};
  \node[dyn] (D2) at (1.6,-\dy) {};
  \node[dyn] (D3) at (2.8,-\dy) {};
  \node       at (3.9,-\dy) {$\cdots$};
  \node[dyn] (Dn2) at (5.2,-\dy) {};
  \node[dyn] (Dn1) at (6.4,-\dy+0.6) {};
  \node[dyn] (Dn)  at (6.4,-\dy-0.6) {};
  \draw (D1)--(D2)--(D3)--(3.6,-\dy);
  \draw (4.2,-\dy)--(Dn2)--(Dn1);
  \draw (Dn2)--(Dn);
  \node[lab,below=3pt] at (D1) {1};
  \node[lab,below=3pt] at (D2) {2};
  \node[lab,below=3pt] at (D3) {3};
  \node[lab,below=3pt] at (Dn2) {$n\!-\!2$};
  \node[lab,right=3pt] at (Dn1) {$n$};
  \node[lab,right=3pt] at (Dn) {$n\!-\!1$};
  \node[anchor=west] at (7.4,-\dy)
    {$\begin{cases}
        \text{for $n$ even: } & \bar a = a,\ a=1,\dots,n,\\
        \text{for $n$ odd: }  & \!\!\!\!\!\begin{cases}\bar a = a,\ a=1,\dots,n\!-\!2,\\ 
         \bar n=n-1\,.
        \end{cases}
      \end{cases}$};

  %==== E_n ====================================================
  \node at (-1.0,-2*\dy) {$E_n :$};
  \node[dyn] (E1) at (0.4,-2*\dy) {};
  \node[dyn] (E2) at (1.6,-2*\dy) {};
  \node[dyn] (E3) at (2.8,-2*\dy) {};
  \node       at (3.9,-2*\dy) {$\cdots$};
  \node[dyn] (En2) at (5.2,-2*\dy) {};
  \node[dyn] (En1) at (6.4,-2*\dy) {};
  \node[dyn] (En)  at (2.8,-2*\dy+0.9) {};
  \draw (E1)--(E2)--(E3)--(3.6,-2*\dy);
  \draw (4.2,-2*\dy)--(En2)--(En1);
  \draw (E3)--(En);
  \node[lab,below=3pt] at (E1) {1};
  \node[lab,below=3pt] at (E2) {2};
  \node[lab,below=3pt] at (E3) {3};
  \node[lab,below=3pt] at (En2) {$n\!-\!2$};
  \node[lab,below=3pt] at (En1) {$n\!-\!1$};
  \node[lab,above=3pt] at (En) {$n$};
  \node[anchor=west] at (7.4,-2*\dy)
    {$\begin{cases}
        \text{for $n=6$: } & \bar 1=5,\ \bar 2=4,\ \bar 3=3,\ \bar 6=6,\\
        \text{for $n=7,8$: } & \bar a=a,\ a=1,\dots,n.
      \end{cases}$};

  %==== T_n ====================================================
  \node at (-1.0,-3*\dy) {$T_n :$};
  \node[dyn] (T1)   at (0.4,-3*\dy) {};
  \node[dyn] (T2)   at (1.6,-3*\dy) {};
  \node[dyn] (T3)   at (2.8,-3*\dy) {};
  \node       at (3.9,-3*\dy) {$\cdots$};
  \node[dyn] (Tnm1) at (5.2,-3*\dy) {};
  \node[dyn] (Tn)   at (6.4,-3*\dy) {};
  \draw (T1)--(T2)--(T3)--(3.6,-3*\dy);
  \draw (4.2,-3*\dy)--(Tnm1)--(Tn);
  % Self-loop for tadpole
  \draw (Tn) .. controls +(0.6,0.6) and +(0.6,-0.6) .. (Tn);
  \node[lab,below=3pt] at (T1) {1};
  \node[lab,below=3pt] at (T2) {2};
  \node[lab,below=3pt] at (T3) {3};
  \node[lab,below=3pt] at (Tnm1) {$n\!-\!1$};
  \node[lab,below=3pt] at (Tn) {$n$};
  \node[anchor=west] at (7.4,-3*\dy)
    {$\;\;\;\bar a=a,\quad a=1,\dots,n$.};

\end{tikzpicture}
\caption{Dynkin diagrams of type $A$, $D$, $E$ and tadpole $T$,
together with their charge conjugation assignments.}
\label{DDs}
\end{figure}
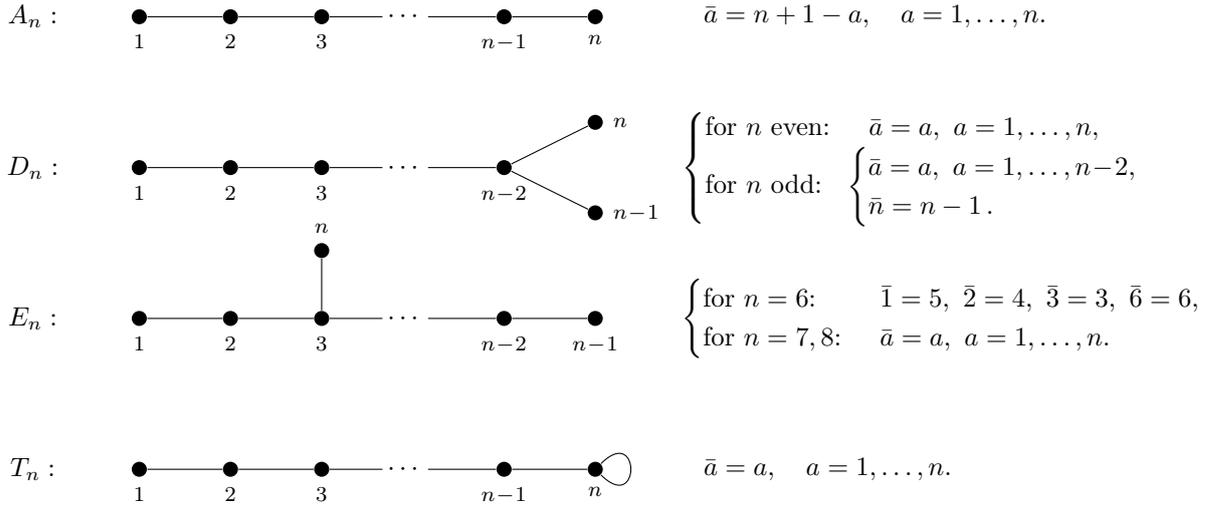

\subsection{Reflectionless theories and minimal amplitudes} 
In \textit{purely elastic scattering theories}, no reflection between particles is allowed. All interactions are transmissive, and the S-matrix becomes diagonal:
\begin{equation}
    S_{ab}^{ cd}(\vartheta) = \delta_a^d\delta_b^cS_{ab}(\vartheta)\,.
\end{equation}
This diagonal form dramatically simplifies the bootstrap program. Starting in the early 1990s, systematic studies of these transmissive S-matrices uncovered large families of minimal solutions to the bootstrap equations \cite{Braden:1989bu, Braden:1989bg, Dorey:1990xa, Klassen:1989ui, Klassen:1990dx, Zamolodchikov:1991et}. These developments led to a partial classification of two-dimensional integrable quantum field theories associated with simply-laced Lie algebras and certain generalisations. The resulting models are collectively known as the ADET scattering theories. The name reflects the correspondence with the $A$, $D$ and $E$ families of simply-laced Dynkin diagrams, with the additional $T$ corresponding to the ``tadpole'' diagram (see Figure \ref{DDs}). A Dynkin diagram $G$ is a finite, undirected graph whose nodes correspond to the simple roots of a Lie algebra. In the ADET theories, all edges are single and unweighted, except for the tadpole diagram, which contains a self-connection. All diagrams in the ADET family are represented in Figure \ref{DDs}. In all such cases, the diagram can be unambiguously represented in terms of its incidence matrix $\mathcal{G}_{ab}$, defined as follows:
\begin{equation}
    \mathcal{G}_{ab} = \begin{cases}1 & \text { if nodes } a \text { and } b \text { are connected by an edge in } G\,, \\ 0 & \text { otherwise\,. }\end{cases}
\end{equation}
In ADET scattering theories, the Dynkin diagram $G$ plays a central role in organising the physical content of the model: its structure dictates the spectrum, interactions, and conserved quantities of the theory, as follows. Each node $a \in G$ corresponds to a distinct particle species in the spectrum. Bound states and fusion rules follow directly from the connectivity of the diagram: two particles $a$ and $b$ can form a bound state if and only if the nodes $a$ and $b$ are connected by an edge, i.e. if $\mathcal{G}_{ab}=1$. Since $\mathcal{G}_{ab}$ is symmetric -- if  node $a$ is connected to node $b$, then $b$ is likewise connected to $a$ -- it can be diagonalised by an orthogonal transformation, and its eigenvalues $\lambda_s$ take the universal form:
\begin{equation}\label{eigenvalues}
    \lambda_s = 2 \cos\left(\frac{s\pi }{h}\right)\,. 
\end{equation}
The integers $\{s_i\}_{i=1,\dots, \dim\mathcal{G}_{ab}}$ are called the exponents of the Lie algebra, and $h$ is the Coxeter number. For ADET Dynkin diagrams, their values are summarised in Table \ref{tab_cox}. The exponents of the algebra are in direct correspondence with the spins $s$ of the conserved charges $Q_s^\pm$ in the theory: in particular, the two sets coincide up to periodicities in $h$. Moreover, ratios among the charge eigenvalues on one-particle states are constrained by the eigenvalue equations:
\begin{equation}
      \mathcal{G}_{ab}q^\pm_{s,b}(\vartheta) = \lambda_s q_{s,a}^\pm(\vartheta)\,.
\end{equation}
 In particular, the case $s=1$ fixes the mass spectrum, up to an overall scale. In these models, the two-body scattering amplitudes can be expressed in a universal form. For each pair of particle species $a$ and $b$, associated respectively with two nodes of the Dynkin diagram $G$, the corresponding S-matrix element takes the form of an exponential Fourier transform:
\begin{equation}\label{minimal}
    S_{a b}(\vartheta)=\exp \int_{\mathbb{R}} \frac{d y}{y} \mathcal{K}_{a b}(y) e^{-i y \vartheta}\,.
\end{equation}
Here, $\mathcal{K}_{a b}(y)$ is a matrix-valued function that captures the analytic structure of the scattering process and is determined entirely by the incidence matrix $\mathcal{G}_{a b}$ of the underlying Dynkin diagram. Explicitly, one has:
\begin{equation}\label{kappa}
   \mathcal{K}_{a b}(y) = 2\cosh\left(\frac{\pi y}{h}\right)\left(  2\cosh\left(\frac{\pi y}{h}\right)-\mathcal{G}   \right)^{-1}_{ab}\,. 
\end{equation}
It is also a general property of these models that the forward-scattering amplitude satisfies $S_{aa}(0) = -1$, which reflects the effective fermionic statistics of these systems \cite{Zamolodchikov:1989cf}. Furthermore, the full two-particle S-matrix can be decomposed into products of elementary building blocks, and their origin can be traced back to the spectral properties of the incidence matrix $\mathcal{G}_{ab}$. After an orthogonal transformation, the functions $\mathcal{K}_{ab}(y)$ can be expressed directly in terms of the eigenvalues \eqref{eigenvalues}. In particular, the denominator in equation \eqref{kappa} admits a standard expansion into a finite sum over trigonometric functions, which reorganises the spectrum of $\mathcal{K}_{ab}(y)$ into a finite series of contributions labelled by integers $x \in \{1,\dots, h-1\}$, and the final result can be expressed schematically as \cite{Braden:1989bg}
\begin{equation}\label{block}
 S_{ab}(\vartheta) = \prod_{x=1}^{h-1} (x,\vartheta)^{\mathcal{X}_{ab}}\,,\quad (x,\vartheta) = \sinh\Big(\frac{\vartheta}{2}+\frac{i\pi x}{2h}\Big)\Big/\sinh\Big(\frac{\vartheta}{2}-\frac{i\pi x}{2h}\Big)\,,   
\end{equation}
where the multiplicities $\mathcal{X}_{ab} \in \mathbb{N}$ follow directly from the spectrum of $\mathcal{G}_{ab}$. For example, the Ising model discussed in the previous sections corresponds to the $A_1$ Dynkin diagram, which consists of a single, edgeless node, with a trivial incidence matrix $\mathcal{G}_{ab}=0$ and a Coxeter number $h=2$. The S-matrix simply encodes the fermionic exchange statistics, with $S(\vartheta) = -1$. Beyond their algebraic elegance, these theories have a clear physical significance, as they describe integrable deformations of two-dimensional conformal field theories by relevant operators. A celebrated example is the scattering theory associated with the $E_8$ algebra, which emerges from the integrable perturbation of the critical Ising model by a magnetic field \cite{Zamolodchikov:1989fp}. More generally, the ADET models provide families of simple yet non-trivial examples of integrable QFTs in two dimensions, where the particle spectrum, scattering amplitudes, and conserved charges are all rigidly determined by algebraic and analytic constraints. In the following sections, we will demonstrate how these minimal models can be naturally extended to incorporate internal symmetries, thereby enriching their structure while preserving integrability.

\renewcommand{\arraystretch}{1.3}

\begin{table}
\centering
\begin{tabular}{|l|l|l|}
\hline
\textbf{Diagram $G$} & \textbf{Exponents $s_i$} & \textbf{Coxeter number $h$} \\
\hline
$A_n$ & $1,\,2,\,\dots,\,n$ & $n+1$  \\ \hline 
$D_n$  & $1,\,3,\,5,\,\dots,\,2n-3,\,n-1$ & $2n-2$\\\hline
$E_6$  & $1,\,4,\,5,\,7,\,8,\,11$ & $12$\\\hline
$E_7$  & $1,\,5,\,7,\,9,\,11,\,13,\,17$& $18$ \\\hline
$E_8$  & $1,\,7,\,11,\,13,\,17,\,19,\,23,\,29$ & $30$\\\hline
$T_n$  & $1,\,3,\,5,\,\dots,\,2n-1$ & $2n+1$\\ \hline
\end{tabular}

\caption{Coxeter numbers and exponents for the ADET Dynkin diagrams.}
\label{tab_cox}
\end{table}

\subsection{Cyclic identities for minimal S-matrices}\label{identity}

We now turn to the scattering description and examine a key functional relation satisfied by the minimal two-particle amplitudes $S_{ab}(\vartheta)$ introduced in equations \eqref{minimal}. These amplitudes capture the analytic structure of the theory, and we will show that, after suitable reparametrisations of rapidity space, they factorise into $n$-fold products. The resulting structure reproduces, at the level of the S-matrix, the same graded deformation of functional relations that we previously uncovered in the Ising model. 

We consider integers $w \geq 0$, $n \geq 1$ such that $\operatorname{gcd}(w, n)=1$, and a sign $\xi= \pm 1$. We use these parameters to define the conformal maps
\begin{equation}\label{eq:reparam_map}
    f_k(\vartheta) = \frac{\xi}{n}(\vartheta-2\pi i w k)\,.
\end{equation}
Each function $f_k(\vartheta)$ parametrises one of the $n$ sheets of a conformal covering of the complex $\vartheta$-plane. The parameter $w$ fixes the winding number of this covering, while the sign $\xi$ determines its overall orientation. Given these maps, the minimal two-particle amplitudes $S_{ab}(\vartheta)$ can be evaluated consistently on all sheets, and one can define the product
\begin{equation}\label{product_P}
    P_{ab}(\vartheta) = \prod_{\ell\in\mathbb{Z}_n} S_{ab}(f_{k-\ell}(\vartheta))\,.
\end{equation}
For generic values of $n$, the quantity \eqref{product_P} does not simplify further. Nevertheless, as we show below, a remarkable closure occurs for specific combinations of parameters satisfying:
\begin{equation}\label{condition_2}
    n = 2wh + \xi\,,
\end{equation}
where $h$ is the Coxeter number of the underlying Lie algebra. In this case, the product \eqref{product_P} reproduces the original amplitude, $P_{ab}(\vartheta) =  S_{ab}(\vartheta)$. To prove this result, we rely on the representation \eqref{block} of the minimal amplitudes, which decomposes each $S_{ab}(\vartheta)$ into a product of fundamental blocks. The proof then reduces to a trigonometric identity that reorganises the product over shifts. Here, we use the relation (see, for example, page 41 of \cite{Gradshteyn:1943cpj})
\begin{equation}\label{trig}
 2\sin{\vartheta}=2^{n} \prod_{\ell\in \mathbb{Z}_n} \sin \left(\frac{\vartheta}{n} - \frac{\pi \ell}{n}\right),
\end{equation}
which mirrors precisely the multiplicative structure \eqref{product_P} appearing in $P_{\alpha\beta}(\vartheta)$. Because the product ranges over all elements of the cyclic group $\mathbb{Z}_n$, any bijective (invertible) reparametrisation of the index $\ell$ modulo $n$ merely permutes the factors and therefore does not change the product's value. Moreover, since $w$ and $n$ are coprime, multiplication by $w$ defines an automorphism of the additive group $\mathbb{Z}_n$, and the substitution $\ell \mapsto w(k-\ell)$ simply permutes the factors in the product without affecting its value. After generalising this identity to the hyperbolic case, one finds that the minimal blocks \eqref{block} can be parametrised as:
\begin{equation}
    (x,\vartheta) =  \prod_{\ell\in \mathbb{Z}_n}\left(\frac{x}{n}, \xi f_{k-\ell}(\vartheta)\right)\,.
\end{equation}
It is important to observe that the factor $1 / n$ appearing in the first argument on the right-hand side modifies the algebraic structure of the building blocks. In particular, this rescaling maps the Coxeter number $h$ to $nh$. However, writing each block explicitly, and introducing the auxiliary label $ \delta = {x (n-\xi)}/({2 w h})$, we can write:
\begin{equation}
    (x,\vartheta) =\prod_{\ell\in \mathbb{Z}_n} \sinh\left(\frac{\xi f_{k-\ell+\delta}( \vartheta)}{2}+\frac{i\pi \xi x}{2 h}\right)\Big/\sinh\left(\frac{\xi f_{k-\ell-\delta}( \vartheta)}{2}-\frac{i\pi \xi x}{2 h}\right)\,.
\end{equation}
When $\delta$ is an integer, the corresponding overall phases can be absorbed by redefining the dummy index $\ell$, without altering the product’s final value. This happens exactly when $n$ satisfies the closure condition \eqref{condition_2}. Multiplying both arguments of the hyperbolic sine by $\xi$, we obtain the equivalent expression:
\begin{equation}\label{flip}
\begin{split}
    (x,\vartheta) =\prod_{\ell\in \mathbb{Z}_n} \sinh\left(\frac{ f_{k-\ell}( \vartheta)}{2}+\frac{i\pi x}{2 h}\right)\Big/\sinh\left(\frac{ f_{k-\ell}( \vartheta)}{2}-\frac{i\pi x}{2 h}\right)  =\prod_{\ell\in \mathbb{Z}_n} (x,f_{k-\ell}(\vartheta))\,.
   \end{split}
\end{equation}
Applying the identity \eqref{flip} to each block appearing in equation \eqref{block}, we conclude that $P_{ab}(\vartheta) = S_{ab}(\vartheta)$. As a side remark, we observe that a completely analogous discussion can be generalised to the sinh-Gordon theory.

\section{The S-matrix bootstrap for graded IQFTs}\label{SEC:SMAT}

The $\mathbb{Z}_n$-graded structure emerging from the analytic decomposition of the two-body S-matrix admits a natural interpretation: it corresponds to lifting the asymptotic one-particle states across the $n$ cyclic sectors of the rapidity plane defined by the reparametrisation maps $f_k(\vartheta)$. Each particle species $a$ in the asymptotic Hilbert space is promoted to an $n$-component multiplet $\{a_k\}_{k\in\mathbb{Z}_n}$, whose elements are related by
\begin{equation}\label{rep_states}
    \ket{\vartheta, a_k} = \ket{f_k(\vartheta), a}.
\end{equation}
With this identification, the scattering of a particle of type $a_k$ (and rapidity $\vartheta_1$) with a particle of type $b_{\ell}$ (and rapidity $\vartheta_2$) is described by the \textit{graded} amplitude:
\begin{equation}\label{graded_ampl}
      S_{a_kb_\ell}(\vartheta) = S_{ab}(f_k(\vartheta_1) - f_\ell(\vartheta_2)) = S_{ab}(f_{k-\ell}(\vartheta))\,,
\end{equation}
where $\vartheta = \vartheta_1-\vartheta_2$. The amplitudes \eqref{graded_ampl} are more than a notational refinement; rather, they reflect a finer structure already present in the original theory. As shown in Section \ref{identity}, the minimal two-body amplitudes $S_{a b}(\vartheta)$ can be factorised into cyclic products of elementary blocks, each evaluated at shifted, rescaled rapidities. This means that every $S_{a b}(\vartheta)$ already contains several intertwined analytic components. The graded amplitudes $S_{a_k b_{\ell}}(\vartheta)$ isolate these components explicitly: they represent the elementary scattering processes resolved within the different $\mathbb{Z}_n$ sectors of the rapidity plane. Finally, because each amplitude $ S_{a_kb_\ell}(\vartheta) $ depends on the discrete labels $k$ and $\ell$ only through their difference, and given that $S_{ab}(f_{k+n}(\vartheta)) = S_{ab}(f_{k}(\vartheta))$, the graded scattering theory is symmetric under the $\mathbb{Z}_n$ group.

Note that construction with a superficially similar spirit was explored in \cite{Fring:2000ng}. There, the grading enters through a generalised CDD factor: the analytic structure and the minimal S-matrix remain unchanged, while the deformation modifies only the nonminimal part via a multiplicative phase. In contrast, the present work introduces the $\mathbb{Z}_n$ grading at a more fundamental level, as it acts already on the minimal scattering block.

\subsection{Consistency conditions, bound states and bootstrap relations}
In graded QFTs, braiding unitarity follows directly from the analytic property $f_{k}(-\vartheta) = -f_{-k}(\vartheta)$. As a result, in each $\mathbb{Z}_n$ sector one finds
\begin{equation}\label{braid_grad}
    S_{a_kb_\ell}(\vartheta) S_{b_\ell a_k}(-\vartheta) = 1\,.
\end{equation}
Similarly, a generalised notion of crossing symmetry (see equation \eqref{cs}) can be formulated by observing that the maps $f_k(\vartheta)$ obey $ i\pi- f_k(\vartheta) =  f_{-k-\xi h}(i\pi-\vartheta)$. This implies that the interchange of a particle with its antiparticle induces a simultaneous relabelling of the discrete index $k$, yielding the generalised crossing relation
\begin{equation}
    S_{a_kb_\ell}(\vartheta) = C_{ac}D_{kj} S_{b_\ell c_j}(i\pi-\vartheta)\,.
\end{equation}
Here, $C_{ab} = \delta_{b\bar a}$, with $\bar a$ denoting the conjugate representation (see Figure \ref{DDs}), whereas the operator $D_{k\ell}$ acts linearly on the graded indices as
\begin{equation}
   D_{k\ell}\ket{\vartheta, a_\ell} = \ket{\vartheta, a_{k+\xi h}}.
\end{equation}
Note that, while the charge-conjugation matrix $C_{a b}$ satisfies $C_{a c} C_{c b}=\delta_{a b}$, the operator $D_{k \ell}$ is in general not an involution. Instead, it generates a cyclic translation on the graded indices, and it satisfies $\left(D^n\right)_{k \ell}=\delta_{k \ell}$. As a result, crossing symmetry in the graded theory intertwines standard charge conjugation with a non-trivial cyclic rotation in the rapidity-sheet index, reflecting the multi-valued analytic structure of the $\mathbb{Z}_n$ covering of the rapidity plane. Although both braiding unitarity and crossing symmetry pull-back along the reparametrisations \eqref{rep_states}, Hermitian analyticity does not directly extend to the graded case. Instead, one finds $(S_{a_kb_\ell}(\vartheta))^* = S_{b_ka_\ell}(-\vartheta^*)$, which differs from the usual relation by the exchange of the $\mathbb{Z}_n$ indices. Since, in addition to \eqref{braid_grad}, Hermitian analyticity is also required for quantum unitarity and probability conservation (see the discussion below equation \eqref{braid}), this observation suggests that the physical Hilbert space should be restricted so that the full many-body scattering amplitudes remain Hermitian analytic. In practice, this amounts to identifying configurations related by the equivalence relation $k \sim n-k$, which enforces an effective reflection symmetry among the graded sectors. While this condition is not imposed a priori, it naturally emerges from the Thermodynamic Bethe Ansatz analysis, where the same equivalence appears in the structure of the pseudoenergy functions.\\

\noindent\textbf{Bound states and the graded bootstrap}. In integrable QFTs, the analytic structure of the two-particle S-matrix encodes detailed information about the spectrum. In particular, simple poles of $S_{ab}(\vartheta)$ in the physical strip $\Im{\vartheta} \in(0, \pi)$ signal the presence of bound states. A pole at $\vartheta = i u^c_{ab}$ indicates that particles of species $a$ and $b$, with rest masses $m_a$ and $m_b$, can form a bound state $c$ whose mass is determined by the relativistic dispersion relation:
\begin{equation}\label{shell}
m_c^2=m_a^2+m_b^2+2 m_a m_b \cos u_{ab}^c \,.    
\end{equation}
In particular, using the decomposition \eqref{block} for the two-body minimal amplitudes, we see that each fundamental block $(x,\vartheta)$ has a simple pole in the physical strip at
\begin{equation}
    \vartheta = iu_{ab}^c= \frac{ i\pi x}{h}  \,,\quad x=1, \ldots, h-1\,,
\end{equation}
with bound-states of species $c$ occurring when in the presence of a simple pole, i.e. precisely at those $x$ for which $\mathcal{X}_{ab}(x)=1$. Near the pole,
\begin{equation}
    (\vartheta-i u_{ab}^c)S_{ab}(\vartheta) \simeq \operatorname{Res}\left\{S_{ab}(\vartheta), \vartheta = i u_{ab}^c\right\}\,,
\end{equation}
with the residue fixing the three-particle coupling. In unitary field theories, positivity of the residue ensures positivity of the intermediate state’s norm and consistency with probability conservation, whereas non-unitary models may feature negative values of the residue, signalling a breakdown of conventional fusion unitarity. In the graded theory, because $S_{a_k b_{\ell}}(\vartheta)=S_{ab}(f_{k-\ell}(\vartheta))$, bound-state poles of the graded amplitudes occur when $f_{k-\ell}(\vartheta) = i u_{ab}^c$, which means :
\begin{equation}
    \vartheta\equiv i\xi n u_{ab}^c-2 \pi i w(k-\ell) \bmod 2 \pi i n\,.
\end{equation}
Requiring that the pole lie in the physical strip selects the $n$ sectors that satisfy $k-\ell \equiv - \xi x \bmod n$. Only these graded components exhibit a physical bound-state pole at the same rapidity value as in the ungraded theory. Moreover, for fixed $k-\ell$, we observe that the residue at the poles scales as
\begin{equation}
\operatorname{Res}\left\{S_{a_kb_\ell}(\vartheta), f_{k-\ell}(\vartheta) = i u_{ab}^c\right\}  = \xi n\operatorname{Res}\left\{S_{ab}(\vartheta), \vartheta = i u_{ab}^c\right\}\,,  
\end{equation}
while the position is determined by the selection rule above. Thus, the integer $n$ rescales the overall coupling strength, while $\xi= \pm 1$ may flip the residue's sign. Naively, one might expect that such a sign change signals a potential violation of one-particle unitarity in certain sectors of an otherwise unitary theory, or conversely, its restoration in non-unitary models, such as the graded extension of the Lee–Yang model. However, the situation is more subtle here, as the physical unitarity condition is generally violated for generic, that is, unrestricted, many-particle scattering states.

Suppose now that two particle species $a$ and $b$ fuse to a bound state $c$. Consistency of all amplitudes with this fusion -- the \textit{bootstrap principle} -- imposes functional relations among two-body S-matrices. In ADET theories, these constraints can be summarised in a compact fusion relation, which implicitly contains the full set of bootstrap equations, and reads
\begin{equation}\label{bootstr}
    S_{ab}(\vartheta^+) S_{ab}(\vartheta^-) =\prod_{c \in G}\left(S_{ac}(\vartheta)\right)^{\mathcal G_{bc}} e^{-2 \pi i  \mathcal  G_{ab} \Theta(\vartheta)}\,, 
\end{equation}
where $\vartheta^\pm = \vartheta \pm i \pi/h$. Here, $\Theta(\vartheta)$ is a smoothed step function interpolating between $0$ and $1$, taking the value $1/2$ at the origin. This regularisation ensures continuity across $\vartheta=0$ while preserving both braiding unitarity (see equation \eqref{braid}) and the correct particle statistics (i.e., $S_{aa}(0)=-1$). The same logic extends naturally to graded models, where particle multiplets are organised into cyclic families $a_k$ related by internal automorphisms of order $n$. Rescaling and shifting the rapidities in \eqref{bootstr}, and tracing carefully how the fusion conditions propagate across sectors, one arrives at the graded generalisation of the bootstrap relation:
\begin{equation}\label{grad_bstr}
    S_{a_{k-\xi} b_{\ell}}(\vartheta^+)S_{a_{k+\xi} b_{\ell}}(\vartheta^-) = \prod_{c\in G} \left(S_{a_{k} c_{\ell}}(\vartheta)\right)^{\mathcal G_{bc}} e^{-2 \pi i \mathcal G_{ab} \Theta(\vartheta)}\,,
\end{equation}
with $\vartheta^{ \pm}$ defined as in \eqref{bootstr}. Equation \eqref{grad_bstr} preserves the self-consistency of the bootstrap while enriching its structure: the incidence matrix $\mathcal{G}_{ab}$ still determines the fusion rules, but each node of the Dynkin diagram now unfolds into a cyclic family of sectors, related to one another through the maps $f_k(\vartheta)$. Closure of the graded bootstrap is ensured precisely when the parameters satisfy $n = 2wh + \xi$, the same condition that guarantees the cyclic factorisation of the two-particle amplitudes. In this way, integrability can accommodate the discrete $\mathbb{Z}_n$ symmetry without compromising its analytic or algebraic coherence. 

\subsection{Fractional-spin charges and generalised CDD deformations}
In $2d$ IQFTs, the existence of an infinite set of conserved charges strongly constrains the structure of the scattering amplitudes.  A typical consequence is that, at large rapidity, the minimal two-particle S-matrix can be systematically expanded in terms of these conserved quantities:
\begin{equation}\label{start_here}
    \log S_{ab}(\vartheta) = \int_\mathbb{R} \frac{\dd y}{y}\mathcal K_{ab}(y)e^{-iy\vartheta} \simeq \text{const.} + \sum_{s\in\mathcal{S}} c_{s,ab} e^{-s\vartheta}\,,
\end{equation}
where $\mathcal{S}$ is the set of Lorentz spins of the conserved charges. For ADET theories, $\mathcal{S}$ is in direct correspondence with the set of exponents of the algebra -- see equation \eqref{eigenvalues} and the comments that follow. The expansion \eqref{start_here} can be easily recovered using
the following spectral representation of the matrix $\mathcal K_{ab}(y)$, which comes as a direct consequence of equation \eqref{eigenvalues}: 
\begin{equation}\label{kappa(y)}
   \mathcal K_{ab}(y) =2\sum_{s\in \mathcal{S}/\sim} \frac{\mathcal{U}_{as} \mathcal{U}_{bs}\cos(\pi y/h)}{2\cos(\pi y/h)-\lambda_s} \,.
\end{equation}
The matrices $\mathcal{U}_{as}$ form an orthogonal basis that diagonalises the incidence matrix $\mathcal{G}_{ab}$, and the equivalence relation $\sim$ identifies indices differing by an integer multiple of the Coxeter number $h$. Each factor in \eqref{kappa(y)} has simple poles where the denominator vanishes, namely, for
\begin{equation}
    y_* \equiv \pm is \bmod 2h\,,
\end{equation}
so that poles of $K_{ab}(y)$ lie on the imaginary axis, with the nearest ones at $\pm i$. In the limit of large $\vartheta$, one integrates \eqref{start_here} in the complex $y$-plane, closing the contour in the lower half-plane. The constant term comes from a small contour deformation around $y=0$, which avoids the $1/y$ singularity, whereas the coefficients $c_{s, ab}$ can be determined explicitly in terms of the exponents of the algebra and the matrices $\mathcal{U}_{as}$. When the integral representation \eqref{start_here} is pulled back along the maps  $f_{k-\ell}(\vartheta)$, and the result is expanded for large values of $\vartheta$, one finds:
\begin{equation}\label{fractional_exp}
    \log S_{a_kb_\ell}(\vartheta) \simeq \text{const.} + \sum_{s\in\mathcal{S}} c_{s,ab}  e^{-2\pi i \xi w (k-\ell)s/n} e^{-s\vartheta/n}\,. 
\end{equation}
The exponential suppression factor $e^{-s \vartheta / n}$ is independent of the sign of $\xi$, while the difference between the cases $\xi=\pm1$ manifests only as a phase multiplying each term in the expansion. As a direct consequence of the grading, the expansion receives non-trivial contributions with fractional effective spins $s_{\text {eff }}=s / n$. At first sight, this suggests that the graded amplitudes are governed by a spectrum of fractional-spin charges, with occasional integer values when $s$ and $n$ are coprime. However, the integer-spin modes that contributed to \eqref{start_here} do not disappear. Because the parameters satisfy $n \equiv \pm 1 \bmod 2 h$, multiplication by $n$ simply permutes the exponents of the Lie algebra. As a result, the graded expansion necessarily contains terms with $s_\text{eff} \in \mathcal{S}$, and in fact the entire tower of integer-spin charges is preserved. Thus, the integer-spin charges of the ungraded theory remain present and protected: the reparametrisations $f_k(\vartheta)$ do not replace them but embed them inside a denser spectrum of fractional spins. Finally, ratios between conserved charges remain consistent, because the incidence matrix eigenvalues satisfy $\lambda_s=\lambda_{n s}$ for $n=2 w h+\xi$ (see equation \eqref{eigenvalues}).

As a final remark, we observe that fractional-spin conserved charges are not an exclusive feature of the graded construction introduced here. In fact, nonlocal integrals of motion with fractional Lorentz spin appear already in conventional integrable quantum field theories. A well-known example occurs in the sine-Gordon model, where the large-rapidity asymptotic expansion of Baxter’s Q-functions reveals the presence of conserved quantities carrying fractional spin \cite{Bazhanov:1994ft, Bazhanov:1996dr}. \\

\noindent \textbf{Generalised $T\overline{T}$ flows from fractional-spin charges}. The fractional-spin expansion suggests that the graded theory supports a richer hierarchy of conserved charges that can be used to generate controlled deformations of the S-matrix. A standard mechanism for deforming integrable QFTs -- without spoiling factorisation, unitarity, or crossing -- is the introduction of dynamical CDD factors. In the standard setting, $T\overline{T}$ deformations \cite{ Cavaglia:2016oda, Smirnov:2016lqw} and their higher-spin generalisations introduce multiplicative phase factors built from bilinears of conserved charges. These preserve unitarity, crossing symmetry and the Yang–Baxter equation, and, in the large-$\vartheta$ expansion, they only modify the coefficients $c_{s, ab}$, while preserving of tower of conserved charges. Since additional contributions naturally emerge in expression \eqref{fractional_exp}, it is natural to consider fractional-spin analogues of these deformations.  We introduce them as CDD factors of the form
\begin{equation}\label{szcdd}
    \mathcal F_{a_kb_\ell}(\vartheta, s) = \exp( i\mu q_{s,a} q_{s,b} \sinh(sf_{k-\ell}(\vartheta)))\,,
\end{equation}
where $q_{s, a}\equiv q_{s, a}^\pm(\vartheta=0)$ is the rest-frame eigenvalue of the spin-$s$ conserved charge for species $a$, and the parameter $\mu$ controls the strength of the deformation. Factors of the form \eqref{szcdd} are bona fide CDD deformations: they are analytic in the physical strip, satisfy generalised notions of braiding unitarity and crossing, and thus preserve factorisation of the S-matrix. Moreover, when all the $\mathbb{Z}_n$ sectors are recombined, the fractional-spin deformation trivialises, since
\begin{equation}
    \prod_{\ell\in\mathbb{Z}_n}F_{a_kb_\ell}(\vartheta, s) = 1\,.
\end{equation}
\section{Thermodynamics of graded scattering theories}\label{SEC:TBA}
In infinite volume, the two-particle S-matrix provides a complete and surprisingly economical description of the Hilbert space of a $2d$ integrable quantum field theory. All interactions are encoded in the exact two-body amplitudes, and every multi-particle process factorises into a sequence of elastic two-particle scatterings. When the theory is placed on a spatial circle of circumference $\beta$, this picture changes drastically. Particles are no longer asymptotically independent: their worldlines wrap around the compact direction and interact with themselves and with other particles. These wrapping events correspond to virtual processes that circle the cylinder. In perturbative language, they appear as exponentially suppressed Lüscher corrections, coming from virtual particles propagating around the compact dimension. Directly quantising the theory in such a setting quickly becomes intractable, because one must sum over infinitely many of these virtual windings. The Thermodynamic Bethe Ansatz (TBA), originally formulated by Yang and Yang in the study of the Lieb–Liniger model \cite{Yang:1968rm}, and later extended by Al. B. Zamolodchikov to relativistic IQFTs \cite{Zamolodchikov:1989cf}, offers a remarkably elegant solution to this problem. Rather than tackling the finite-size system directly, one exploits the equivalence between the spectrum of a theory on a cylinder of circumference $\beta$ and the thermodynamics of the same theory in infinite volume at temperature $T=1 / \beta$. In this mirror picture, energy levels on the cylinder correspond to free-energy densities of a thermal ensemble, while virtual particles propagating around the spatial circle appear as thermal excitations winding along the Euclidean time direction. 
\subsection{TBAs and GGEs for $2d$ IQFTs}
The logic behind the TBA construction goes as follows. For simplicity, let us start by considering a single particle species of mass $m$; the generalisation to multiple species is straightforward. We work in the mirror channel, where the roles of space and Euclidean time are interchanged, and take the spatial circumference $L$ to be large. The theory is assumed to be integrable with diagonal scattering, with a two-body S-matrix $S(\vartheta)$ encoding the phase shift between particles as a function of their rapidity difference $\vartheta$.
In the infinite-volume limit, multi-particle states are labelled by continuous rapidities. When the system is compactified on a circle of size $L$, however, particle momenta become quantised. Each particle, when circling the system, accumulates the dynamical phase $e^{i p_j L}$ from free propagation, together with additional phases from elastic collisions with all other particles.
Requiring the many-body wavefunction to remain single-valued leads to asymptotic Bethe equations of the form:
\begin{equation}\label{bethe_eq}
    e^{i m L \sinh\vartheta_j} \prod_k S(\vartheta_j-\vartheta_k) = 1\,.
\end{equation}
In the thermodynamic limit, the discrete set of solutions to these equations becomes dense and can be described by continuous rapidity distributions. These distributions correspond to equilibrium configurations minimising the free energy at inverse temperature $\beta$. The result is a self-consistent non-linear integral equation for the pseudoenergy function $\varepsilon(\vartheta)$, whose solution completely determines the thermodynamic properties of the system. Once $\varepsilon(\vartheta)$ is known, the free energy density follows directly. For ADET-type integrable models \cite{Zamolodchikov:1991et, Klassen:1990dx, Ravanini:1992fi}, the TBA equations take the general form:
\begin{equation}\label{TBA}
   \varepsilon_{a}(\vartheta) = m_a\beta\cosh\vartheta - \sum_{b\in G}\int_\mathbb{R} \frac{\dd y}{2\pi}\varphi_{ab}(\vartheta-y) \log\big(1+e^{-\varepsilon_b(y)}\big)\,,
\end{equation}
where the index $a$ labels different particle species, ranging over the nodes of the Dynkin diagram $G$, and the scattering kernels $\varphi_{ab}(\vartheta)=-i \partial_\vartheta\log S_{ab}(\vartheta)$ encode the effective two-body interactions between particles of type $a$ and $b$. The energy term $\nu_a(\vartheta) = m_a\beta \cosh\vartheta$ is called the driving term, and it reflects the underlying relativistic Gibbs ensemble. One immediately observes that \eqref{TBA} generalises the familiar case of the Ising model, where $G$ has a single, disconnected node, the S-matrix $S(\vartheta) = -1$ is trivial, and the pseudoenergy reduces to the free-particle form. If the system of coupled non-linear integral equations \eqref{TBA} can be solved  -- typically relying on iterative numerical methods, where the pseudoenergies $\varepsilon_a(\vartheta)$ are sequentially updated until convergence -- the free energy of the theory at inverse temperature $\beta$ follows directly from the equilibrium distribution. It is given by:
\begin{equation}
    f(\beta) = -\sum_{a \in G}\frac{m_a}{2\pi\beta}\int_\mathbb{R}\dd\vartheta  \cosh\vartheta \log \big(1+e^{-\varepsilon_a(\vartheta)}\big)\,.
\end{equation}
Finally, returning to the original channel, the ground-state energy of the theory on a cylinder of circumference $\beta$ and length $L$ is given by $E(\beta)=\beta f(\beta)$. Similarly to the free case (see equations \eqref{en_gr}–\eqref{c_e_is}), it is convenient to introduce a dimensionless parameter $r$ that controls the physical length scales at which the theory is probed. Since several particle species contribute to the ground-state energy, we select a reference mass $m_*$ (typically, $m_*=m_1$) and define $r=m_* \beta$. Introducing the reduced masses $\hat{m}_a=m_a / m_*$, the ground-state scaling function can then be expressed as
\begin{equation}\label{ceTBA}
      c_\text{eff}(r) = \frac{3r}{\pi^2} \sum_{a\in G}\int_\mathbb{R}\dd\vartheta \,\hat{m}_a  \cosh\vartheta \log \big(1+e^{-\varepsilon_a(\vartheta)}\big)\,.
\end{equation}
Finally, when combining the non-linear integral equations \eqref{TBA} with the bootstrap relations \eqref{bootstr}, one obtains a set of functional relations for the quantities $Y_a(\vartheta) = e^{\varepsilon_a(\vartheta)}$, known collectively as the Y-system. For ADET-type theories, these relations take the universal form:
\begin{equation}\label{Y-system-ade}
    Y_a(\vartheta^+)Y_a(\vartheta^-) = \prod_{b\in G}(1+Y_b(\vartheta^+))^{\mathcal{G}_{ab}}\,,
\end{equation}
where the rapidity shifts $\vartheta^{ \pm}=\vartheta \pm i \pi / h$ are determined by the Coxeter number $h$ of the corresponding Lie algebra, and $\mathcal{G}_{a b}$ is the adjacency matrix of its Dynkin diagram. The functions $Y_a(\vartheta)$ satisfy the  periodicities 
\begin{equation}\label{period_ade}
    Y_a(\vartheta + i \pi P) = Y_{\bar a}(\vartheta)\,,\quad P =\frac{h+2}{h}\,,
\end{equation}
where $\bar a$ is the antiparticle of $a$. A formal proof of \eqref{period_ade} can be found in \cite{keller2012periodicityconjecturepairsdynkin}. The above periodicity property can be shown to be related to the conformal dimension $\Delta$ of the operator perturbing the UV conformal field theory,
\begin{equation}\label{period_delta}
  \Delta=\begin{cases}
      1-1/P & \text{if }G = A_n,D_n,E_n \,,\\
      1-2/P & \text{if }G= T_n\,.
  \end{cases} 
\end{equation}
In complete analogy with the graded free theories discussed in Section \ref{Majorana:SEC}, the thermodynamics of interacting, integrable QFTs admits a natural generalisation in which the standard Gibbs weight is replaced by a generalised Gibbs ensemble. In this framework, the conserved charges $Q_{s}^\pm$ of spin $s \in \mathcal{S}$ enter the equilibrium density matrix coupling to generalised inverse temperatures $\beta_s$. In the absence of net fluxes, one has
\begin{equation}
    \nu_a(\vartheta) = \sum_{s\in\mathcal{S}} \frac{\beta_s}{2} (q^+_{s,a}(\vartheta) + q_{s,a}^-(\vartheta))\,,
\end{equation}
with $q^\pm_{s,a}(\vartheta)$ denoting the eigenvalue $Q^\pm_{s}$ when acting on the asymptotic one-particle states $\ket{\vartheta,a}$. For the usual energy operator, corresponding to $s=1$, the associated inverse temperature $\beta_1=\beta$ coincides with the standard thermodynamic temperature. For higher-spin charges, relativistic invariance and scaling arguments imply that their eigenvalues behave as $q^\pm_{s,a}(\vartheta) \propto m_a^s e^{\pm s\vartheta}$. Introducing the reference values $q_{s, a}=q^\pm_{s, a}(\vartheta=0)$ and defining the dimensionless ratios $\hat{q}_{s, a}= q_{s, a} / q_{s, *}$ with respect to a reference charge $q_{s, *}$, we can express the driving terms in a manifestly dimensionless form. Setting $\gamma_s=\beta_s / \beta^s$ and reabsorbing an overall factor $q_{s, *} / m_*^s$ into the definition of $\gamma_s$, we finally obtain:
\begin{equation}
      \nu_a(\vartheta) = \sum_{s\in\mathcal{S}} \gamma_s \hat q_{s,a} r^s \cosh(s\vartheta)\,,
\end{equation}
which generalises the familiar thermal driving term to the full hierarchy of conserved quantities. Adding these extra driving terms to the TBA equations \eqref{TBA} gives:
\begin{equation}\label{TBA_GGE}
      \varepsilon_{a}(\vartheta) = \sum_{s\in\mathcal{S}}\gamma_s\hat q_{s,a} r^s\cosh(s\vartheta) - \sum_{b\in G}\int_\mathbb{R} \frac{\dd y}{2\pi}\varphi_{ab}(\vartheta-y) \log\big(1+e^{-\varepsilon_b(y)}\big)\,,
\end{equation}
and the generalised ground-state scaling function $c_\text{eff}(r,\{\gamma_s\})$  is obtained after plugging the solution to \eqref{TBA_GGE} into definition \eqref{ceTBA}. It was shown in \cite{Castro-Alvaredo:2022pgz} that the Y-system \eqref{Y-system-ade} remains unchanged in the presence of a generalised Gibbs ensemble, as the generalised driving terms do not alter the functional relations themselves, but rather fix the asymptotic behaviour of their solutions. 
\subsection{Graded scattering: a thermodynamic perspective}
The TBA framework offers a natural way to lift a theory to its graded counterpart. One considers the pseudoenergies $\varepsilon_a(\vartheta)$ on a multi-sheeted rapidity plane, obtained by pulling them back along the reparametrisation maps 
\begin{equation}
    f_k(\vartheta) = \frac{\xi}{n}(\vartheta-2\pi i wk)\,,
\end{equation}
with $n = 2wh+\xi$. Doing this, the graded S-matrix elements of Section \ref{SEC:SMAT} emerge directly from the TBA construction, with the reparametrised kernels reproducing the same cyclic pattern that underlies the graded scattering theory. In practice, one defines the graded pseudoenergies $\varepsilon_{a_k}(\vartheta) = \varepsilon_{a}(f_k(\vartheta))$, and equation \eqref{TBA_GGE} yields:
\begin{equation}\label{gge+fk}
    \varepsilon_{a_k}(\vartheta) = \sum_{s\in\mathcal{S}}\gamma_s\hat q_{s,a} r^s\cosh(f_k(\vartheta)) - \sum_{b\in G}\int_\mathbb{R} \frac{\dd y}{2\pi}\varphi_{ab}(f_k(\vartheta)-y) \log\big(1+e^{-\varepsilon_b(y)}\big)\,. 
\end{equation}
The discussion of the driving-term contributions proceeds in close analogy with Section \ref{Majorana:SEC}. The charge of spin $s=n$ generates the energy-type term in the theory, thereby restoring the correct relativistic dispersion relation. Furthermore, since the eigenvalues $\lambda_s$ of the adjacency matrix $\mathcal{G}_{ab}$ satisfy the relation $\lambda_s= \lambda_{ns}$, the ratios among the spin-$s$ charges coincide with those among the spin-$n s$ charges. As a result, the consistency of each source term is automatically preserved. To restore the standard conventions, we can define the scaling parameter $r = \beta_n m^n$, and introduce the quantities $\alpha_s = \beta_s /\beta_n^{s/n}$, so that the new energy-type term scales linearly with $r$. The inverse temperatures can then be tuned so that the resulting driving term reproduces the statistical weights of a generalised Gibbs ensemble with additional fractional-spin contributions, each depending on the $\mathbb{Z}_n$ index $k$. Taking all these considerations into account, the graded TBA equations are sourced by terms of the form
\begin{equation}\label{def_nu_tba}
    \nu_{a_k}(\vartheta) = \sum_{s\in\mathcal{S}}\alpha_s\hat q_{s,a} r^{s/n}\cosh\!\left(\frac{s\vartheta}{n}-\frac{2 \pi i swk}{n}\right)\!\,.
\end{equation}
The interactions, controlled by terms appearing in the convolution on the right-hand side of \eqref{gge+fk}, can be more conveniently handled in Fourier space, where convolutions transform into standard multiplications. Thanks to the compact expression \eqref{minimal} for two-body amplitudes, the scattering kernels admit a simple Fourier representation in terms of Lie algebraic data, 
\begin{equation}
   \tilde \varphi_{ab}(y)= \delta_{ab} - \mathcal{K}_{ab}(y) = \delta_{ab}- 2\cosh\left(\frac{\pi y}{h}\right)\left(  2\cosh\left(\frac{\pi y}{h}\right)-\mathcal{G}   \right)^{-1}_{ab}\,.
\end{equation}
Translating a function introduces a rescaling of the integration variable in Fourier space (together with an overall Jacobian), while rescaling its argument produces a phase. We also observe that any pair of complex-valued functions $A(\vartheta)$ and $B(\vartheta)$ satisfy:
\begin{equation}
    \int_\mathbb{R} \dd yA(f_k(\vartheta)-y)B(y) = n\int_\mathbb{R} \dd y e^{-iy\vartheta} e^{2\pi ywk} \tilde A(n\xi y)\tilde B(n\xi y)\,.
\end{equation} 
Writing $e^{2\pi ywk} = e^{2\pi yw(k-\ell)} e^{2\pi yw\ell}$ and averaging over the $n$ images of the map $f_\ell(\vartheta)$ leads to the graded TBA equations:
\begin{equation}\label{grad-tba}
     \varepsilon_{a_k}(\vartheta) = \nu_{a_k}(\vartheta) - \sum_{b\in G}\sum_{\ell \in \mathbb{Z}_n}\int_\mathbb{R} \frac{\dd y}{2\pi}\varphi_{a_kb_\ell}(\vartheta-y) \log\big(1+e^{-\varepsilon_{b_\ell(y)}}\big)\,,
\end{equation}
where the kernels $\varphi_{a_kb_\ell}(\vartheta)$ appearing in \eqref{grad-tba} coincide with those obtained by differentiating the amplitudes $S_{a_kb_\ell}(\vartheta) = S_{ab}(f_{k-\ell}(\vartheta))$ introduced in Section \ref{SEC:SMAT}:
\begin{equation}\label{grad_ker}
  \varphi_{a_kb_\ell}(\vartheta)=  -i \partial_{\vartheta} \log S_{a_kb_\ell}(\vartheta)=\frac{\xi}{n} \varphi_{ab}(f_{k-\ell}(\vartheta))\,.
\end{equation}
When all the $\alpha_s$ are set to zero with the exception of $\alpha_n=1$, the driving term does not depend on the discrete $\mathbb{Z}_n$ index. Assuming a $k$-independent solution yields a self-consistent equation, with $\varepsilon_{a_k}(\vartheta) \equiv \varepsilon_a(\vartheta)$ and hence $L_{b_\ell}(\vartheta) = L_b(\vartheta)$. The sum over kernel collapses thanks to the cyclic identity $P_{ab}(\vartheta) = S_{ab}(\vartheta)$ (see equation \eqref{product_P} and below), and each sector independently obeys equation \eqref{TBA}. The graded system trivialises: one finds that the full set of equations reduces to $n$ decoupled copies of the original ungraded equations, and the total ground-state scaling function at fixed radius is simply $n$ times that of the ungraded theory. Under the reparametrisations $f_k(\vartheta)$, the ADET Y-system in equation \eqref{Y-system-ade} is mapped into the form:
\begin{equation}\label{graded_y_system_ADET}
Y_{a_{k-\xi}}(\vartheta^+)Y_{a_{k+\xi}}(\vartheta^-) = \prod_{b\in G}(1+Y_{b_k}(\vartheta^+))^{\mathcal{G}_{ab}}\,,
\end{equation}
with $\vartheta^\pm = \vartheta \pm i \pi/h$. Closure of the Y-system follows from $ f_k(\vartheta) \pm i\pi/h = f_{k\mp\xi}(\vartheta \pm i\pi/h)$, an identity that relies crucially on the condition $n = 2wh + \xi$. Moreover, as in the analysis below \eqref{gge+fk}, the grading acts so that all charge ratios -- and consequently the associated mass ratios -- remain the same. Under these mappings, the periodicity property of the original system \eqref{period_ade} becomes
\begin{equation}
  Y_{a_{k}}(\vartheta+i\pi P) = Y_{\bar a_{k+\xi h}}(\vartheta)\,,\quad   P = \frac{h+2}{h}\,.
\end{equation}
Note that, up to a permutation of the lower indices, the set of functions $\{Y_{a_k}\}$ for $a \in G$ and $k \in \mathbb{Z}$ is mapped into itself under $\vartheta \mapsto \vartheta+i\pi P$. Moreover, the appearance of the $\mathbb{Z}_n$-conjugate index $k+\xi h$ precisely matches the considerations of Section \ref{SEC:SMAT} concerning the behaviour of particles under crossing symmetry in the graded setting. \\

\noindent\textbf{Fourier decomposition}. As a final comment, we observe that a key simplification of the graded TBA system comes from exploiting the discrete $\mathbb{Z}_n$ symmetry carried by the index $k$. For any family of functions $X_k(\vartheta)$ with $k \in \mathbb{Z}_n$, we introduce the discrete Fourier transform
\begin{equation}\label{dft}
    \bar{X}^{(q)}(\vartheta)=\frac{1}{n} \sum_{k\in\mathbb{Z}_n}e^{-2 \pi i q k / n} X_k(\vartheta)\,,
\end{equation}
with $q \in \mathbb{Z}_n$. This representation diagonalises the action of the kernels. Since $\varphi_{a_k b_\ell}(\vartheta)$ depends only on the difference $k-\ell$, its Fourier transform is block-diagonal, and convolutions take the simple form:
\begin{equation}\label{f.c.}
    \sum_{\ell\in\mathbb{Z}_n} \int_\mathbb{R} \frac{\dd y}{2\pi} \varphi_{a_k b_\ell}(\vartheta-y) X_\ell(y)=\sum_{q\in\mathbb{Z}_n} e^{2 \pi i q k / n} \int_\mathbb{R} \frac{\dd y}{2\pi} \bar{\varphi}_{ab}^{(q)}(\vartheta-y) \bar{X}^{(q)}(y)\,.
\end{equation}
Thus, each Fourier component of the TBA equations evolves independently: the integral equations decouple into $n$ disjoint sectors labelled by $q$. In other words, the kernels conserve $\mathbb{Z}_n$ charge. This implies that if the input of the equations belongs to a definite Fourier sector, so does the output. In particular, the integer-spin GGE-type source terms $\propto \cosh(s\vartheta)$ contribute only to the neutral block, whereas the graded terms decompose into left- and right-moving phases of the form $e^{\pm 2\pi i wsk/n}$, and therefore reside entirely in the $q\equiv \pm sw \bmod n$ sectors. Note that the ground-state energy (and similarly the effective central charge) is controlled by the neutral sector $q=0$. Nevertheless, the charged sectors still matter dynamically, encoding excitations, possible twisted boundary conditions, and the manner in which conserved charges are dressed. 
\subsection{The graded scaling Lee–Yang model}\label{lygr:SEC}
Among all the theories in the ADET classification of two-dimensional integrable quantum field theories, the Lee–Yang model stands out as the simplest and most fundamental example. It corresponds to the $T_1$ tadpole diagram, consisting of a single, self-connected node. Despite this apparent simplicity, it captures many of the essential features of the TBA formalism in a remarkably transparent way. In the ultraviolet limit, the Lee–Yang model flows to a non-unitary conformal field theory that describes the Lee–Yang edge singularity \cite{Yang:1952be, Lee:1952ig, Fisher:1978pf, Cardy:1985yy} -- originally introduced in the context of statistical mechanics to characterise the distribution of zeros of the partition function in the complex magnetic field (or fugacity)  plane -- with effective central charge $c_{\mathrm{eff}}=2/5$. 
A closed form for the two-body S-matrix of the theory was first proposed in \cite{Cardy:1989fw}. The TBA equations capture the flow away from criticality induced by the unique relevant operator of the theory, and consist of a single non-linear integral equation of the form:
\begin{equation}\label{LY}
  \varepsilon(\vartheta)=r \cosh \vartheta-\int_\mathbb{R} \frac{\dd y}{2 \pi} \varphi\left(\vartheta-y\right) \log \big(1+e^{-\varepsilon(y)}\big) \,,\quad \varphi(\vartheta) = -\frac{4\sqrt{3}\cosh\vartheta}{1+2\cosh(2\vartheta)}\,. 
\end{equation}

\begin{figure}
    \centering
    \begin{subfigure}[b]{0.495\textwidth}
        \centering
        \includegraphics[width=\textwidth]{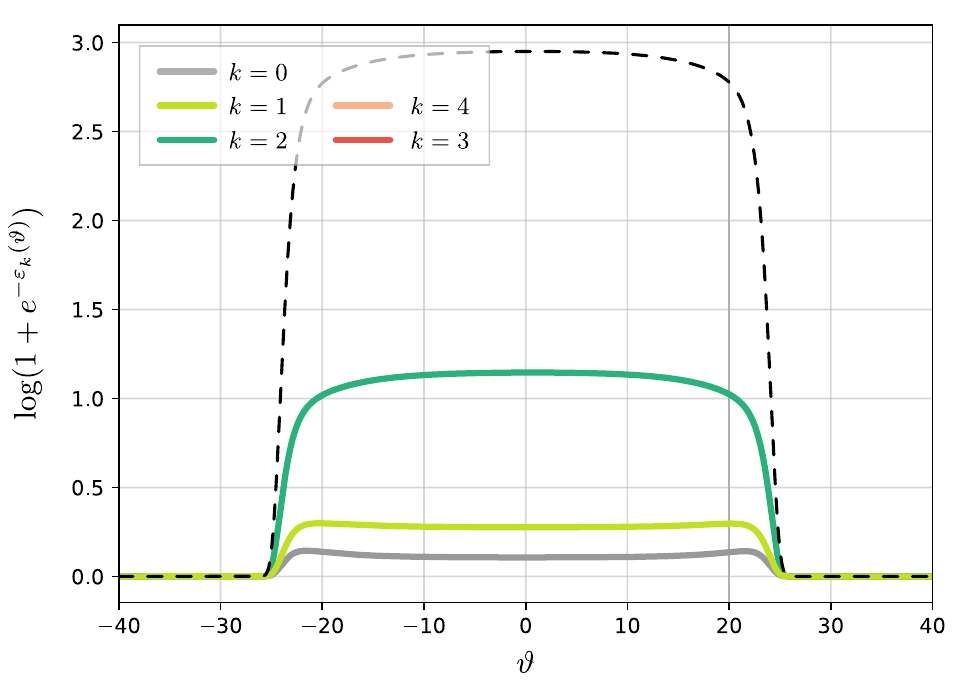}
        \caption{Real part.}
        \label{fig:12a}
    \end{subfigure}
    \hfill
    \begin{subfigure}[b]{0.495\textwidth}
        \centering
        \includegraphics[width=\textwidth]{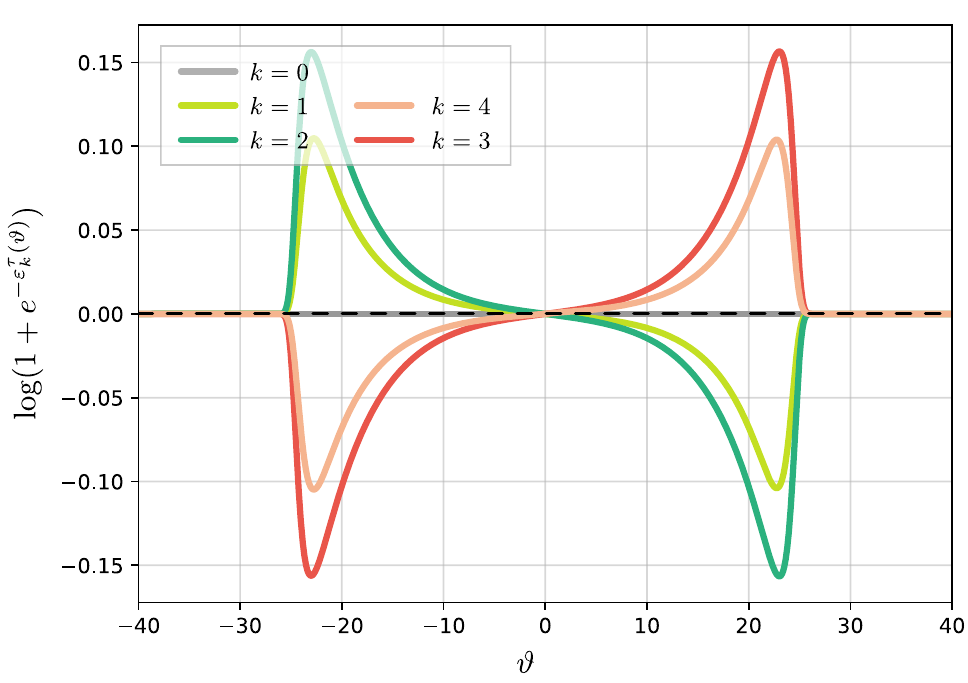}
        \caption{Imaginary part.}
        \label{figdddddd:1b}
    \end{subfigure}
    \caption{Real and imaginary parts of the density functions $\text{log} (1+e^{-\varepsilon_k(\vartheta)})$ in the $\mathbb{Z}_5$-graded Lee–Yang model, for $k=0,1,\dots, 4$, evaluated at $r = 10^{-11}$. The total sum across all sectors is represented by a dashed black line, and the plateau structure is clearly visible. For the real parts, the sectors $k$ and $n-k$ coincide, while for the imaginary parts, they appear with opposite signs.}
    \label{fiwg:both}
\end{figure}

\noindent In the graded setup, we consider the case in which only the spin-$1$ and spin-$n$ currents contribute to the GGE. Pulling back each $\varepsilon(\vartheta)$ along the maps $f_k(\vartheta)$, we obtain
\begin{equation}\label{lee_yang_graded_tba}
   \varepsilon_k(\vartheta)=r \cosh \vartheta+\alpha r^{1/5} \cosh \left(\frac{\vartheta}{n}+\frac{2 \pi i w k}{n}\right)-\sum_{\ell \in \mathbb{Z}_n} \int_\mathbb{R} \frac{\dd y}{2 \pi} \varphi_{k-\ell}\left(\vartheta-y\right) \log \big(1+e^{-\varepsilon_\ell(y)}\big)  \,.
\end{equation}
Here, $n = 6w+ \xi$, with $\xi =\pm 1$, and the graded kernels are obtained from the Lee–Yang kernel \eqref{LY} using the definition \eqref{grad_ker}. As in the free Majorana fermion case, our aim for the Lee–Yang model is to compute the graded ground-state scaling function $c_{\text {eff }}(r, \alpha)$ in selected regimes of interest. Owing to the theory's interactive nature, its analytic structure is less tractable, and one must instead rely on linear-response methods. We begin by observing that when $\alpha=0$, the graded theory reduces to $n$ decoupled copies of the original Lee–Yang model. In particular, $  \varepsilon_k(\vartheta) = \varepsilon(\vartheta)$. This simple observation implies $ c_{\text {eff }}(r, \alpha=0) = 2n/5$. At linear order in $\alpha$, we instead find $\varepsilon_k(\vartheta) = \varepsilon(\vartheta) + \delta\varepsilon_k(\vartheta)$. Taking a discrete Fourier transform in $k$ space as in \eqref{dft}, we see that $\varepsilon(\vartheta) = \bar\varepsilon^{(0)}(\vartheta)$, so that $\delta\bar\varepsilon^{(0)}(\vartheta) =0$. On the other hand, to study genuinely charged sectors, we can expand the graded TBA equations around $\alpha=0$, and obtain
\begin{equation}\label{dfttba}
    \delta \bar\varepsilon^{(q)}(\vartheta) =\frac{\alpha}{2}r^{1/n}e^{\vartheta / n} \delta_{q, w}+\frac{\alpha}{2}r^{1/n}e^{-\vartheta / n} \delta_{q,-w} - \int_\mathbb{R} \frac{\dd y}{2\pi} \bar \varphi^{(q)}(\vartheta-y)\frac{ \delta \bar\varepsilon^{(q)}(y)}{1+e^{\varepsilon(y)}}\,.
\end{equation}
One observes that only the charges $q=\pm w$ are directly sourced by grading, whereas the physical energy drives the neutral block $q=0$. To invert equation \eqref{dfttba}, we introduce the operators
\begin{equation}
    W_q[g(\vartheta)]=\int_\mathbb{R} \frac{\dd y}{2 \pi} \bar{\varphi}^{(q)}(\vartheta-y) \frac{ g(y)}{1+e^{\varepsilon(y)}}\,,
\end{equation}
so that $(1+W_q)\delta\bar \varepsilon^{(q)}(\vartheta) =   \bar\nu^{(q)}(\vartheta)$, with $\nu_{a_k}(\vartheta)$ defined as in \eqref{def_nu_tba}. Finally, defining the resolvent $R_q =(1+W_q)^{-1}$, we can invert this last relation to obtain
\begin{equation}\label{plat1}
  \delta \bar{\varepsilon}^{(\pm w)}(\vartheta)=\frac{\alpha}{2}r^{1/n} R_{\pm w}(e^{\pm \vartheta / n})\,, 
\end{equation}
while $\delta \bar{\varepsilon}^{(q)}(\vartheta) = 0$ in all other cases. In the ultraviolet regime, the left and right edges $\vartheta = \pm \log r$ dominate the integral in rapidity space. In the region between these two, the pseudoenergies are essentially constant, and the resolvent $R_q$ acts almost diagonally,
\begin{equation}\label{plat2}
 R_{\pm w}(e^{\pm \vartheta / n})\simeq \mathcal{C}_\pm e^{\pm\vartheta / n}\,,    
\end{equation}
where the numbers $\mathcal{C}_{ \pm}$ are finite, non-zero, and encode the dressing on the plateau. Using the Fermi–Dirac representation \eqref{fermi_dirac}, and splitting $\varepsilon_k(\vartheta) = \varepsilon(\vartheta) + \delta\varepsilon_k(\vartheta)$,  we write the effective central charge as
\begin{equation}\label{ref1}
    c_{\mathrm{eff}}(r,\alpha)=\frac{6}{\pi^2} \sum_{k\in\mathbb{Z}_n} \sum_{y = 1}^\infty \frac{(-1)^{y-1}}{y} \sum_{p = 0}^\infty \frac{(-y)^p}{p!}\int_0^\infty \dd \vartheta \,r \cosh \vartheta e^{-y \varepsilon(\vartheta)}\left(\delta \varepsilon_k(\vartheta)\right)^p\,.
\end{equation}
From the earlier discrete Fourier analysis of the graded source, the only non-zero charged components near $\alpha = 0$ are
\begin{equation}\label{splitt}
    \delta \varepsilon_k(\vartheta)=\delta \varepsilon_k^{(+)}(\vartheta)+\delta \varepsilon_k^{(-)}(\vartheta)\,,\quad  \delta \varepsilon_k^{( \pm)}(\vartheta) = \frac{1}{n}e^{ \pm 2 \pi i w k / n} \delta \bar{\varepsilon}^{( \pm w)}(\vartheta)\,.
\end{equation}

\begin{figure}
    \centering
    \begin{subfigure}[b]{0.495\textwidth}
        \centering
        \includegraphics[width=\textwidth]{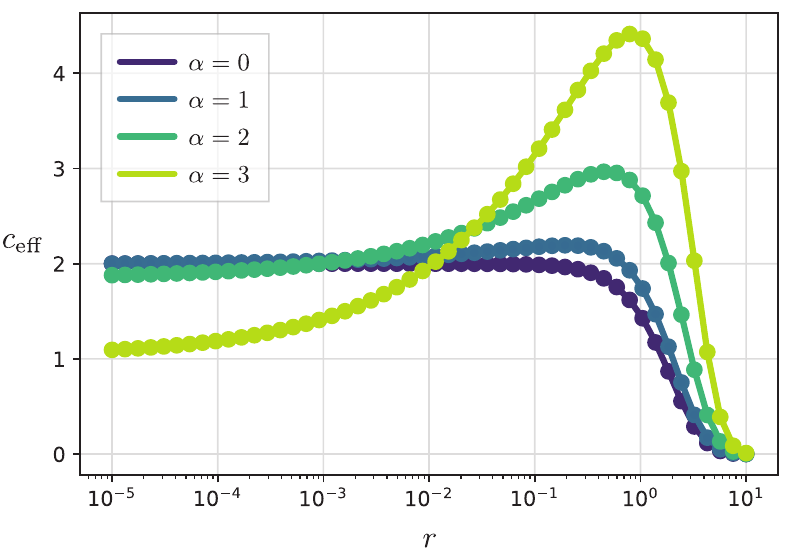}
        \caption{$n=5$.}
        \label{fig:11a}
    \end{subfigure}
    \hfill
    \begin{subfigure}[b]{0.495\textwidth}
        \centering
        \includegraphics[width=\textwidth]{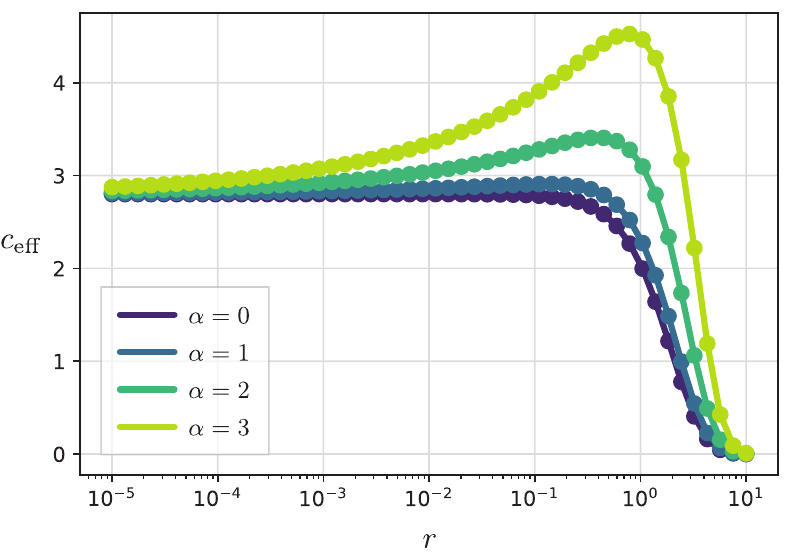}
        \caption{$n=7$.}
        \label{ffig:1b}
    \end{subfigure}
    \caption{Ground-state scaling function of the graded Lee–Yang model for sufficiently small values of $\alpha$. In all cases, only the trivial case $\alpha = 0$ flows to  $n$ independent copies of the Lee–Yang CFT. When  $n=5$, the UV corrections are clearly visible. For $n\geq7$, these corrections are strongly suppressed, but still observable (see Figure \ref{z7vsa}).}
    \label{figr:both}
\end{figure}

\begin{figure}
    \centering
    \includegraphics[width=.7\textwidth]{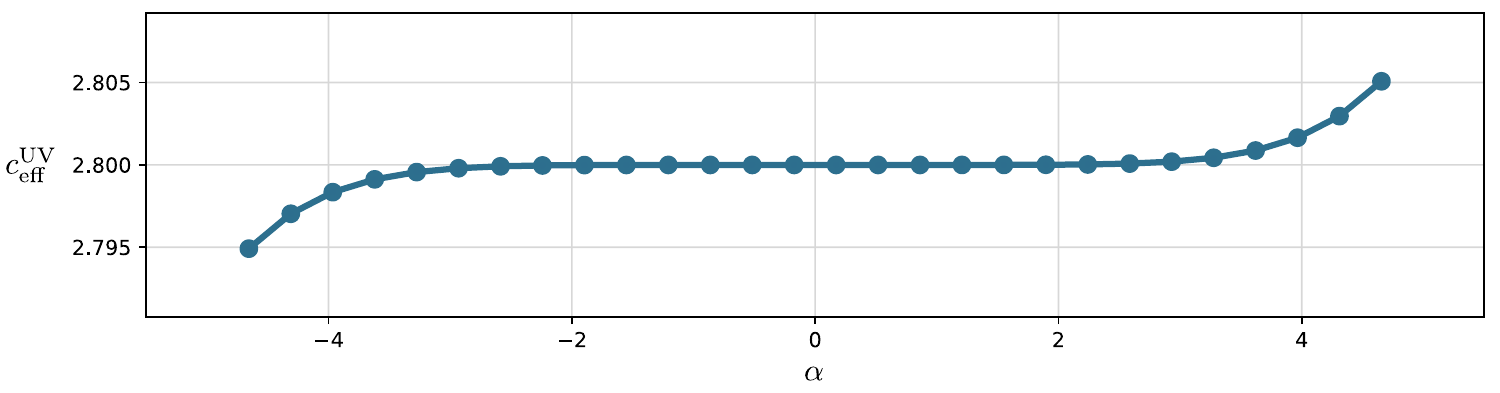}
 \caption{Plot of the ultraviolet ground-state scaling function $c_{\mathrm{eff}}^{\mathrm{UV}}(\alpha)$ for the $n=7$ graded Lee-Yang model. The curve shows a small but systematic dependence on the parameter $\alpha$.}
\label{z7vsa}
\end{figure}

\noindent At a fixed order $p$, let $j$ denote the number of $\delta\varepsilon_k^{(+)}$ factors,
and $p-j$ the number of $\delta \varepsilon_k^{(-)}$ factors. By the binomial theorem, 
\begin{equation}\label{prod_simp}
   \left(\delta \varepsilon_k(\vartheta)\right)^p = \sum_{j=0}^p\binom{p}{j}\big(\delta \varepsilon_k^{(+)}(\vartheta)\big)^{j}\big(\delta \varepsilon_k^{(-)}(\vartheta)\big)^{p-j} \,.
\end{equation}
Using equation \eqref{splitt} and collecting the $k$-dependent phases, the product in \eqref{prod_simp} simplifies to
\begin{equation}\label{ref2}
    \big(\delta \varepsilon_k^{(+)}(\vartheta)\big)^{j}\big(\delta \varepsilon_k^{(-)}(\vartheta)\big)^{p-j} = n^{-p} e^{2\pi i w k(2j-p)/n}\big(\delta \varepsilon^{(+w)}(\vartheta)\big)^{j}\big(\delta \varepsilon^{(-w)}(\vartheta)\big)^{p-j}\,.
\end{equation}
When summing over $k \in \mathbb{Z}_n$, the root of unity projects over the $2j-p \equiv 0 \bmod n$ sector. Combining expressions \eqref{plat1} and \eqref{plat2}, we  conclude that the effective central charge must admit the following nested expansion as $r\ll 1$:
\begin{equation}\label{d-sum}
\begin{split}
        c_\text{eff}(r,\alpha)\simeq & \,\frac{3}{\pi^2} \sum_{y=1}^\infty \left[ \frac{(-1)^{y-1}}{y} \sum_{p= 0}^\infty \left(\frac{(-y \alpha r^{1/n})^p}{p!} (2n)^{1-p}\right.\right.\\ &\left.\left.\times\sum_{j=0}^p \binom{p}{j} \mathcal{C}_+^j\mathcal{C}_-^{p-j} \delta_{2j-p, 0 \bmod n}U_{(2j-p)/n}(r,y)\right)\right]\,,
\end{split}
\end{equation}
where the quantities $U_j(r,y)$ are formally the same integrals that formally appear when expanding the Lee–Yang ground-state scaling function in the UV regime, 
\begin{equation}  \label{vj}
   U_m(r,y) =  \int_0^\infty \dd \vartheta\, r \cosh\vartheta e^{-y\varepsilon(\vartheta)} e^{m\vartheta}\,.
\end{equation}
The scaling behaviour of the integrals $U_m(r, y)$ determines which powers of $\alpha$ can contribute a finite term in the ultraviolet limit. In the regime $r \ll 1$, the integrals are dominated by the two edges of the pseudoenergy plateau, $\vartheta \simeq \pm \log r$. Evaluating the exponential factor $e^{m \vartheta}$ on the edges gives $U_m(r, y) \sim r^{-|m|}$ as $r \rightarrow 0$. Each contribution in the double sum of equation \eqref{d-sum} therefore carries an overall factor $r^{p / n} U_{(2 j-p) / n} \simeq r^{(p-|2 j-p|) / n}$. For $0<j<p$, this exponent is positive, so the term vanishes in the ultraviolet; only the endpoint values $j=0$ and $j=p$ can survive with a finite limit. Imposing at the same time the root-of-unity projector $2 j-p \equiv 0\bmod n$ restricts these endpoints to $p \equiv 0\bmod n$. Hence, as $r \to 0$, all intermediate contributions are suppressed by powers of $r^{1 / n}$, and the ultraviolet expansion of the scaling function organises into a series in $\alpha^n$:
\begin{equation}\label{cds_ly}
        c_\text{eff}^\text{UV}(\alpha) = \frac{2n}{5}+\sum_{j=1}^\infty T_j(n)\alpha^{jn}\,,
\end{equation}
with $T_j(n)$ determined by the plateau constants $\mathcal{C}_\pm$ and the edge integrals. Differently from the graded Ising case, where all such coefficients vanish identically for $n>3$, in the interacting case, the $T_{j}(n)$ are generically non-zero.
However, except for the minimal grading $n=5$, where the first non-trivial correction is comparatively large, these coefficients remain numerically very small, indicating that the ultraviolet response to the deformation is present but extremely weak (see Figures \ref{figr:both} and \ref{z7vsa}). As $\alpha$ is tuned away from zero, we observe indications of phase transitions and level crossings in the graded Lee–Yang model as well. A detailed analysis of these phenomena, including their analytic continuation and physical interpretation, is left for future work.
\section{Chemical potentials, twisted sectors and cyclic orbifolds}\label{SEC:TWIST}
So far, we have constructed graded extensions of known Y-systems by pulling back the Y-functions along suitable reparametrisation maps.
These graded systems replicated the original functional relations across $n$ sheets, while preserving the overall periodicity up to a cyclic permutation of the Y-functions. We now wish to go a step further and introduce chemical potentials. In general, chemical potentials can be incorporated without spoiling the structure of the Y-system \cite{Martins:1991hw, Fendley:1991xn}. In the present context, however, we want them to play a more explicit role: rather than preserving the original periodicity, they are introduced to \textit{distinguish} among the different $\mathbb{Z}_n$ sectors. As a result, the chemical potentials modify the analytic continuation properties of the system and alter its internal periodicity, leading to a genuinely new class of twisted Y-systems.

\subsection{Chemical potentials in the Ising model}
In this section, we briefly comment on graded, free fermionic theories sourced by the driving terms
\begin{equation}\label{twisted_e_ising}
     \varepsilon^\tau_k(\vartheta)=\sum_{s \in 2\mathbb{N}+1} \alpha_s r^{s/n} \cosh\left(\frac{s\vartheta}{n}-\frac{2\pi i wsk}{n}\right) + \frac{2\pi i \tau k}{n}\,,
\end{equation}
where $\tau,k \in \mathbb{Z}_n$, and we restrict to odd values of $n$ for internal consistency (see the discussion below equation \eqref{resc.}). Here, we call $\tau$ the twist parameter. The $s=n$ term in \eqref{twisted_e_ising} reproduces the standard energy contribution, which scales linearly with $r$, whereas the remaining terms correspond to graded, GGE-type sources. If one consider the Y-functions $Y_k(\vartheta) =  e^{\varepsilon_k(\vartheta)}$, the following functional relation holds: 
\begin{equation}
    Y_{k-\xi}(\vartheta+i\pi/2)Y_{k+\xi}(\vartheta-i\pi/2) = \lambda^{2\tau}_k\,,
\end{equation}
with $\lambda_k = e^{2 \pi i k /n}$. The above equation provides a direct generalisation of \eqref{ysys_graded_ising} for non-vanishing values of $\tau$, and it crucially modifies the periodicity property \eqref{ising_periodicity}. In particular, one can verify that $P = 2n$ unless $\tau = 0 \bmod n$. To study \eqref{twisted_e_ising}, we restrict to the case where only $\alpha_1 = \alpha$ and $\alpha_n = 1$ are non-vanishing,
\begin{equation}
    \varepsilon^\tau_k(\vartheta)=r \cosh \vartheta+\alpha r^{1/n} \cosh \left(\frac{\vartheta}{n}+\frac{2 \pi i w k}{n}\right)+\frac{2\pi i \tau k}{n}\,.
\end{equation}
In particular, we are interested in studying the $\tau$-twisted ground-state scaling functions:
\begin{equation}\label{twisted_c}
    c^\tau_{\text{eff}}(r,\alpha) = \frac{6r}{\pi^2} \sum_{k\in\mathbb{Z}_n}\int_0^{\infty}\dd\vartheta \cosh\vartheta \log\big(1+e^{-\varepsilon_k^\tau(\vartheta)}\big)\,.
\end{equation}
For simplicity, we begin by considering the case $\alpha=0$. We use the representation \eqref{fermi_dirac} for the logarithm, which yields:
\begin{equation}
    \log\big(1+e^{-r\cosh\vartheta -2\pi i \tau k/n}\big) = \sum_{y=1}^\infty\frac{(-1)^{y+1}}{y}e^{-yr\cosh\vartheta}e^{-2\pi iy\tau k/n}\,.
\end{equation}
The integral over rapidities can then be evaluated using standard representations for the modified Bessel function of the second kind, and one obtains

\begin{equation}
     c^\tau_{\text{eff}}(r,\alpha=0)=\frac{6r}{\pi^2} \sum_{k\in\mathbb{Z}_n}\sum_{y=1}^\infty\frac{(-1)^{y+1}}{y}  K_1(y r) e^{-2\pi i y\tau k/n}\,.
\end{equation}
\begin{figure}
    \centering
    \begin{subfigure}[b]{0.495\textwidth}
        \centering
        \includegraphics[width=\textwidth]{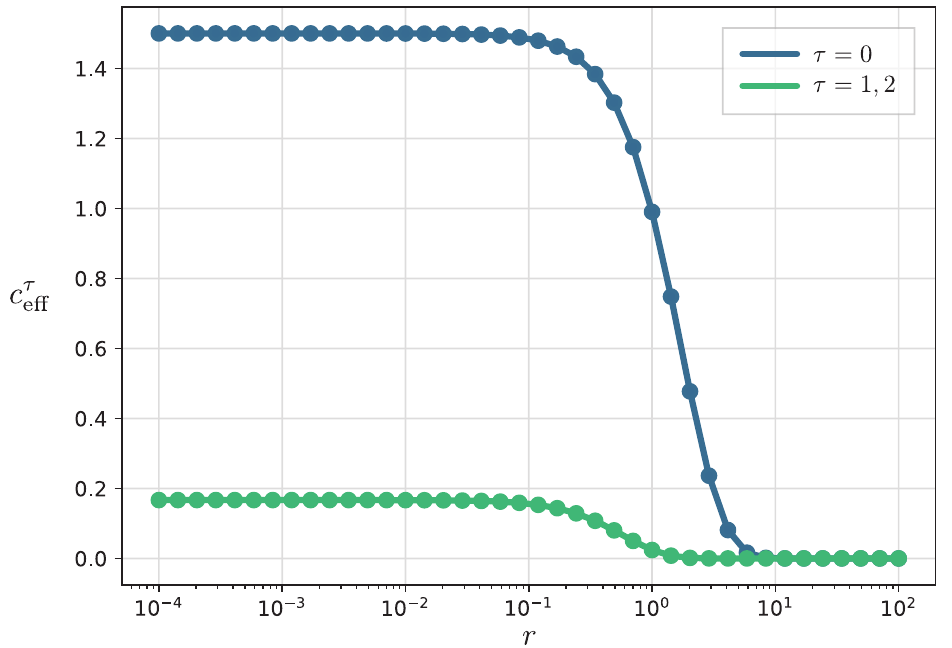}
        \caption{$n=3$.}
        \label{fig:111a}
    \end{subfigure}
    \hfill
    \begin{subfigure}[b]{0.495\textwidth}
        \centering
        \includegraphics[width=\textwidth]{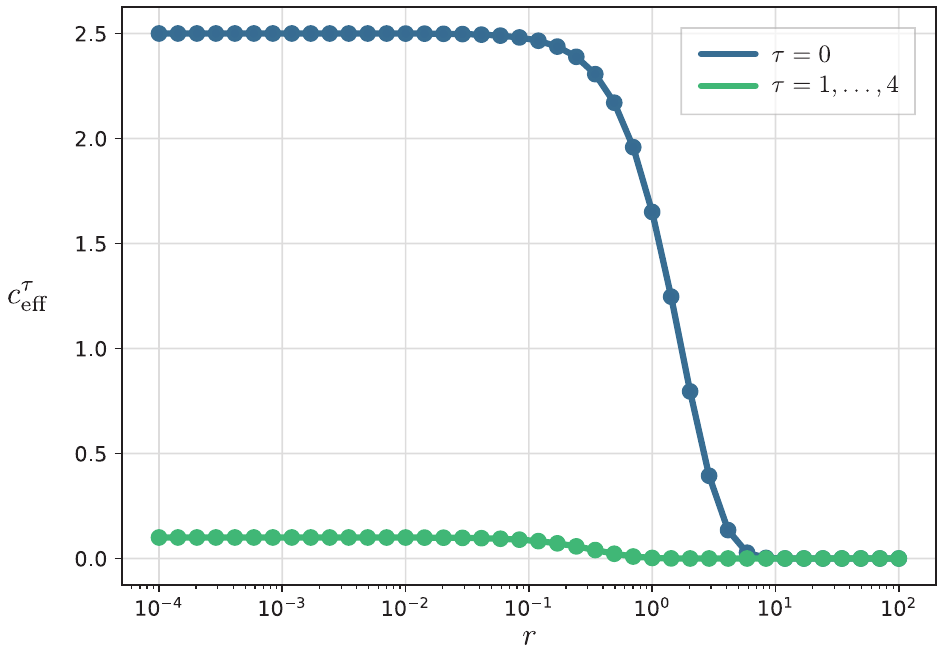}
        \caption{$n=5$.}
        \label{fizg:1b}
    \end{subfigure}
    \caption{Twisted and untwisted scaling functions for the graded Ising model with various $n$ at $\alpha = 0$. In the untwisted sector, the scaling function approaches $n$ times the Ising central charge in the ultraviolet limit, while each twisted sector flows to $1/n$ of that value. At fixed $n$, all genuinely twisted sectors are mutually isomorphic, yielding identical scaling functions.}
    \label{fig:btoth}
\end{figure}

\noindent The sum over $k\in\mathbb{Z}_n$ is a finite geometric sum, and can now be carried off explicitly using orthogonality of the roots of unity. In particular, the sum is non-zero if and only if $ y\tau/n$ is an integer. For simplicity, we assume $ \text{gcd}(\tau,n) = 1$. Then the exponential sum equals $n$ whenever $y$ is a multiple of $n$, and the scaling function reduces to:
\begin{equation}\label{c^tau_2}
   c^\tau_{\text{eff}}(r,\alpha=0)=\frac{6 n r}{\pi^2} \sum_{\ell=1}^{\infty} \frac{(-1)^{\ell n+1}}{\ell n} K_1(\ell n r)\,.
\end{equation}
In the ultraviolet regime, for $r \ll 1$, we can use the leading order $rK_1(\ell nr ) \simeq 1/\ell n$ to recast \eqref{c^tau_2} into an alternating series which evaluates to (see Figure \ref{fig:btoth}):
\begin{equation}\label{plateau}
    c^{\tau,\text{UV}}_\text{eff}(\alpha=0) = \frac{1}{2n}\,.
\end{equation}
The value reached by the UV plateau in the $\tau$-twisted scaling function strongly suggests an orbifold interpretation. Indeed, for $\tau = 1,\dots, n-1$, the pattern $c^{\tau,\text{UV}}<c^{0,\text{UV}}$ is naturally accounted for if the $\tau$-sector corresponds to the insertion of a twist operator that implements a cyclic identification of $n$ copies of the original theory. In this picture, the twist field produces a branch cut that cyclically permutes the replicas, effectively realising a $\mathbb{Z}_n$-orbifold of the parent conformal field theory. The UV free energy difference between the twisted and untwisted sectors can then be interpreted as the contribution of a primary field with conformal dimension
\begin{equation}\label{orbifold_dim}
    \Delta_\text{tw} = \frac{1}{48}\left(n-\frac{1}{n}\right)\,,
\end{equation}
reproducing known results from the literature on twist fields in orbifold CFTs \cite{Knizhnik:1987xp, Calabrese:2009qy}. Equation \eqref{orbifold_dim} supports the idea that the twisted sectors generated by the $\mathbb{Z}_n$-grading correspond to orbifold sectors of a replicated CFT, each labelled by a distinct cyclic intertwining of the replicas. Interestingly, the modified periodicity of the associated Y-system, which changes from $P=2$ in the untwisted case to $P=2 n$ in the $\tau$-twisted one, reinforces this interpretation. The enlarged periodicity implies that the perturbing operator in the twisted sector effectively carries a fractionalised conformal dimension, in agreement with the presence of fractional Virasoro modes in the orbifold CFT. In this sense, the Y-system periodicity encodes the same topological information as the branch structure introduced by the twist field, linking the analytic continuation of the TBA equations to the operator content of the underlying orbifold theory.

We now turn to the case of non-vanishing $\alpha$. Since the $\tau \equiv  0 \bmod n$ case has already been extensively discussed in Section \ref{Majorana:SEC}, we implicitly exclude it from the present analysis. Moreover, because the set of $n$-th roots of unity is invariant under inversion, the resulting ground-state scaling function is the same whether the twist enters with either a positive or a negative phase (modulo $n$), $c^{\tau}_\text{eff}(r,\alpha) =  c^{n-\tau}_\text{eff}(r,\alpha)$.
Expanding the logarithm as in \eqref{fermi_dirac}, we use the Bessel representation \eqref{bess_rep} and the identity \eqref{t_non_0} obtained in Appendix \ref{app:root} to write:
\begin{equation}\label{twist_c}
\begin{split}
     c^{\tau}_\text{eff}(r,\alpha) &= \frac{6 n r}{\pi^2} \sum_{y=1}^\infty\frac{(-1)^{y+1}}{y}  I_0(y\alpha r^{1/n}) K_1(y r) \delta_{y\tau, 0 \bmod n} \\&+\frac{12}{\pi^2} \sum_{j=0}^\infty \frac{(-1)^{y \tau+j n}}{y}(y \tau+j n) I_{y \tau+j n}(y\alpha r^{1/n}) K_{y \tau/n+j }(y r)\,.   
\end{split}
\end{equation}
\noindent \textbf{Infrared regime}. At small, fixed $\alpha$, the large-$r$ behaviour is controlled by the asymptotics of the Bessel functions of the second kind. In particular, $K_\nu(z)\simeq z^{-1/2} e^{-z}$ for any positive real $z$ and $\nu \in \mathbb{R}$, and the dominant term in the expansion comes from the smallest allowed value of $y$. For the first term in \eqref{twist_c}, this corresponds to $y = p$. In contrast, the second term already contributes at $y=1$, and thus controls the entire infrared tail. Keeping only this leading sector, the scaling function reduces to:
\begin{equation}
c_{\text{eff}}^{\tau,\text{IR}}(\alpha) =  \frac{12}{\pi^2} \frac{\tau}{n} I_\tau(\alpha r^{1/n}) K_{\tau/n}(r)\,.
\end{equation}
In particular, we see that the effective central charge vanishes exponentially fast, independently of the value of $\tau$. Physically, the infrared theory is described by a diluted gas of massive excitations: the fine $\mathbb{Z}_n$-grading becomes invisible, and the scaling function vanishes.\\

\begin{figure}
    \centering
    \begin{subfigure}[b]{0.495\textwidth}
        \centering
        \includegraphics[width=\textwidth]{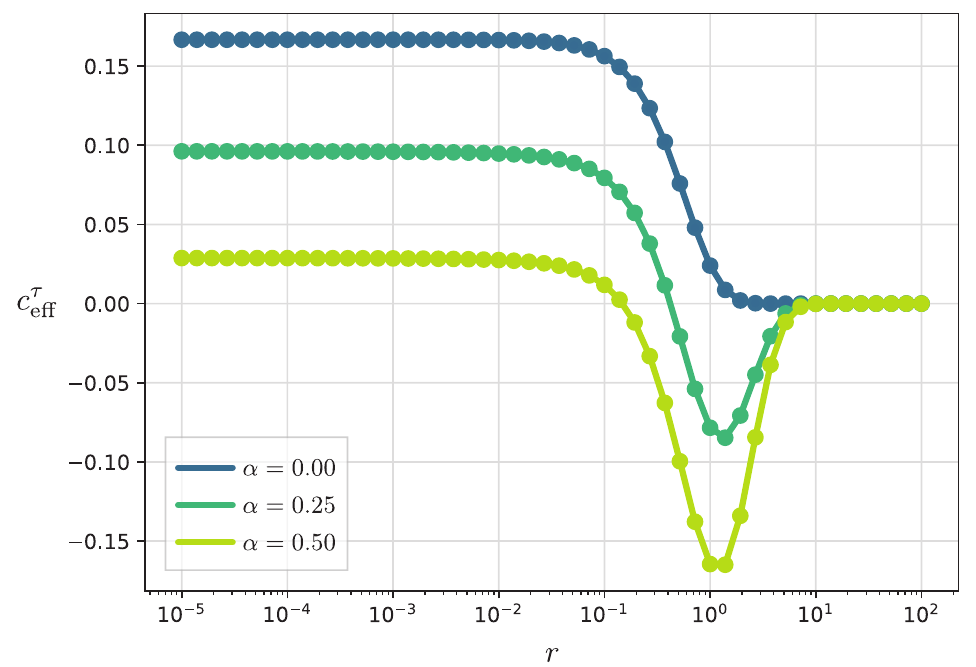}
        \caption{$n=3$.}
        \label{fig:19a}
    \end{subfigure}
    \hfill
    \begin{subfigure}[b]{0.495\textwidth}
        \centering
        \includegraphics[width=\textwidth]{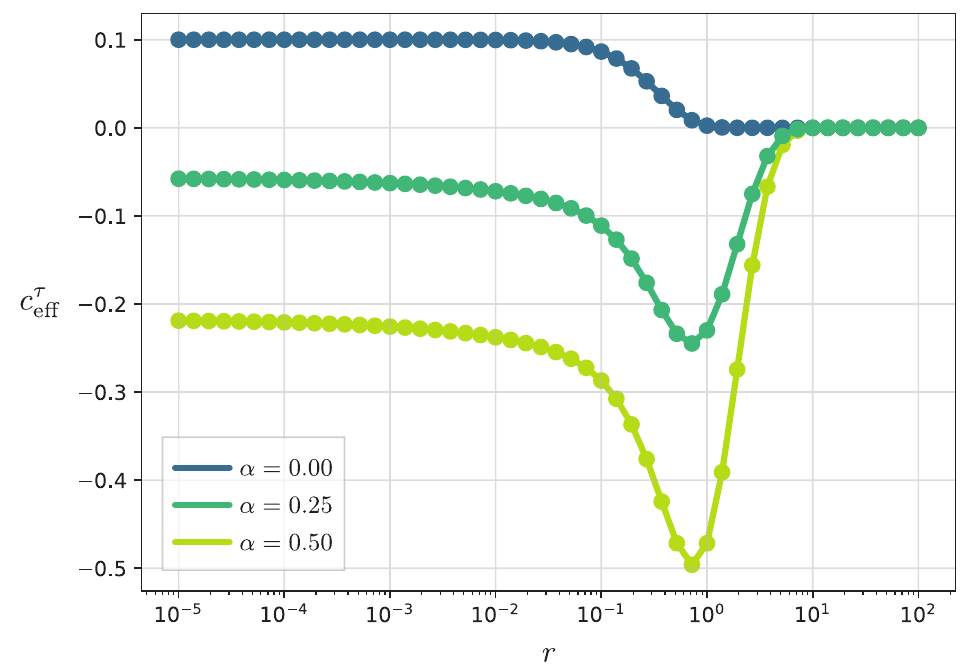}
        \caption{$n=5$.}
        \label{fwig:1b}
    \end{subfigure}
    \caption{Twisted scaling functions for the graded Ising model at various values of $\alpha$ for $\tau = 1,\dots,n-1$. Unlike the untwisted case, all values of $n$ display visible modifications in the ultraviolet regime, while the infrared behaviour consistently flows to a gapped phase.}
    \label{fig:bogth}
\end{figure}

\noindent \textbf{Ultraviolet regime}. We observe that all the $\alpha$-dependent pieces in the terms proportional to $I_0(y \alpha r^{1/n}) K_1(y r)$ carry extra positive powers of $r^{1/n}$, and they vanish as $r\to 0$. Overall, the only $\mathcal{O}(1)$ term contributes to the constant, ungraded plateau \eqref{plateau}. On the other hand, in the second term of \eqref{twist_c}, the powers $r^{m/n}$ and $r^{-m/n}$ -- coming from $I_m(y\alpha r^{1/n})$ and $K_{m/n}(yr)$ respectively -- cancel out reciprocally. Here, $m = jn+\tau n$. We thus observe that the admissible powers of $\alpha$ contributing to the ultraviolet expansion organise in the sum:
\begin{equation}
c_\text{eff}^{\tau,\text{UV}} = \frac{1}{2n} + \sum_{m \in M} T_m(n) \alpha^m\,,\quad M = \{m = y\tau + jn\,\,|\,\, y = 1, 2,\dots n\,\text{ and } j \in \mathbb{N}\}\,.
\end{equation}
The coefficients $T_m(n)$ can be obtained from the exact Bessel series. For a fixed $m \in M$, there is a unique $y \in\{1, \ldots, n\}$ such that $m-y \tau$ is a nonnegative multiple of $n$. For such $y$, one finds:
\begin{equation}
   T_m(n) = \frac{6}{\pi^2} \frac{m\Gamma\left(\frac{m}{n}\right)}{m!} 2^{m/n-m}(-1)^{m+y+1} y^{m-m/n-1}\,.
\end{equation}
As shown in Figure \ref{fig:bogth}, non-trivial corrections are present in the UV even when $n>3$.
\subsection{Generalised Y-systems and twisted interacting theories}
Similarly to the free case, interacting graded Y-systems can be extended to include chemical potentials that distinguish between the individual $\mathbb{Z}_n$ sectors. Apart from the Ising field theory -- whose degenerate incidence matrix, $\mathcal{G}_{ab} \equiv 0$ for the underlying Dynkin diagram $G=A_1$, makes its analysis qualitatively different -- one can formulate systems of coupled functional relations that generalise the Y-systems of graded, interacting QFTs. In particular, we consider the set of equations
\begin{equation}\label{chemical_pot-Y}
    Y_{a_{k-\xi}}\left(\vartheta^{+}\right) Y_{a_{k+\xi}}\left(\vartheta^{-}\right)=\prod_{b \in G}\left(\lambda_k^\tau+Y_{b_k}\left(\vartheta^{+}\right)\right)^{\mathcal{G}_{a b}}\,,
\end{equation}
where $\vartheta^\pm = \vartheta \pm i\pi /h$. As for the free case, here $\lambda_k = e^{2\pi i k/n}$, and $\tau$ is a twist parameter. The extra factors $\lambda_k$  appearing in \eqref{chemical_pot-Y} represent the minimal modification compatible with the analytic and periodicity constraints imposed by the grading. Moreover, the modified Y-system \eqref{chemical_pot-Y} exhibits an extended periodicity. While at $\tau=0$ the set of functions $\left\{Y_k(\vartheta)\right\}$ is invariant under the shift $\vartheta \mapsto \vartheta+i \pi P$, the introduction of these chemical potentials enlarges the periodic structure, and invariance now occurs only under transformation of the type $\vartheta \mapsto \vartheta+i n \pi P$. A natural question concerns the relation between the present twisted construction and the appearance of dilogarithm identities in the ultraviolet limit of TBA systems. In the graded setting, the additional periodic structure induced by the chemical potentials suggests that the corresponding Rogers dilogarithm sum rules may admit a refined version that is sensitive to the twist. It would therefore be interesting to investigate whether the modified Y-system admits an uplift to a full CFT partition function -- possibly along the lines proposed in references \cite{Nahm:1992sx, Terhoeven:1992zf}.
\begin{figure}
    \centering
    \includegraphics[width=0.6\textwidth]{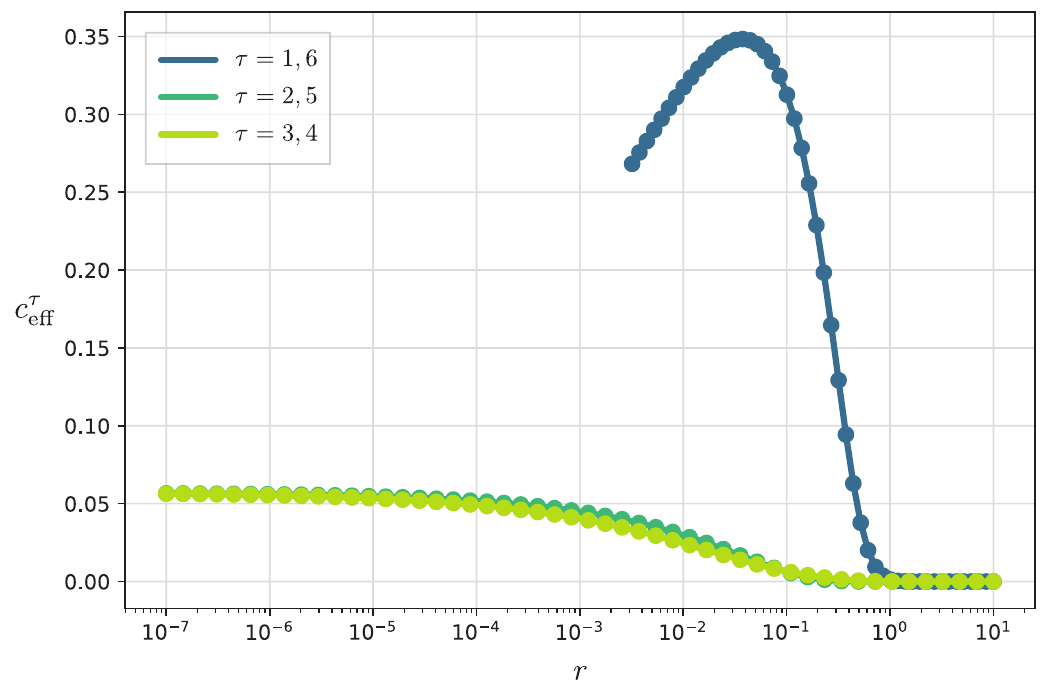}
   \caption{
Twisted scaling functions (for $\alpha=0$) of the graded Lee–Yang model with $n=7$. 
The curves $\tau=1,6$ terminate at points beyond which the numerical algorithm becomes unstable, 
due to logarithmic singularities approaching the integration contour (see Figure \ref{ff11}). 
A proper modification of the TBA equations would be required to smoothly continue these energy levels into the deep UV. 
A detailed analysis of this exact treatment is deferred to future work. In Figure \ref{ff12}, a numerical estimate of the full scaling behaviour in the $\tau =1,6$ sectors is presented. The result we obtain is compatible with analytic considerations. The remaining curves, which are not affected by such singularities, converge smoothly to the expected ultraviolet value. 
Similar behaviours are observed for other values of $n$.
}
    \label{ff9}
\end{figure}
In the corresponding TBA equations, these chemical potentials manifest as imaginary shifts of the driving term, effectively adding an extra phase $2 \pi i \tau k / n$ to each component of the $\mathbb{Z}_n$ multiplet. For simplicity, we focus on the case where only the spin-$1$ and spin-$n$ currents contribute to the GGE, yielding the following expression for the pseudoenergies:
\begin{equation}
\begin{split}
          \varepsilon^\tau_k(\vartheta)=&\, r \cosh \vartheta+\alpha r^{1/n} \cosh \left(\frac{\vartheta}{n} - \frac{2\pi i wk}{n}\right)+\frac{2\pi i \tau k}{n}\\ &-\sum_{\ell \in \mathbb{Z}_n} \int_\mathbb{R} \frac{\dd y}{2 \pi} \varphi_{k-\ell}\left(\vartheta-y\right) \log \big(1+e^{-\varepsilon^\tau_\ell(y)}\big)  \,.
\end{split}
\end{equation}
For the purposes of this section, we will later restrict to the case $\alpha=0$, thereby reducing the parameter space and isolating the essential features of the deformation. Expanding the logarithmic term, the effective central charge can be expressed as:
\begin{equation}
    c_{\mathrm{eff}}^{\tau}(r,\alpha)=\frac{6 r}{\pi^2} \sum_{k \in \mathbb{Z}_n} \sum_{y = 1}^\infty \frac{(-1)^{y+1}}{y} \int_0^{\infty} \dd \vartheta \cosh \vartheta e^{-y \varepsilon^\tau_k(\vartheta)} \,.
\end{equation}

\begin{figure}
    \centering
    \includegraphics[width=1\textwidth]{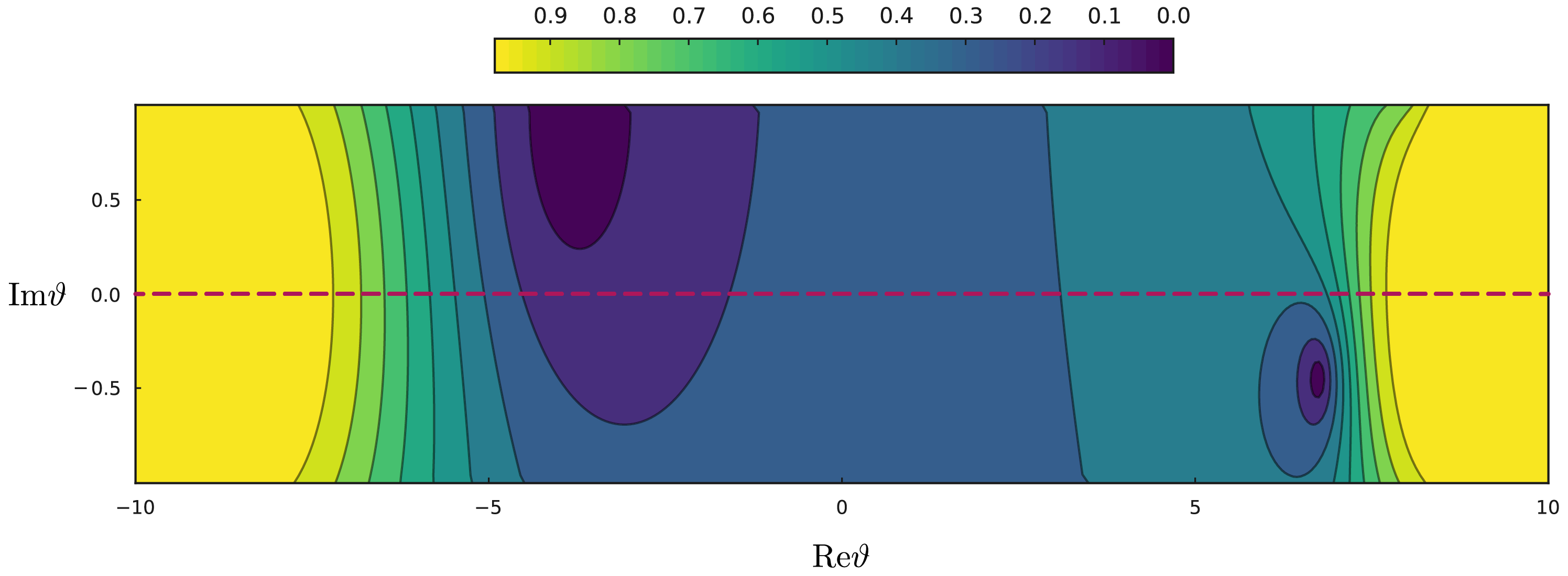}
 \caption{
The numerical instability observed for the $n=7$, $\tau=1,6$ levels arises from 
$Y = -\lambda$ singularities -- located at the centres of the two darker regions -- 
approaching the integration contour. 
The figure shows a plot of $|\lambda^{1}_{3} + Y_3(\vartheta)| / (1 + |\lambda^{1}_{3} + Y_3(\vartheta)|)$ in the range 
$\mathrm{Re}\,\vartheta \in (-10,10)$ and $\mathrm{Im}\,\vartheta \in (-\pi/3,\pi/3)$. 
The picture corresponds to the last convergent point, $c_\text{eff}(r) \simeq 0.26819$, obtained for $r = 0.0031989$. Determining whether these singularities actually cross the real axis at some smaller values of $r$ would require a more detailed analysis, which lies beyond the scope of this preliminary investigation.
}
\label{ff11}
\end{figure}

\begin{figure}[h]
    \centering
    \includegraphics[width=0.6\textwidth]{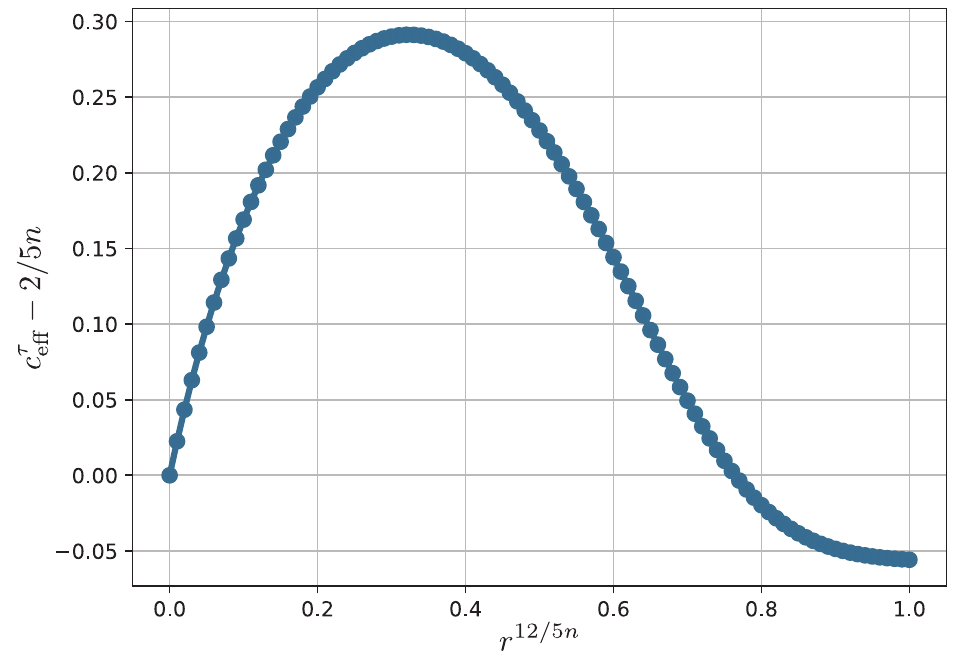}
     \caption{
Interpolation/extrapolation of the effective central charge 
$c_{\mathrm{eff}}(r) - 2/35$ versus $r^{12/35}$ for $n=7$, $\tau=1,6$. 
The estimated value at $r=0$ is $c_{\mathrm{eff}}(0)- 2/35  = (6.7 \pm 1.5) \times 10^{-6}$. 
The analysis was performed using a \textit{Mathematica} implementation of the 
Rational Function Interpolation and Extrapolation algorithm, as described in 
\textit{Numerical Recipes: The Art of Scientific Computing} (see \cite{press2007numerical}). The above uncertainty reflects the difference between extrapolations performed using $N$ and $N-1$ points in the interpolation/extrapolation algorithm. Given the level of precision of our TBA numerical procedure — affected by discretisation, numerical instabilities, and truncation errors — additional systematic effects are likely to dominate over the statistical spread. The result, therefore, appears compatible with zero within the combined numerical uncertainty.}
\label{ff12}
\end{figure}

\noindent To extract the UV behaviour of the theory, we follow the logic of Section \ref{lygr:SEC}. There, we wrote each graded pseudoenergy $\varepsilon_k(\vartheta) = \varepsilon(\vartheta)+\delta\varepsilon_k(\vartheta)$, with factor $\delta \varepsilon_k$ carrying the graded phase $e^{ \pm 2 \pi i w k / n}$ in its charged components, and after binomial bookkeeping one gets an overall $k$-phase $e^{2 \pi i k(2 j-p) w / n}$ at order $p$ (see \eqref{ref1}--\eqref{ref2}). If one includes the $\tau$-phase from the chemical potential, each Fermi–Dirac harmonic contributes an extra $e^{-2 \pi i y \tau k / n}$. The resulting net phase is
\begin{equation}
    \exp \left[\frac{2 \pi i k}{n}((2 j-p) w-y \tau)\right]\,,
\end{equation}
and when summing over $k \in \mathbb{Z}_n$ this imposes the root-of-unity selection rule $(2 j-p) w \equiv y \tau \bmod n$. At $\alpha = 0$, the only surviving term in the Fermi–Dirac expansion is $p=0$, and the selection rule reduces to $y\tau\equiv 0 \bmod n$. For simplicity, we again assume $\operatorname{gcd}(\tau, n)=1$, so that the sum in non-zero if and only if $y = \ell n$ for some integer $\ell$, and
\begin{equation}
    c_{\mathrm{eff}}^{\tau}(r, \alpha=0)=\frac{6 n r}{\pi^2} \sum_{\ell=1}^\infty \frac{(-1)^{\ell p-1}}{\ell p} \int_0^{\infty} \dd \vartheta  \cosh \vartheta e^{-\ell p \varepsilon(\vartheta)}\,.
\end{equation}
The UV analysis for the scaling Lee–Yang model shows that the above integral has a finite plateau limit, and yields $c_{\text {eff }}^{\mathrm{UV}}=2 n / 5$ at $\tau=\alpha=0$. If only one in every $p$ harmonics survives (by the projector above), the same edge analysis yields a reduction by the universal factor $1 / n$ -- the same arithmetic factor found in the free Ising calculation -- with the Ising central charge $c= 1 / 2$ replaced by the Lee–Yang effective central charge  $c= 2/5$:
\begin{equation}\label{result_uv_LY}
    c_{\mathrm{eff}}^{\tau, \mathrm{UV}}(\alpha=0)=\frac{2}{5n}\,.
\end{equation}
Since this argument relies only on the block-diagonal (charge-conserving) structure of the graded kernels, the same reduction is expected to hold universally for any graded integrable theory: the UV effective central charge of the twisted sector should gain an extra $1/n$ factor whenever the twist is primitive. However, this analysis is purely perturbative; it neglects possible non-perturbative effects, such as logarithm branch cuts crossing the integration contour.\\
\begin{figure}
    \centering
    \begin{subfigure}[b]{0.495\textwidth}
        \centering
        \includegraphics[width=\textwidth]{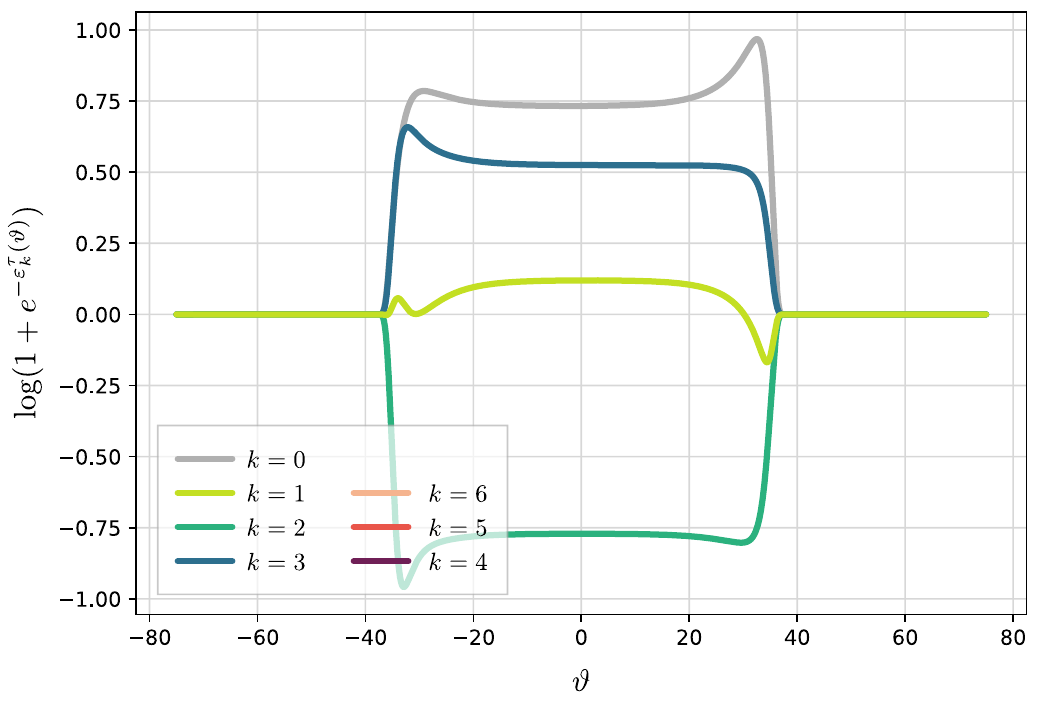}
        \caption{Real part.}
        \label{fig1:1a}
    \end{subfigure}
    \hfill
    \begin{subfigure}[b]{0.495\textwidth}
        \centering
        \includegraphics[width=\textwidth]{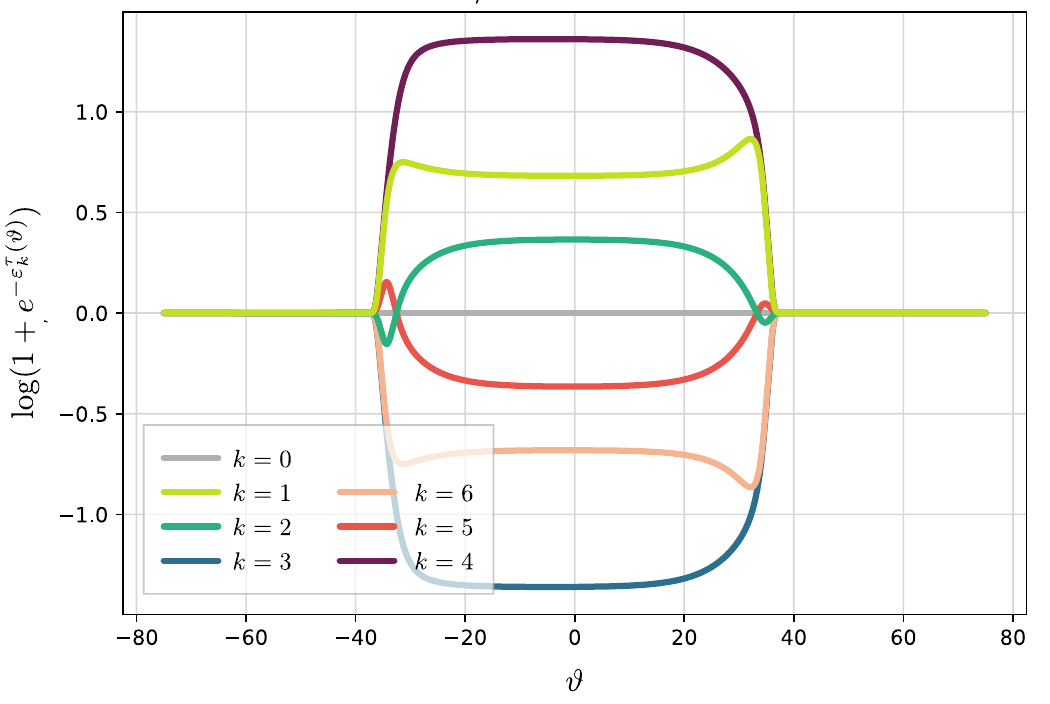}
        \caption{Imaginary part.}
        \label{fig:11b}
    \end{subfigure}
    \caption{Real and imaginary parts of the density functions $\text{log} (1+e^{-\varepsilon^\tau_k(\vartheta)})$ in the twisted, $\mathbb{Z}_7$-graded Lee–Yang model, for $k=0,1,\dots , 6$ and $\tau = 2$, evaluated at $r = 10^{-15}$. For the real parts, the sectors $k$ and $n-k$ coincide, while for the imaginary parts, they appear with opposite signs.}
    \label{fig:bocth}
\end{figure}
\begin{figure}[h!]
    \centering
    \begin{subfigure}[b]{0.495\textwidth}
        \centering
        \includegraphics[width=\textwidth]{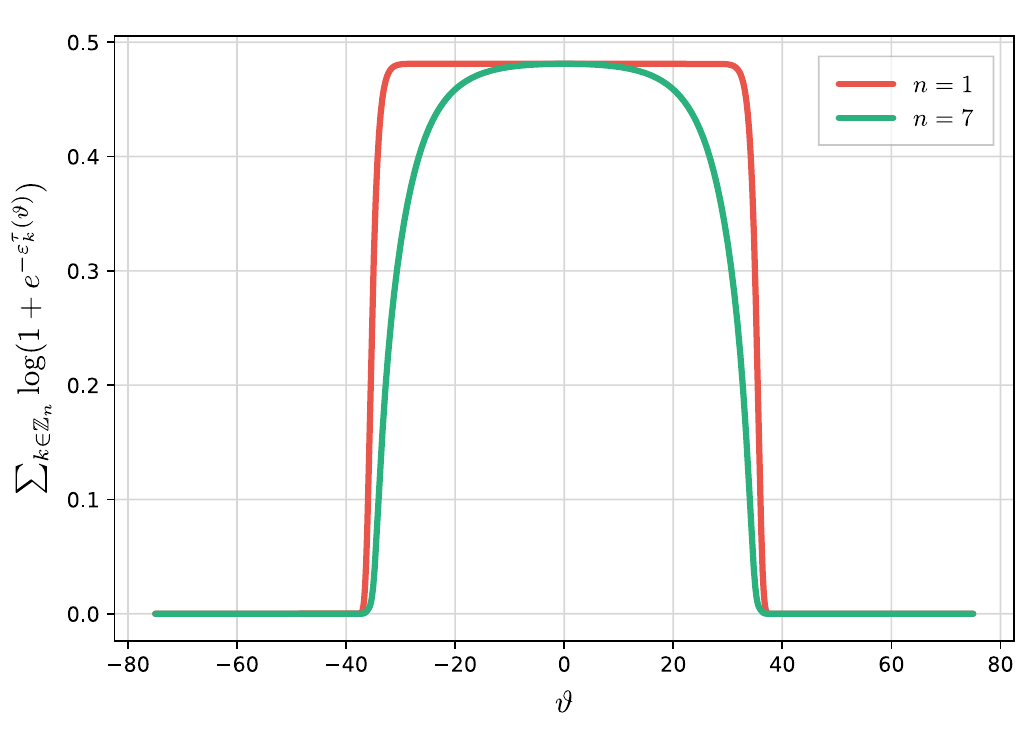}
        \caption{Central plateau region.}
        \label{fig:1a_cc}
    \end{subfigure}
    \hfill
    \begin{subfigure}[b]{0.495\textwidth}
        \centering
        \includegraphics[width=\textwidth]{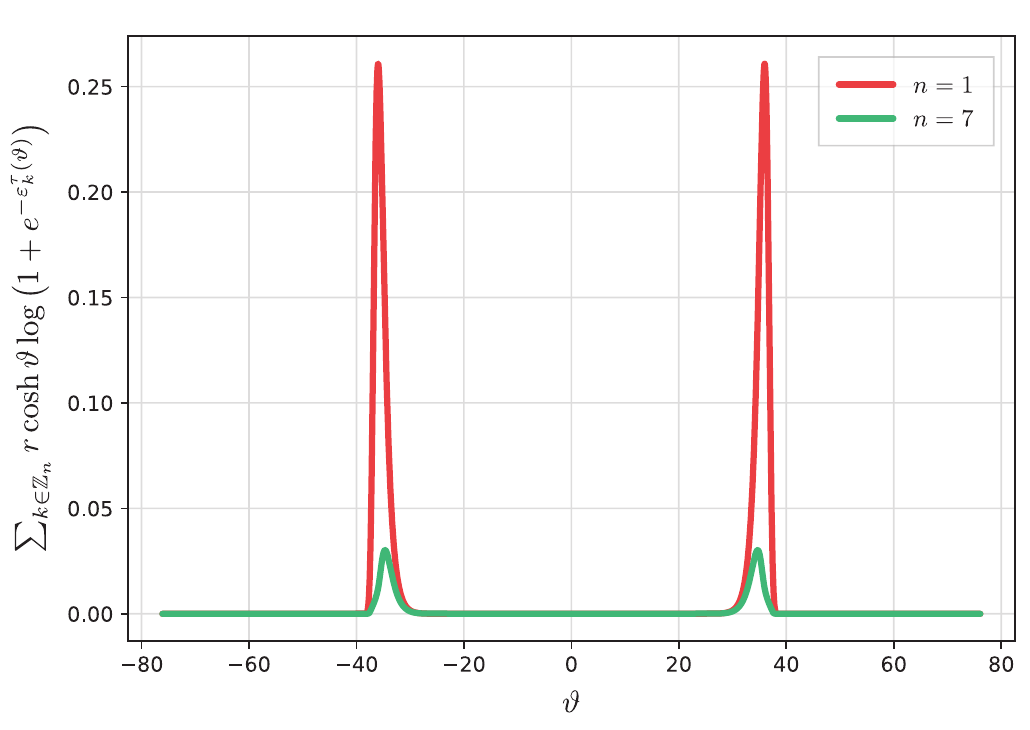}
        \caption{Integrand in $c_{\text{eff}}^{\tau}$ in the UV.}
        \label{fig:1b_cc}
    \end{subfigure}
    \caption{On the left (see Figure \ref{fig:1a_cc}), the sum of the density functions $\log \big(1+e^{-\varepsilon_k^\tau(\vartheta)}\big)$ for the scaling Lee–Yang model (formally, $n=1$) and for its twisted analogue with $n=7$, evaluated at $\tau=2$. Both curves are plotted at $r=10^{-15}$. After summing over the $k$-components, the height of the plateau is unchanged, but its width is reduced by a factor $1 / n$. Since the plateau must still reach the same height while fitting into a region $n$ times narrower, the average growth toward the flat UV region becomes $n$ times slower. Multiplication by $r\cosh\vartheta$ suppresses the contribution of the central region, and isolates the UV edges around $\vartheta = \pm \log r$. As a result, the effective area under the curve plotted on the right (see Figure \ref{fig:1b_cc}) -- corresponding to the integrand in $c_\text{eff}^\tau$ -- is effectively reduced by a factor $1 / n$, producing the suppression in the UV scaling function.}
    \label{fig:11}
\end{figure}

\noindent \textbf{Numerical remarks}. We solved the twisted TBA equations numerically for several values of $n$ and for different twisted sectors. Representative results are displayed in Figures~\ref{ff9}–\ref{fig:11}, where we plot the corresponding scaling functions $c^{\tau}_\text{eff}(r)$, which illustrate how the UV value is approached, as well as other quantities of interest in the ultraviolet regime. A general feature of the computation is that UV convergence in the twisted sectors is \textit{significantly slower} than in the untwisted case. When $\tau = 0$, values as large as $r = 10^{-15}$ are already sufficient to resolve the UV central charge with $\sim 14$ correct digits in double precision. Moreover, we find
\begin{equation}
    c^{0}_\text{eff}(r,\alpha=0)=\frac{2n}{5}+\mathcal{O}(r^{y_0})\,,
\end{equation}
where the exponent $y_0$ is related to the periodicity of the $T_1$ Y-system. In contrast, in genuinely twisted sectors ($\tau = 1, \dots, n-1$), the UV expansion acquires a dominant correction of the form
\begin{equation}
    c^{\tau}_\text{eff}(r,\alpha=0)=\frac{2}{5n}+\mathcal{O}(r^{y_0/n})\,.
\end{equation}
Therefore, the leading corrections are parametrically larger than in the untwisted theory. This delays the onset of the UV plateau and makes the numerics much more sensitive to round-off errors.

An additional complication, present for all values of $n$ we tested, is that in some twisted sectors the TBA pseudoenergies develop a pair of complex singularities that move towards the real axis. As the singularities approach the contour, the TBA iterations become unstable, and the algorithm cannot be reliably continued to smaller values of~$r$. A similar behaviour is observed in the $n=5$ graded Lee–Yang model, where the instability appears at significantly larger values of $r$, in a region where $c_\text{eff}(r)$ is still negative and there is no clear numerical indication of a change in concavity towards the UV plateau. A particularly interesting case is $n=7$. For $\tau = 1,6$, the singularities again approach the contour as in Figure \ref{ff11}, but in this case, the available data are sufficiently deep in the UV regime to allow for a numerical extrapolation, as in Figure \ref{ff12}. Using rational interpolation/extrapolation techniques, we find that the scaling function converges to a UV value compatible with \eqref{result_uv_LY}, as expected from the analytic prediction for the graded TBA. The remaining twisted sectors for $n=7$ are not affected by singularities and converge smoothly to the same UV value without the need for extrapolation. Understanding the behaviour of these singularities in detail -- and possibly modifying the integral equations to track them across the contour -- would require a substantial reformulation of our numerical approach, which we leave to future work.

\subsection{Potential relation with cyclic orbifolds}
An intuitive way to understand the emergence of twisted sectors in graded models is to recall that the reparametrisations of rapidity space introduced in Section \ref{Majorana:SEC} can be viewed as multi-sheeted coverings of the rapidity plane. In the free-fermion benchmark, these coverings unfold a single copy of the Ising model into $n$ interacting replicas, cyclically connected along branch cuts in the complex rapidity plane. This geometric picture closely parallels the construction of orbifold conformal field theories, where multiple copies of a parent theory are glued along branch points, and twisted boundary conditions encode the non-trivial monodromy between sheets. Similar multi-copy and branched-manifold interpretations have been emphasised in \cite{Cardy:2007mb} in the context of entanglement entropy and twist fields. From this perspective, the graded reparametrisation can be seen as a field-theoretic realisation of the same idea: the $\mathbb{Z}_n$ grading acts as a discrete rotation in rapidity space, generating a cyclic identification among $n$ replicas of the parent theory. Each branch of the covering corresponds to one copy, while the twisted boundary conditions between adjacent sheets define the twisted sectors of the theory.
\begin{figure}
    \centering
    \begin{subfigure}[b]{0.495\textwidth}
        \centering
        \includegraphics[width=\textwidth]{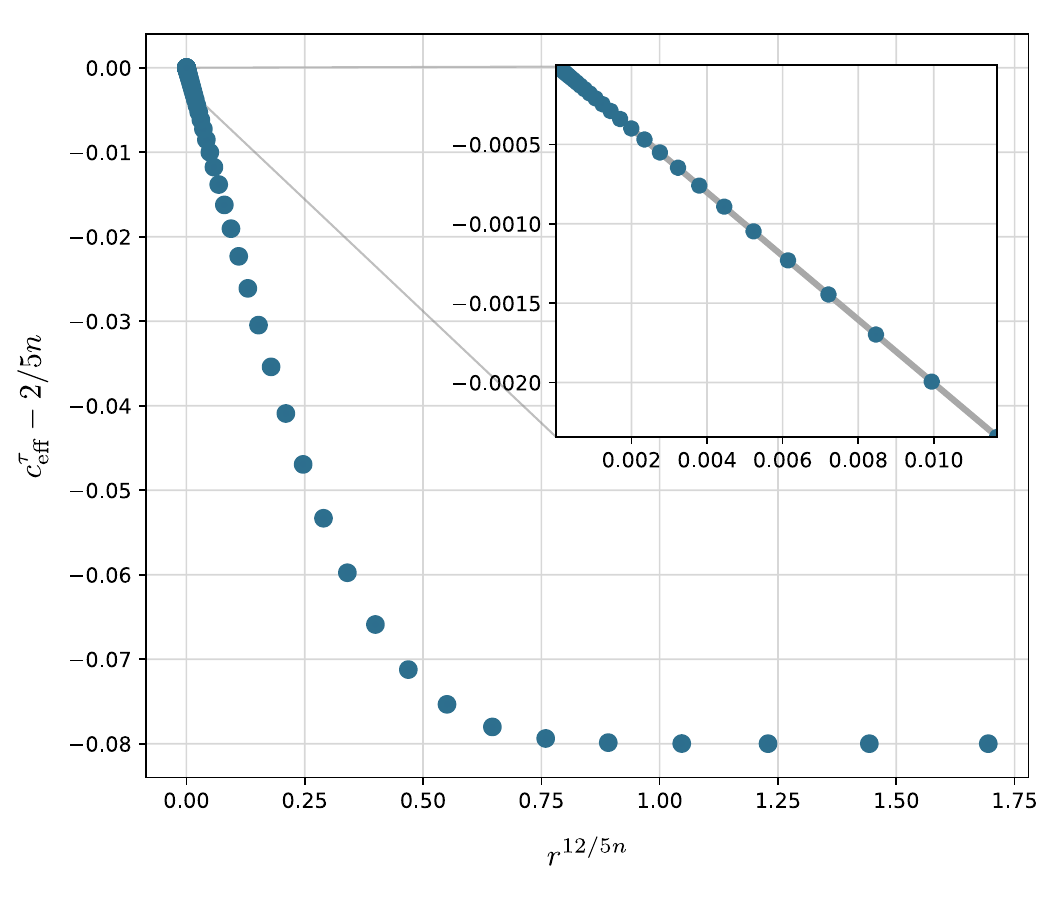}
        \caption{$n=5$.}
        \label{fig:1aa}
    \end{subfigure}
    \hfill
    \begin{subfigure}[b]{0.495\textwidth}
        \centering
        \includegraphics[width=\textwidth]{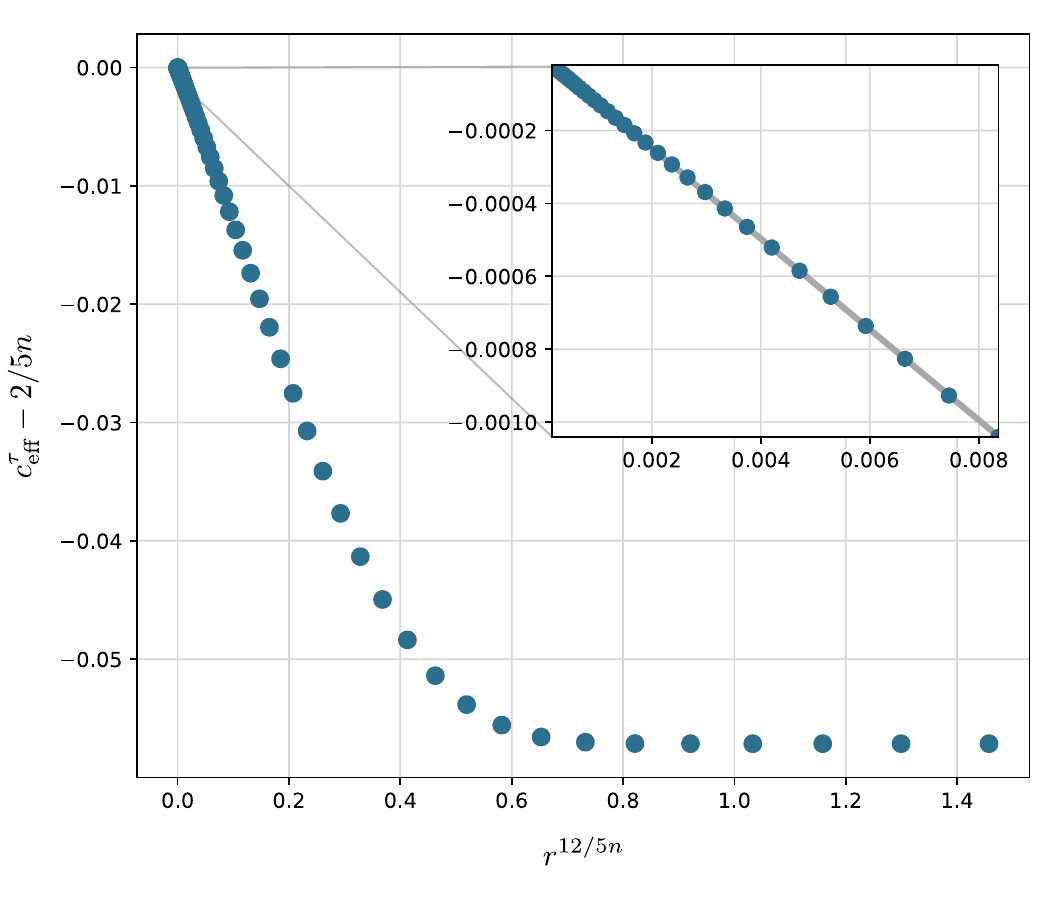}
        \caption{$n=7$.}
        \label{figg:1b}
    \end{subfigure}
    \caption{Scaling analysis of the twisted TBA equations for the twisted Lee–Yang model at $\tau  = 2$. The effective central charge is used to extract the scaling dimension of the perturbing operator in each sector.}
    \label{fig:both_scaling}
\end{figure}
In both free and interacting settings, the TBA analysis supports this interpretation, at least when the energy term acts as the sole source (corresponding to the case $\alpha=0$ in our analysis). For $\tau=0$, corresponding to the untwisted sector, the effective central charge converges to $n$ times that of the parent theory, as expected for $n$ decoupled replicas. In contrast, each twisted sector flows to $1 / n$ of the parent central charge in the ultraviolet limit, matching the behaviour of cyclic orbifolds where the vacuum of a twisted sector corresponds to the insertion of a branch-point twist field of conformal dimension
\begin{equation}
    \Delta_{\mathrm{tw}}=\frac{c_\text{eff}}{24}\left(n-\frac{1}{n}\right)\,.
\end{equation}
From a renormalisation-group-like perspective, the TBA naturally arises as a trajectory connecting a UV fixed point to a massive infrared theory, generated by perturbing the CFT with a relevant operator.  Consider a CFT on the cylinder, deformed by a scalar primary operator $\varphi$ of dimension $\Delta_\tau$. This perturbation introduces a finite correlation length and a mass scale in the partition function. Doing perturbation theory around the UV conformal fixed point, one finds that when $\alpha = 0$, the twisted scaling function $c_\text{eff}^{\tau}(r)$ admits the expansion
\begin{equation}\label{c_expansion_tw}
    c_\text{eff}^{\tau}(r,\alpha=0) - c_\text{eff}^{\tau,\text{UV}}(\alpha=0) \sim B(r) + \sum_{\ell=1}^\infty C^\tau_\ell r^{\ell y_\tau} \,,
\end{equation}
with $y_\tau =2-2\Delta_\tau$. The coefficients $C^\tau_\ell $ are determined by integrated correlators on the complex plane. The term $B(r)$ is known as the \textit{bulk term}. It is either proportional to $r^2$ or to $r^2 \log r$  (depending on the model), and its presence in the small-$r$ expansion is such that $c_\text{eff}^{\tau}(r,\alpha=0)/r$ has a finite large-$r$ limit. Moreover, we find that the bulk contribution to the twisted scaling function in the $\mathbb{Z}_n$-graded theory is simply $n$ times the bulk term of the original (ungraded) model. Expression \eqref{c_expansion_tw} suggests a practical method to compute the scaling dimension of the perturbing operator: since corrections appear as powers of $r^{y_\tau}$, fitting the small-$r$ behaviour of $c_{\text {eff }}^{\tau}(r,\alpha=0)$ allows extracting $\Delta_\tau$ from TBA numerics. For $\tau=0$, the perturbing operator retains the same scaling dimension as in the conventional, ungraded theory. In particular, one has $y_0 = 12/5$ in the scaling Lee–Yang model. In this case, $\Delta_0 = \Delta$ is related to the periodicity of the Y-system as in \eqref{period_delta}. Conversely, in genuinely twisted sectors, we find that $y_\tau = y_0/n$, and the corresponding scaling dimension is:
\begin{equation}
    \Delta_\tau = \frac{\Delta_0}{n} + \frac{n-1}{n}\,.
\label{Dtau}
\end{equation}
For $n=5$ and $n=7$, the near-ultraviolet scaling behaviour of the effective central charge is shown in Figure \ref{fig:both_scaling}. Because the twisted configuration corresponds to inserting a defect operator along the spatial circle, one should also account for an additional contribution coming from the scaling dimension $\Delta_\text{tw}$ of that operator. The appearance of fractional scaling is compatible with the existence of fractional Virasoro modes in orbifold conformal field theories, reinforcing the geometric interpretation discussed above.

Despite these promising preliminary results, a full identification between the graded construction and the cyclic orbifold framework will require several additional ingredients. First, a detailed characterisation of the perturbation away from criticality is still missing -- in particular, the precise form of the perturbing operator, its fusion rules, and its coupling to the twisted sectors. Complementary evidence could be obtained from a \textit{Truncated Conformal Space Approach} (TCSA) analysis \cite{Yurov:1989yu, Yurov:1991my}, allowing a direct comparison between the finite-size spectrum of the graded theory and that of the corresponding orbifold. Moreover, a systematic exploration of the remaining excited-state TBA equations is necessary to complete the spectrum. Finally, a more transparent interpretation of the GGE would shed light on how conserved charges reorganise across sectors and whether they admit an orbifold counterpart. \\
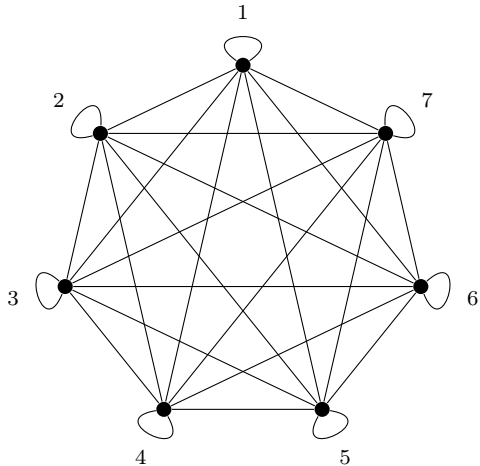
\begin{figure}
\centering
\begin{tikzpicture}[every node/.style={font=\small}]
  %===== styles =====
  \tikzset{pt/.style={circle,fill=black,inner sep=2pt}, lab/.style={font=\scriptsize}}

  \def\r{2.4}       % radius of the heptagon
  \def\loopr{0.85}  % larger self-loop size (tadpole radius)

  %===== place 7 nodes, perfectly symmetric =====
  \foreach \i in {1,...,7}{
    \pgfmathsetmacro{\ang}{90 + (\i-1)*360/7}
    \node[pt] (v\i) at (\ang:\r) {};
    \node[lab] at (\ang:\r+0.70) {\i};  % label distance
  }

  %===== draw all edges pairwise (complete graph K7) =====
  \foreach \i in {1,...,7}{
    \foreach \j in {1,...,7}{
      \ifnum\i<\j
        \draw (v\i) -- (v\j);
      \fi
    }
  }

  %===== draw larger self-loops (tadpoles) =====
  \foreach \i in {1,...,7}{
    \pgfmathsetmacro{\ang}{90 + (\i-1)*360/7}
    \draw (v\i)
      .. controls +({\ang+55}:\loopr) and +({\ang-55}:\loopr) ..
      (v\i);
  }

\end{tikzpicture}

\caption{Infrared vacuum structure for $n=7$ at $\alpha = 0$. The graded, twisted TBA produces $n$ degenerate infrared vacua labelled by $\tau \in \mathbb{Z}_n$. The lightest excitations are kinks interpolating between any ordered pair of vacua, resulting in a fully connected interpolation graph whose adjacency matrix has all entries equal to one. The spectrum of this matrix contains a single non-zero eigenvalue, $n$, corresponding to the symmetric infrared sector selected by the TBA, while the remaining $n-1$ directions are suppressed at large $r$.}
\label{kinks:figure}
\end{figure}

\noindent \textbf{Infrared vacua and interpolating kinks}. For $\alpha=0$, the graded TBA equations admit a particularly transparent infrared expansion. In the large-$r$ regime, the effective central charge in the $\tau$-twisted sector reduces to
\begin{equation}
   c_\text{eff}^{\tau}(r,\alpha=0)\simeq \frac{6 n }{\pi^2} \delta_{\tau,0} \sum_{a \in G} \hat{m}_a r  K_1(\hat{m}_a r)\,.
\end{equation}
The Kronecker symbol $\delta_{\tau, 0}$ indicates that only the untwisted component survives at leading order, while all other twisted sectors are exponentially suppressed. This feature allows for a direct physical interpretation of the infrared limit, where the grading index $\tau \in \mathbb{Z}_n$ labels $n$ distinct infrared vacua \cite{Dorey:1996he}. Their degeneracy follows from the fact that the TBA retains only a single infrared contribution: the theory effectively projects onto the completely symmetric combination of the $\mathbb{Z}_n$-graded pseudoenergies, and the lightest excitations are kinks interpolating between these vacua, as we depicted in Figure \ref{kinks:figure}. Since every vacuum can be reached from any other vacuum with the same infrared weight, the adjacency (or interpolation) matrix that describes kink connectivity is simply the $n \times n$ matrix with all entries equal to one. Its spectrum consists of a single non-zero eigenvalue, equal to $n$, and $n-1$ vanishing eigenvalues. This matches precisely the infrared structure extracted from the TBA: one surviving contribution and $n-1$ directions that become irrelevant at large $r$. The kink interpretation above relies on a natural choice of kink basis, suggested by the graded scattering data. However, the TBA analysis alone does not exclude the possibility that a different basis of kink states could lead to the same infrared vacuum structure and the same TBA equations, while simultaneously restoring properties that seem obscured in our current parametrisation, such as physical unitarity or parity symmetry of the S-matrix.
\section{Conclusions and future directions}
In this work, we showed how  $\mathbb{Z}_n$-graded integrable QFTs can be realised by exploiting the analytic structure of the S-matrix. A set of conformal maps reorganises the rapidity plane into an $n$-sheeted domain, and evaluating the same scattering amplitude on different sheets produces a cyclic family of amplitudes related by analytic continuation. Embedding the construction in a generalised Gibbs ensemble ensures the correct relativistic scaling and yields a consistent graded TBA with new Y-systems. In special cases, these coincide with the deformed functional relations arising from the monodromy analysis of the cubic oscillator in the ODE/IM correspondence.

Graded free theories serve as an exactly solvable setting in which the construction can be explored in full detail. In these models, the graded lift of the TBA equations yields closed scaling functions with controlled UV and IR limits, and already captures non-perturbative behaviour, such as level crossings, in appropriate regimes.

For interacting QFTs, we built $\mathbb{Z}_n$-graded S-matrices by pulling back minimal ADET amplitudes. Braiding unitarity remains intact, while crossing symmetry is extended to a transformation that pairs charge conjugation with a cyclic shift of the sheet index. The bootstrap closes via a graded cyclic identity, preserves the fusion geometry, and selects which graded components host physical bound-state poles. The residues scale linearly with $n$, with a definite sign pattern emerging from the graded structure.

Large-rapidity asymptotics reveal a tower of fractional effective spins coexisting with the original integer-spin charges. This motivates fractional-spin CDD factors that preserve factorisation and crossing, and that become trivial when all $\mathbb{Z}_n$ sectors are combined.

By combining the reparametrisation of rapidity space with the Generalised Gibbs Ensemble, we obtain a graded Thermodynamic Bethe Ansatz for graded, interacting QFTs, consisting of $n$ coupled non-linear integral equations. The resulting graded Y-system generalises the ordinary functional relations. The graded Lee–Yang model provided a concrete test case.

Chemical potentials split the $\mathbb{Z}_n$ sectors and generate genuinely twisted Y-systems. In the UV, untwisted sectors flow to $n$ replicas of the parent theory, while primitive twists reduce the effective central charge by a universal $1/n$, in agreement with the cyclic orbifold picture via branch-point twist fields.

Finally, both analytical and numerical TBA analysis validate the framework and provide access to finite-size data. 

\vspace{0.5cm}
\noindent\textbf{Outlook}. There are several natural directions in which the present work can be extended. Here, we list a few natural directions for future investigation.
\begin{itemize}
    \item Generalise the graded TBA construction to a wider class of integrable theories, such as the sinh-/sine-Gordon \cite{Tateo:1995sz} and the non-diagonal ADET/RSOS models \cite{Ahn_1998}. See \cite{Hegedus:2025yhr} for recent progress on NLIEs and generalised Gibbs ensembles in the sine-Gordon model. In the same spirit, it would be interesting to explore boundary scattering, graded reflection matrices, and the associated boundary TBAs \cite{GHOSHAL_1994, Fendley_1994, Dorey2004IntegrableQF, Jiang:2021jbg, Brizio:2024doe}.
    \item Understand where the grading comes from at a more fundamental level. One promising route is to look for a Lagrangian realisation, for example, in affine Toda theories, where discrete symmetries or topological charges could naturally generate the graded sectors \cite{Fring:2005vg}. A complementary perspective could also come from the $4d$ Chern–Simons construction of integrable models, where Toda systems emerge from line and surface defects and appropriate boundary conditions \cite{Costello:2017dso, Costello:2018gyb, Costello:2019tri} -- see also \cite{Yamazaki:2025yan} for a brief introduction. Moreover, the framework of \cite{Sakamoto:2025hwi} could provide a complementary perspective on the emergence of CDD deformations at fractional spin.
    \item Understand the structure of correlators in the graded setting and to investigate possible connections with the results of \cite{Cardy:2007mb}, as well as \cite{Fendley:1991xn}. Establishing this link would clarify how the orbifold interpretation extends beyond the spectrum and into correlation functions. The same structure suggests a connection to RG interfaces or topological defects \cite{Gaiotto:2012np}, which still needs to be explored.
    \item Systematically analyse fractional-spin CDD deformations.  It remains an open problem to study their analyticity, locality, and UV behaviours, as well as to understand how they affect TBA/GGE flows. 
    \item Make the link with the ODE/IM correspondence more explicit and universal, obtaining graded NLIE/Y-systems directly from monodromy data in differential equations \cite{Bridgeland_2022}, and comparing them with the graded constructions introduced here. Within this framework, it seems natural to explore whether fractional-spin CDD deformations can be characterised similarly to the $T\overline{T}$ case as discussed in \cite{Aramini:2022wbn}. It would also be important to understand whether the graded construction also fits within the \textit{massive} ODE/IM correspondence \cite{Lukyanov:2010rn, Dorey:2012bx, Ito:2013aea, Lukyanov:2013wra, Bazhanov:2013cua, Adamopoulou_2014, Ito:2015nla, Negro:2016yuu}. Work on related topics is in preparation with Hongfei Shu. See also \cite{Suzuki:2000fc} and \cite{Ito:2025sgq} for possibly related results.
    \item  Investigate possible connections with cyclotomic Gaudin models, where a natural $\mathbb{Z}_n$ symmetry also appears \cite{Lacroix:2016mpg}. This may offer an alternative algebraic interpretation of the graded structure.
    \item Explore whether the graded functional relations may offer an alternative perspective on the {Dubrovin conjecture} regarding the pole-free sector of the tritronquée solution Painlevé I, which was first proven in \cite{Costin_2014}.
    \item Clarify the geometric structure underlying  $T\overline{T}$-like deformations at fractional-spin \cite{Conti:2018tca, Cardy:2018sdv, Tolley:2019nmm, Caputa:2020lpa}. This may also point toward classical extensions to higher dimensions and their gravitational counterparts \cite{Conti:2018jho, Taylor:2018xcy, Bonelli:2018kik, Conti:2022egv, Ferko:2023wyi, Morone:2024ffm, Babaei-Aghbolagh:2024hti, Ferko:2024yua}.
    \item Compare our graded construction with integrable orbifolds and twisted sectors in the holographic setup (see, for example, \cite{Frolov:2023pjw, Fabri:2025rok}). 
\item 
Finally, a natural direction for future work is a systematic Truncated Conformal Space Approach (TCSA) study of the graded theories and of their possible cyclic orbifold counterparts, aimed at testing and clarifying any potential connection between these two descriptions. In particular, we expect that achieving a quantitative match with the numerical TBA spectra in the twisted sectors will require switching on a perturbing operator with the scaling dimension given by equation (\ref{Dtau}), which should act non-trivially precisely in those sectors.
\end{itemize}

\noindent \textbf{Acknowledgements}. We are grateful to Patrick Dorey, Sergei Frolov, Katsushi Ito, Stefano Negro, Alessandro Sfondrini, Hongfei Shu, István M. Szécsényi, Benoît Vicedo, Gerard Watts, and Robert Weston for their helpful discussions, and to Alessandro Cortassa for his involvement in the early stages of this project.
This work was partially supported by the INFN project SFT and by the PRIN project No. 2022ABPBEY, with the title “Understanding Quantum Field Theory Through its Deformations.” This work was partially supported by the INFN project ST$\&$FI and by the CARIPARO Foundation Grant under grant No. 68079. This work benefited from networking activities supported by COST Action CA22113, ``Fundamental challenges in theoretical physics (Theory-Challenges).'' 
NB and RT participate in the project HORIZON-MSCA-2023-SE-01-101182937-HeI. NB, TM, and RT acknowledge fruitful discussions, partially related to this project, with Nicola Baglioni, Daniele Bielli, Christian Ferko, Jue Hou, Michele Galli, Gabriele Tartaglino-Mazzucchelli, and all participants of the workshop “Deformations of Quantum Field and Gravity Theories,” held at the University of Queensland (January 30 – February 6, 2025). We also thank the University of Queensland for the financial support provided through the Ethel Raybould Visiting Fellowship, which facilitated participation in this meeting during the early stages of the project. NP and RT also thank the organisers of the workshop “Higher-$d$ Integrability” held in Favignana, Italy (June 3–13, 2025) for their invitation, financial support, and stimulating atmosphere during the final stages of this work. RT is also grateful to the organisers of ``ICFT 2025 – UK Meetings on Integrable and Conformal Field Theory'' (June 19-20, 2025, King’s College, London) for their invitation and financial support. 
Finally, TM and RT would also like to express their gratitude to Ningbo University for the warm hospitality. Part of this work was completed while attending the conference ``Exploring $T\overline{T}$ Deformations, Integrable Models, and String Theory'' (October 12–17, 2025), whose stimulating scientific environment greatly contributed to the progress of this research.

\appendix
\section{The graded Lee–Yang model and the ODE/IM correspondence}\label{masoero_appendix}
The analysis of \cite{Masoero:2010is} focuses on the monodromy problem for the cubic anharmonic oscillator,
\begin{equation}
    -\dv[2]{\Psi(x)}{x} + V(x)\Psi(x) = E\Psi(x)\,,\quad V(x) = 4x^3 - \alpha x\,,
\end{equation}
where $\alpha \in \mathbb{R}$ is a deformation parameter that tilts the potential away from the symmetric cubic form, while $E \in \mathbb{C}$ plays the role of a spectral parameter. The coefficient 4 in front of the cubic term is a normalisation choice: by rescaling $x$, one may always arrange the leading behaviour at infinity into this form, which simplifies the asymptotic analysis. A natural way to probe the equation at infinity is through the WKB (Wentzel–Kramers–Brillouin) approximation. One writes the solution in exponential form,
\begin{equation}
    \Psi_\pm(x) \simeq Q(x)^{-1/4} \exp(\pm\delta(x))\,,\quad  Q(x) = V(x)-E\,,
\end{equation}
where the phase $ \delta(x)$ is governed by the integral 
\begin{equation}
    \delta(x) =\int^x\dd y \sqrt{Q(y)}\,.
\end{equation}
At large $|x|$, the cubic term dominates, and this phase grows like $x^{5/2}$. The rays where this quantity is purely imaginary split the plane into five angular regions, each of opening angle $2\pi/5$. In each such region -- called a Stokes sector -- one can define a canonical WKB solution $\Psi_k(x)$ that decays exponentially as $|x|$ grows towards infinity. As one moves across a boundary between sectors, their behaviour is described by Stokes multipliers, constants that encode how a solution in sector $k$ is expressed as a linear combination of solutions in neighbouring sectors. 
In the case $\alpha=0$, the cubic potential is invariant under a fivefold rotation, and the multipliers repeat in a simple cyclic pattern, so that the consistency conditions reduce to a single functional relation. With $\alpha \neq 0$, however, the pattern is broken, and one might expect the global structure to become complicated. Once the WKB analysis has singled out the five Stokes sectors, the question becomes: how do we encode the global analytic behaviour of solutions as one moves around infinity? The traditional way is through Stokes multipliers, but these depend on the normalisation of each sectorial solution and are not the most natural language for capturing the underlying geometry. An alternative idea is to replace them with projective data that are independent of arbitrary choices. To do this, recall that the cubic oscillator equation is a second-order linear ODE. Its solution space is two-dimensional: any solution can be written as a linear combination of two linearly independent ones. Let us fix such a pair, $\psi_1(x)$ and $\psi_2(x)$. The quantity
\begin{equation}
    F(x) = \frac{\psi_1(x)}{\psi_2(x)}
\end{equation}
is then a well-defined function taking values in the Riemann sphere. Different choices of $\psi_{1,2}$ yield different ratios, but only up to Möbius transformations:
\begin{equation}
F(x) \mapsto \frac{A F(x)+B}{C F(x)+D}\,,\quad
{\setstretch{1}\begin{pmatrix} A & B \\ C & D \end{pmatrix}}\in \mathrm{SL}(2,\mathbb{C})\,.
\end{equation}
This means that while $F(x)$ itself is basis-dependent, its projective geometry is intrinsic to the differential equation. As one lets $|x |\to \infty$ within a given Stokes sector, the exponential hierarchy of the WKB solutions forces one of them to dominate, and the ratio $F(x)$ converges to a definite limit in the $k$-th sector. We denote this asymptotic value by $w_k$. In this way, the differential equation is associated with five points $\left\{w_0, \ldots, w_4\right\}$ on the Riemann sphere, arranged cyclically around infinity. These five points contain the same information as the Stokes multipliers, but in a more geometric and Möbius-invariant form. To make this invariance explicit, one introduces cross-ratios, 
\begin{equation}
    R_k=\frac{\left(w_{k+1}-w_{k-1}\right)\left(w_{k+2}-w_{k-2}\right)}{\left(w_{k+1}-w_{k-2}\right)\left(w_{k+2}-w_{k-1}\right)}\,,
\end{equation}
that are invariant under the $\mathrm{SL}(2,\mathbb{C})$ group, and thus provide canonical coordinates on the monodromy data. The remarkable discovery is that these invariants are not free: the fivefold arrangement of the asymptotic values imposes the consistency condition
\begin{equation}
    R_{k+2}R_{k-2} = 1-R_k\,,\quad k\in\mathbb{Z}_5\,.
\end{equation}
At this point, the connection with integrable structures begins to emerge. At this stage, each $R_k$ is just a number determined by the data of the ODE (namely, the deformation parameter $a$ and the spectral parameter $E$). However, they are all expressible in terms of a single generating function. To explain this, recall that the leading cubic term in the potential is invariant under a five-fold rotation of the coordinate $x$. This induces a $\mathbb{Z}_5$ action on the pair $(\alpha, E)$, under which the asymptotic values $w_k$ (and hence the cross-ratios) are permuted. Concretely, one finds the covariance relation:
\begin{equation}
    R_k(\alpha,E) = R_0(e^{2\pi ik/5} \alpha,e^{-2\pi ik/5} E)\,,
\end{equation}
which shows that all five cross-ratios can be generated from a single one, $R_0$, evaluated at rotated arguments. In other words, the $\mathbb{Z}_5$ symmetry collapses the full set $\left\{R_k\right\}_{k\in\mathbb{Z}_5}$ onto the orbit of $R_0$. To arrive at a structure reminiscent of integrable Y-systems, one introduces the functions
\begin{equation}
    Y_k(\vartheta) = -R_0(e^{-2\pi i k/5} \alpha, e^{6\vartheta/5})\,,
\end{equation}
where we have parametrised the spectral variable exponentially as $E=e^{6\vartheta/5}$. The reason for this choice is that shifts in $\vartheta$ by multiples of $i \pi / 3$ correspond precisely to analytic continuations across the Stokes boundaries. With this definition, the algebraic constraints satisfied by the cross-ratios reorganise into the functional relations
\begin{equation}\label{deformed_Y:2}
    Y_{k+1}(\vartheta^+)Y_{k-1}(\vartheta^-) = 1+ Y_k(\vartheta)\,,
\end{equation}
where $\vartheta^\pm = \vartheta \pm i\pi/3$, which is the deformed Y-system deduced in \cite{Masoero:2010is}. In the symmetric case $\alpha=0$, all five functions $Y_k$ coincide, and the system reduces to a single relation, the familiar Y-system of the scaling Lee–Yang model:
\begin{equation}\label{LY_Y:2}
     Y(\vartheta^+)Y(\vartheta^-) = 1+ Y(\vartheta) \,.
\end{equation}
Equation \eqref{deformed_Y:2} can be recast into a set of coupled, non-linear integral equations, which correspond to the UV limit of a TBA system. The same set of equations can be recovered by taking the $r\to 0$ limit of \eqref{lee_yang_graded_tba}, which constitutes a massive/UV-regulated extension of Masoero’s setup for the $n=5$ graded Lee–Yang model. In particular, one recognises that the explicit convolution kernels presented in \cite{Masoero:2010is} correspond to the graded Lee–Yang kernels in the discrete Fourier basis \eqref{f.c.}.

\section{Root of unity average and modified Bessel functions}\label{app:root}
In this appendix, we derive a useful closed-form expression for the discrete average:
\begin{equation}
    X^{(\tau)}_n(u, z)=\frac{1}{n} \sum_{k\in\mathbb{Z}_n} \exp \left[-z \cosh \left(u+\frac{2 \pi i k}{n}\right)+\frac{2\pi i \tau k}{n}\right]\,,\quad z>0\,,\quad n \in \mathbb{N}\,,
\end{equation}
which appears in the expansion of the effective central charge for the graded Ising model. A standard starting point is the generating series for the modified Bessel functions of the first kind,
\begin{equation}
    \sum_{m=-\infty}^{\infty} I_m(z) y^m = \exp \left(\frac{yz}{2}+\frac{z}{2y}\!\right)\,.
\end{equation}
Choosing $y=e^{u+iv}$, the above equation becomes:
\begin{equation}
    \exp (z \cosh (u+i v))=\sum_{m=-\infty}^{\infty} I_m(z) e^{m u} e^{i m v}\,.
\end{equation}
If we now replace $z \mapsto-z$, and recall the simple sign relation $I_m(-z)=(-1)^m I_m(z)$, we obtain the useful formula:
\begin{equation}\label{eq1}
\exp (-z \cosh (u+i v))=\sum_{m=-\infty}^{\infty}(-1)^m I_m(z) e^{m u} e^{i m v}\,.
\end{equation}
We now substitute $v = 2 \pi i k/n$, and average over $k \in \mathbb{Z}_n$. Orthogonality of the roots of unity yields:
\begin{equation}\label{eq2}
    \frac{1}{n} \sum_{k\in\mathbb{Z}_n} e^{\frac{2 \pi i (m+\tau) k}{n}}= \begin{cases}1 & \text { if }m-\tau\text{ divides } n\,,\\ 0 & \text { otherwise\,.}\end{cases}
\end{equation}
Applying \eqref{eq2} to equation \eqref{eq1}, only terms with $m -\tau = jn$ survive. Using the symmetry $I_{-m}(z)=I_m(z)$, we can regroup positive and negative modes to obtain the manifestly real form:
\begin{equation}\label{t=0}
    X^{(\tau \equiv0 \bmod n)}_n(u, z)=I_0(z)+2 \sum_{j=1}^\infty(-1)^{n j} I_{n j}(z) \cosh (n j u)\,,
\end{equation}
and
\begin{equation}\label{t_non_0}
  X^{(\tau \equiv 1, \dots, n-1\bmod n)}_n(u, z)=2 \sum_{j=1}^\infty(-1)^{n \ell-\tau} I_{n j-\tau}(z) \cosh ((n j-\tau) u)\,.
\end{equation}

\bibliographystyle{JHEP}
\bibliography{Biblio2.bib}

\end{document}